# Roadmap on structured waves

**Konstantin Y. Bliokh[1], Ebrahim Karimi[2], Miles J. Padgett[3], Miguel A. Alonso[4,5], Mark R. Dennis[6], Angela Dudley[7], Andrew Forbes[7], Sina Zahedpour[8], Scott W. Hancock[8], Howard M. Milchberg[8], Stefan Rotter[9], Franco Nori[1,10], Şahin K. Özdemir[11], Nicholas Bender[12], Hui Cao[12], Paul B. Corkum[13], Carlos Hernández-García[14], Haoran Ren[15], Yuri Kivshar[16], Mário G. Silveirinha[17], Nader Engheta[18], Arno Rauschenbeutel[19], Philipp Schneeweiss[19], Jürgen Volz[19], Daniel Leykam[20], Daria A. Smirnova[16], Kexiu Rong[21], Bo Wang[21,22], Erez Hasman[21], Michela F. Picardi[23,24], Anatoly V. Zayats[23], Francisco J. Rodríguez-Fortuño[23], Chenwen Yang[25], Jie Ren[25], Alexander B. Khanikaev[26], Andrea Alù[27], Etienne Brasselet[28], Michael Shats[15], Jo Verbeeck[29], Peter Schattschneider[30], Dusan Sarenac[31], David G. Cory[31], Dmitry Pushin[31], Michael Birk[32], Alexey Gorlach[32], Ido Kaminer[32], Filippo Cardano[33], Lorenzo Marrucci[33], Mario Krenn[34], and Florian Marquardt[34]**

[1]Theoretical Quantum Physics Laboratory, RIKEN, Wako-shi, Saitama 351-0198, Japan

[2]Department of Physics, University of Ottawa, Ottawa, Ontario K1N 6N5, Canada

[3]Department of Physics and Astronomy, University of Glasgow, Glasgow G12 8QQ, Scotland

[4]Aix-Marseille Univ, CNRS, Centrale Marseille, Institut Fresnel, Marseille 13013, France

[5]The Institute of Optics, University of Rochester, Rochester, New York 14627, USA

[6]School of Physics and Astronomy, University of Birmingham, Birmingham, UK

[7]School of Physics, University of the Witwatersrand, Private Bag 3, Johannesburg 2050, South Africa

[8]Institute for Research in Electronics and Applied Physics, University of Maryland, College Park, Maryland 20742, USA

[9]Institute for Theoretical Physics, Vienna University of Technology (TU Wien), Vienna A-1040, Austria

[10]Department of Physics, University of Michigan, Ann Arbor, Michigan 48109-1040, USA

[11]Department of Engineering Science and Mechanics, and Materials Research Institute (MRI), The Pennsylvania State University, University Park, Pennsylvania 16802

[12]Department of Applied Physics, Yale University, New Haven, CT, USA

[13]Joint Attosecond Science Laboratory, National Research Council and University of Ottawa, Ottawa, Ontario, Canada

[14]Grupo de Investigación en Aplicaciones del Láser y Fotónica, Departamento de Física Aplicada, Universidad de Salamanca, E-37008 Salamanca, Spain

[15]MQ School of Physics and Astronomy, Monash University, Clayton, Victoria 3800, Australia

[16]Research School of Physics, Australian National University, Canberra, ACT 2601, Australia

[17]Instituto de Telecomunicações, Instituto Superior Técnico-University of Lisbon, Lisboa, Portugal

[18]Department of Electrical and Systems Engineering, University of Pennsylvania, Philadelphia, PA 19104, USA

[19]Department of Physics, Humboldt-Universität zu Berlin, 12489 Berlin, Germany

[20]Centre for Quantum Technologies, National University of Singapore, 3 Science Drive 2, 117543, Singapore, Singapore

[21]Russell Berrie Nanotechnology Institute, and Helen Diller Quantum Center, Technion–Israel Institute of Technology, Haifa 3200003, Israel

[22]Shanghai Jiao Tong University, 800 Dongchuan RD. Minhang District, Shanghai, China

[23]Department of Physics and London Centre for Nanotechnology, King's College London, Strand, London WC2R 2LS, United Kingdom

[24]ICFO – Institut de Ciencies Fotoniques, The Barcelona Institute of Science and Technology, Castelldefels (Barcelona) 08860, Spain

[25]Center for Phononics, School of Physics Science and Engineering, Tongji University, Shanghai 200092, China

[26]Department of Electrical Engineering, City College of the City University of New York, 160 Convent Avenue, New York, NY 10031, USA

[27]City University of New York, Photonics Initiative, Advanced Science Research Center, New York, USA

[28]Université de Bordeaux, CNRS, LOMA, UMR 5798, F-33400 Talence, France

[29]Electron Microscopy for Materials Science (EMAT), University of Antwerp, 2020 Antwerp, Belgium

[30]Vienna University of Technology (TU Wien), Vienna A-1040, Austria

[31]University of Waterloo, Waterloo, ON N2L3G1, Canada

[32]Department of Physics, Technion–Israel Institute of Technology, Haifa 3200003, Israel

[33]Dipartimento di Scienze Fisiche, Universita di Napoli 'Federico II', Complesso di Monte S Angelo, 80126 Napoli, Italy

[34]Max Planck Institute for the Science of Light, Erlangen, Germany

**Abstract**

Structured waves are ubiquitous for all areas of wave physics, both classical and quantum, where the wavefields are inhomogeneous and cannot be approximated by a single plane wave. Even the interference of two plane waves, or a single inhomogeneous (evanescent) wave, provides a number of nontrivial phenomena and additional functionalities as compared to a single plane wave. Complex wavefields with inhomogeneities in the amplitude, phase, and polarization, including topological structures and singularities, underpin modern nanooptics and photonics, yet they are equally important, e.g., for quantum matter waves, acoustics, water waves, etc. Structured waves are crucial in optical and electron microscopy, wave propagation and scattering, imaging, communications, quantum optics, topological and non-Hermitian wave systems, quantum condensed-matter systems, optomechanics, plasmonics and metamaterials, optical and acoustic manipulation, and so forth. This Roadmap is written collectively by prominent researchers and aims to survey the role of structured waves in various areas of wave physics. Providing background, current research, and anticipating future developments, it will be of interest to a wide cross-disciplinary audience.

**Contents**

# 1. Introduction

*Konstantin Y. Bliokh[1] and Ebrahim Karimi[2]*

[1]RIKEN
[2]University of Ottawa

As it often happens with phenomena and discoveries in science, it is difficult to indicate a single starting point for the study of structured waves. They have been in front of our eyes from the very beginning of history: as waves on the sea surface, scattered sunlight and rainbows in the sky, etc. In modern optics, seminal works by Nye and Berry [1–3], Baranova et al. [4], Soskin et al. [5, 6], and Allen et al. [7, 8] stimulated the development of the fields of "singular optics" and "optical angular momentum". The development of these closely related areas was surveyed five years ago in the "Roadmap on structured light" [9]. However, phase singularities (analysed by Nye and Berry for ultrasonic pulses) previously appeared in the context of quantum matter waves in pioneering papers by Fock [10], Dirac [11], Aharonov and Bohm [12], and Hirschfelder et al. [13, 14]. Furthermore, such singularities can be found in the 19th century maps of tidal ocean waves [15]. Also, the orbital angular momentum in localized wave states with vortices (described by Allen et al. for optical beams) appear in the form of quantum electron states in atoms or in an external magnetic field as described in textbooks on quantum mechanics [16]. This evidences that structured waves and their main properties are universal phenomena across various wave systems, independently of their nature.

In this roadmap, we aim to review recent achievements related to structured waves in optics, plasmonics, metamaterials, acoustics, electron and neutron optics, and quantum information. We tried to avoid overlaps with the earlier "Roadmap on structured light" [9]. Therefore, the main focus of this roadmap is shifted towards emerging directions which have been rapidly developing in the past five years and to phenomena involving structured waves in non-optical fields. The areas addressed in this roadmap include: non-Hermitian and topological wave systems; plasmonics, metasurfaces, and near-zero index materials; quantum information and artificial intelligence; light-matter interactions and waves in random media; electromagnetic, electron, neutron, acoustic and water waves. One can notice the rapidly growing interest in such directions as: temporal structures, including time-dependent media [17], space-time wavepackets [18], and spatiotemporal vortices [19–24]; novel functionalities involving complex waves in non-Hermitian [384-386,124], topological [122,387], and random [338,388] media; 3D topological polarization structures including polarization Mobius strips [25–29], skyrmions [94,29–35,389], and structured polychromatic fields with commensurable frequencies [36–40]; spin and polarization properties of sound [41–47], elastic [48–53], and water waves [54, 55]; structured neutron and atomic waves [369,375,378,374]; simulating complex quantum systems [56–58], high-dimensional communication [59–61], quantum cryptography [62–63], and sensing [64–67]. Unfortunately, not all invited authors were able to contribute to this project, and therefore, some areas were not covered. The most apparent omission is the absence of sections on structured quantum condensed-matter waves, including BEC, cold atoms, exciton-polaritons, and various quasiparticles in solids [68–75]. This topic deserves a special roadmap project.

This roadmap provides background, state-of-the-art, and perspectives for the interdisciplinary physics of structured waves. We hope that it will illuminate the universality of structured-wave phenomena across various areas of physics, highlight emergent directions involving structured waves, and thus stimulate further development of this exciting field.

## 2. Phase structure matters

*Miles J. Padgett*

University of Glasgow

**Status**

In a traditional sense, the importance of the spatial phase structure of light beams is inherent in any image projection system. However, our eyes are insensitive to phase, perceiving only intensity and hence for most user cases this image projection requires only the shaping of intensity. This shaping is easily accomplished by transmission through an appropriate transparency, or with modern technology in the form of a digital micromirror device normally associated with the projection of our slides in scientific talks. In contrast to this intensity structuring, much of the current research in shaped light is for those situations where the properties or application of the light arise from the phase structuring of the beam.

One well known example of where it is the spatial phase structure of light that is the defining feature is holography. In holography, the light scattered from the object is recorded by interfering this light with a spatially coherent, plane-wave, reference beam. The constructive and destructive interference with the reference captures not only the intensity but the phase of the scattered light too. Traditionally, this interference pattern was captured using high resolution photographic film which, once developed, could be illuminated by the same reference light to create the original light beam as scattered by the object. This recreation of the scattered light creates a visual replica of the object itself. It is interesting to reflect on the fact that despite the advances in the digital replacement of film, these digital devices still lack the pixel count required to fully implement a realistic holographic projector that can recreate the full 3D image of an everyday object. However, despite this technical limitation, the modern advent of digital technologies acting as diffractive optical elements, i.e., digital holography, has led to a multitude of related, albeit simpler, applications.

Beyond holography, the present-day interest in the phase structuring of light beam probably dates to 1992, when the seminal paper published by Allen and co-workers reasoned that a light beam with a helical phase structure, such as Laguerre-Gaussian laser modes, carried an "orbital angular momentum" that was independent of, and additional to, light's spin angular momentum [8]. This postulate was rapidly experimentally verified by Rubinsztein-Dunlop and co-workers [76] along with several other groups, transferring this angular momentum to microscopic particles held in optical tweezers, causing the particles to spin. This early work was based on photographic film acting as holograms, which when illuminated with a Gaussian reference beam, produced a diffracted beam with the required helical phase [5]. While still focused on optical tweezers, Grier and co-workers replaced the film with an interactive, pixelated liquid-crystal phase modulator for the holographic generation of helically-phased and other beams [78]. The pioneering of these spatial light modulators and the ease with which complex beams could be generated spawned work far beyond optical tweezers. In the twenty years since [78], these spatial light modulators have driven an explosion in the study and application of non-Gaussian laser beams and their use and application in areas ranging from the study of optical phenomenology, imaging and sensing to quantum science.

**Current and Future Challenges**

From a technical perspective, there are two key spatial light modulator technologies in widespread use for the generation of complex phase and intensity structured beams: those based on liquid crystal and those based on digital micromirrors, both of which have video resolution. The liquid crystal devices have phase-only modulation and, when used as a diffractive optical element, can reach well over 50% conversion efficiency, albeit only at a video-frame rate switching. The micromirror devices are intrinsically intensity modulators but can be used to create amplitude gratings and hence elements with a low diffraction efficiency (< 5%), but at >10 kHz frame rate [79]. A clear challenge for the development of new technologies is to simultaneously increase the diffraction efficiency while maintaining or increasing further the frame rate. The combination of high efficiency and high speed create new opportunities in real-time aberration correction and quantum information processing.

From a science perspective, most work to date on phase structured beams has focused on specific beam types ranging from beams described by various polynomials—Laguerre, Bessel, Gegenbauer, Airy—which often form complete orthonormal sets (i.e., a set of beams from which any other beam can be synthesized). Typically this work has followed a logic of "what can beams of type X be used for?", whereas an alternative logic is "what design of beams will have the optimum performance in application Y?" One example of these two logical approaches lies in single-pixel imaging systems wherein a sequence of patterns is used to illuminate an object, and the backscattered light for each projected pattern is measured. Summing the patterns, each weighted by the corresponding backscattered single, reveals an image of the object. The majority of the work uses Hadamard or more sophisticated patterns (e.g. [77]), but as an alternative, machine learning and related techniques can be used to define a bespoke pattern set to enable a compressed sensing approach to emphasise particular image properties or distinguish between specific object types [80], see Figure 1.

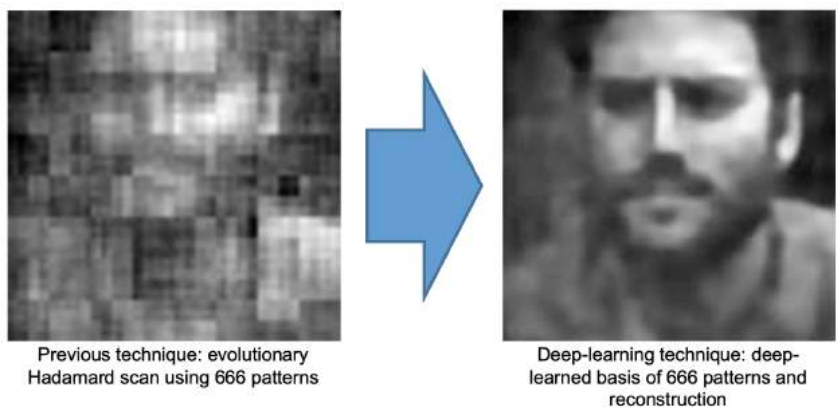

Previous technique: evolutionary
Hadamard scan using 666 patterns

Deep-learning technique: deep-
learned basis of 666 patterns and
reconstruction

**Figure 1.** Within single-pixel imaging, it is possible to use deep learning to define an optimum measurement set based upon a library of typical images. This approach to compressed sensing improves the quality of an image for a fixed number of under-sampled

Rather than applying machine learning techniques to the optimum beam design, these machine learning techniques can be applied to design a sequence of diffractive elements, creating an optical implementation of a neural network [81], or the lossless transformation of one complicated modal set into another simpler set [82]. All these transformations rely upon the phase structure of the beams.

As pioneered by Vellekoop and Mosk, another aspect of general complex beam design is in the creation of light beams that deal with complex aberrations [83]. By precise shaping of the incident light beam, it is possible to create a focused spot after transmission through highly scattering or complex media such as a multimode optical fibre [84]. If the medium is characterised in terms of a transmission matrix, then the inversion of this matrix defines the input beam required to produce any spot at the output. If this spot is then raster scanned, the backscattered or similar signal reveals an image from within or behind the media.

**Advances in Science and Technology to Meet Challenges**

As discussed above, there are clear technical requirements and associated challenges for high-speed spatial light modulators with high diffraction efficiency. Overdriving liquid crystal devices can perhaps bring a speed which approaches that of a digital micromirror device, but progressing beyond a few tens of kilohertz would seem to require a new, as yet unrealised, solid-state technology. If such technology is to be developed it will possibly arise from image-based communication where the high-dimensionality of the image state would bring a similar increase in communication bandwidth.

When it comes to the beams themselves, rather than those beams forming well-known complete sets, future applications are likely to be based upon arbitrary beams optimised to a particular function. Taking the example of aberration correction based upon the inversion of a transmission matrix, the scale of the computation problem is clear. A system, or medium, of N modes is described by a N x N matrix, the measurement time of which alone makes any real-time adaptation unlikely. A route to solving these problems is possibly to identify a subspace so that the modified matrix can be inferred from a relatively small set of additional measurements. The key is perhaps not to apply machine learning to the recovery of the image data but to apply the compressed sensing to the measurement of the transmission matrix itself. Whether a meaningful subspace exists will most likely depend on the system type—a multi-mode fibre would seem a good place to start.

**Concluding Remarks**

There are perhaps two main challenge and opportunity types ahead, both technical and conceptual. A key technical challenge is for developing spatial light modulators with improved diffraction efficiency and/or faster switching and/or much higher resolution. The last of these will undoubtedly be driven by the consumer display market which might potentially enable full holographic projection of images, whereas the first two are more likely to be addressed by specialist development. A key conceptual challenge is to move away from specific beam types and their corresponding diffractive elements to use machine learning and related techniques to create bespoke beams optimised for a particular need. For example, the correction of the aberrations associated with complex media, creating the optimum set of measurement patterns in imaging, or the creation of arbitrary diffractive elements in the optical implementations of neural networks and optical mode transformation. This will be a fruitful ground for the collaboration of optical engineers and computer scientists.

**Acknowledgements**

This work is funded by the Royal Society and EPSRC under the grant number EP/M01326X/1.

# 3. The topology of 3D polarisation

*Miguel A. Alonso[1,2] and Mark R. Dennis[3]*

[1]Aix Marseille Univ, CNRS, Centrale Marseille, Institut Fresnel
[2]University of Rochester
[3]University of Birmingham

**Status**

Among the earliest applications of polarised light, going back to Brewster and Talbot in the 1800s, was to explore structures of materials through crossed polarisers. Polarised light microscopy reveals the optic axes of crystalline minerals, and Schlieren patterns reveal textures in liquid crystals, organised by their topological defects [85]. In the modern study of structured light, where the state of polarisation varies with position, light itself shows complex topological textures and structure [25, 86].

The theory of polarisation for collimated light has a well-established formalism. When light travels in a definite direction, only the two components of the electric field vector perpendicular to the propagation direction are significant. For monochromatic light, the electric field vector at each point traces an ellipse, whose eccentricity, handedness, and orientation constitute what we call polarisation. The overall size (amplitude) and instantaneous position on the ellipse (phase) are usually disregarded. The Poincaré sphere (Fig. 1(b)) is an elegant, abstract geometric representation, where each polarised state corresponds to a point on the surface of a unit sphere. This point's latitude and longitude encode, respectively, the polarisation ellipse's orientation and ellipticity. Its Cartesian coordinates on the sphere, on the other hand, correspond to the Stokes parameters normalised by the total intensity, which are easily measured.

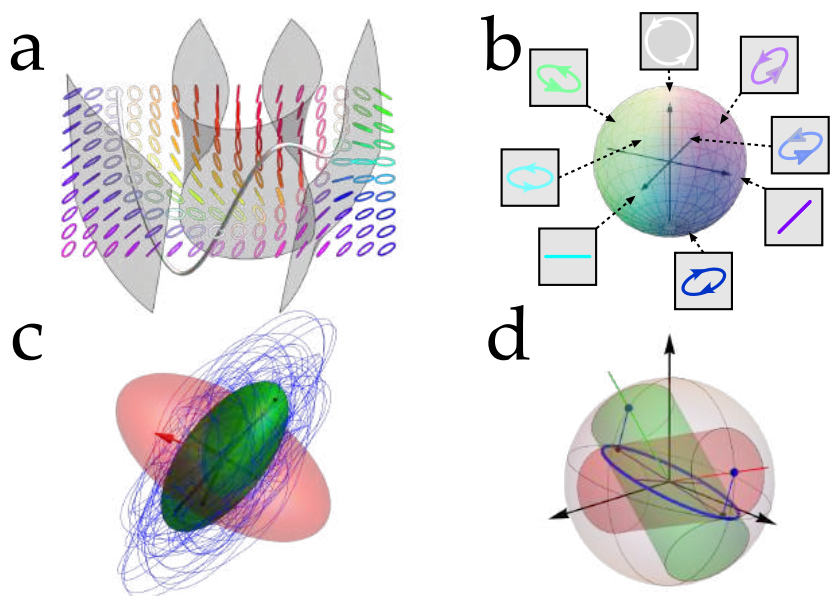

**Figure 1.** (a) 3D transverse polarisation texture with polarisation ellipses in one transverse plane. A C-line (white curve) of circular polarisation passes through this plane several times, as does the L-surface on which the polarisation is linear. (b) Poincaré sphere of transverse polarisations, indicating 2D polarisation states using a Runge colour scheme: polarisation azimuth by hue, and spin by brightness, from black (left-handed circular) to white (right-handed circular). (c) Representation of mixed polarisation state where the E vector traces a path in 3D (blue curve) that is not a simple ellipse: statistical E field described by the moment of inertia ellipsoid, and average spin vector which must lie within the dual ellipsoid [89]. (d) Majorana sphere representation of polarisation as two points on the sphere, projecting directly onto the polarisation ellipse as circles, or as the foci of the ellipse circumscribed by the sphere [90,91].

The Poincaré sphere construction is more than just a map of two physical parameters onto a surface. Amongst its physical properties, it represents:

- the action of polarising and transforming optical elements as geodesic projections and rotations over the sphere through the use of Jones calculus;
- the geometric (Pancharatnam-Berry) phase due to cyclic changes in polarisation as half the solid angle enclosed on the sphere by the path of polarisation transformations [87,88];
- partially polarised fields as points inside the sphere, with the radial coordinate giving the degree of polarization.

The standard polarisation formalism and the Poincaré sphere are not appropriate when light is nonparaxial—i.e., the plane of the ellipse varies—and the normal to the polarisation ellipse is not tied to the propagation direction. Various generalisations have been proposed which extend to the richer structure of nonparaxial polarisation, albeit without the unifying simplicity of the Poincaré sphere. Even for polarisation at a single point, the geometry and topology of the description for the nonparaxial case are quite different.

**Current and Future Challenges**
The topology textures in position-dependent polarisation fields strongly mirror topological textures in liquid crystals; both can be described by headless vectors (directors), representing the major axes of the polarisation ellipses or the axes of the rod-like liquid crystal molecules. Topological singularities organise these patterns within a volume: C-lines of circular polarisation around which the director rotates by ±180°, explored by Nye in the 1980s in Bristol where Frank had discovered disclinations—the liquid crystal counterpart—30 years earlier. These singular filaments occur in transverse and non-transverse polarisation fields. Their 3D structure in a volume can be controlled by holographic manipulation, most often by liquid-crystal-based computer-controlled holograms (spatial light modulators). This has allowed for the production and polarimetric measurement of beams with knotted C-lines.

Determining the 3D state of polarisation directly, including the longitudinal component, is nontrivial even in free space. Nowadays, the most common approach samples the field by scanning with a nano-scatterer; this has allowed for the measurement of the fine structure around a C-line, revealing Mobius band-like configurations [26].

In addition to eccentricity and orientation in its plane, a nonparaxial polarisation ellipse has a varying plane orientation (perpendicular to the direction of spin), requiring four parameters in total. Based on Majorana's spin formalism, Penrose proposed a representation in terms of not one but two indistinguishable points over the surface of a unit sphere in 3D, with a simple geometric meaning: they indicate the two directions for which the ellipse traced by the electric field projects onto a right-handed circle (Fig. 1(d)) [90,91]. Mathematically, this corresponds to the symmetric product of two 2-spheres, a topological representation of the complex projective plane $CP^2$, corresponding to complex 3-vectors and disregarding an overall complex factor (corresponding to intensity and phase). This representation allows for the optical geometric phase to be calculated when the plane of polarisation varies, corresponding to a quantum spin-one geometric phase, as demonstrated, for instance, in the Tomita-Chiao experiment for light in a twisted fibre [92].

For partially polarised fields—that is, for fields that are polychromatic but whose frequency components are uncorrelated—the electric field at a point does not trace an ellipse but a more complex, generally nonplanar and nonperiodic path. A generalisation of the Stokes parameters has been proposed based on an expansion of the polarisation matrix using the 3x3 Gell-Mann matrices, instead of the 2x2 Pauli matrices that yield the standard Stokes parameters [93]. This description encodes polarisation as eight parameters, i.e., the number of real quantities needed to specify a trace-normalized 3x3 Hermitian matrix. These parameters define an eight-dimensional hypervolume, where—like for the Poincaré sphere—complete polarisation corresponds to points at unit distance from the origin, and the remaining points correspond to partial polarisation states. However, without a direct generalisation of optical elements or the Jones calculus to 3D, the power of this high-dimensional description is yet to be realised. In the special case when the polychromatic field is coherent and includes only a small discrete set of mutually rational frequencies—generated, for example, by nonlinear harmonic generation—the electric field at each point traces a periodic path that is not an ellipse but a Lissajous curve [36] that can be knotted in 3D [39].

Recent experiments have revealed topological features in monochromatic polarisation distributions not only along curves (e.g. Möbius strips) but over surfaces or volumes. For example, Skyrmions are distributions that fully wrap around a parameter space corresponding to an n-dimensional sphere. "Baby Skyrmions" for n=2 have been realised in different ways:

- as full Poincaré beams, whose polarisations at any transverse plane cover the Poincaré sphere's surface [94] (Fig. 2(b));
- where, in the evanescent field over a planar metal surface, the linearly-polarised electric field vector points in all 3D directions [30] (Fig. 2(a));
- where ellipses in a plane have spins (ellipse normal) in every 3D direction [31,83] (Fig. 2(c)).

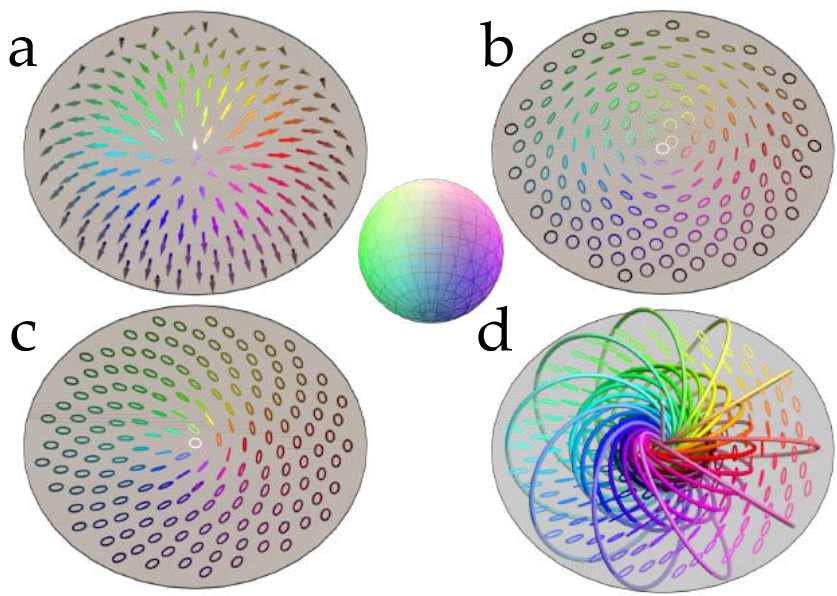

**Figure 2.** Different representations of Skyrmion-like configurations in optical polarisation. (a) E-field points in all directions; (b) full Poincaré beam (realising all points on Poincaré sphere); (c) circular polarised realising all directions of spin; (d) Skyrmionic Hopfion in 3D, realising all polarisations and phases (on optical hypersphere). In all cases, the colour scheme is provided by the Runge sphere.

A richer example of a 3D topological state (n=3) uses the fact that a normalised, transverse polarisation state, including phase (Jones vector), corresponds to a point on a unit hypersphere in 4-dimensional space. These 3-spheres admit the Hopf fibration into interlinking circles (on which the phase varies) parametrised by the Poincaré sphere. Such a structure—a Skyrmionic Hopfion (Fig. 2(d))—realising all polarisations and phases in an entwined texture was recently designed, synthesised experimentally, and measured [34]. The topological complexity of these polarisation distributions increases when more degrees of freedom are considered, as is the case of polychromatic light [36,39].

**Concluding Remarks**

Light polarisation at a point presents a simple yet elegant topological structure, which becomes more complex when the paraxial limit is abandoned. The topological features are much richer when considering the spatial variation of polarisation, echoing the structured topological states studied in many forms of condensed matter. Because light can interact with matter, the structure of each can shape that of the other; most prominently, light fields can be used to interrogate and shape liquid crystals, and conversely. Modern microscopy techniques use light with varying polarisation to probe the configurational 3D structure of biological matter (e.g. actin filaments). 3D topological polarisation structures affect the orbital and spin dynamics of trapped particles. Topology and polarisation have played a central role in the rapid development of the science of structured light over the past few decades. However, the technological potential for many applications is still in its infancy.

**Acknowledgements**

MAA acknowledges funding from the Excellence Initiative of Aix Marseille University – A*MIDEX, a French 'Investissements d'Avenir' programme, and from the Agence Nationale de Recherche (ANR) through project ANR-21-CE24-0014-01. MRD acknowledges support from the EPSRC Centre for Doctoral Training in Topological Design (EP/S02297X/1).

## 4. Shaping Light

*Angela Dudley and Andrew Forbes*

University of the Witwatersrand

**Status**

Light can be shaped in all of its degrees of freedom (DoFs), in time and space, for so-called Structured Light [96]. Traditionally, the spatial DoFs have been exploited—for example, amplitude, phase and polarisation—first in 2D (the transverse plane) and later in 3D (all three components of the electric field), while the time and frequency domains offer the potential for 4D control. Recently, there has been a concerted effort to identify and control new DoFs, and to harness this control for emerging applications, including communications, microscopy, imaging, optical trapping and tweezing, quantum state engineering, and laser machining processes, to name but a few. The challenge is to identify which DoFs can be controlled, to what extent, and with what toolkit [97]. For instance, azimuthal phase control gives rise to orbital angular momentum (OAM) modes, which are easily controlled both as scalar and vectorial superpositions (whose phase and polarization profiles are non-separable). In this sense, OAM can be treated as an easily controllable DoF. In contrast, amplitude and phase shaping to create radially structured modes (the p indexed modes in the Laguerre-Gaussian basis, for example) is common-place [82], but this DoF is not easily controlled with our existing toolkit, limiting its applicability. Path has long been associated with quantum states of light, but less so in classical light, while ray-wave duality in classical fields (see Figure 4) is yet to be fully exploited. Given the many DoFs of light, how can we realise and control them?

Optical cycles are much too fast to allow direct temporal light shaping, and so such light shaping is done spatially. For example, to shape light temporally, the frequency components are usually path separated by a dispersive element, mapping frequency to space, subsequently shaped in amplitude and phase before a return mapping to reconstruct the desired temporal pulse [98]. In the spatial domain, we may control the amplitude and phase of each polarisation component, the latter by propagation and geometric phase, both of which can be made polarisation specific. The ubiquitous lens is a simple example of a beam-shaping optic, adjusting the propagation phase by material thickness and refractive index to shape light by refraction. Recent developments in free-form optics has given new impetuses to refractive shaping of light, with unprecedented control possible [99]. In the early 1990s, there was an explosion of activity in diffractive optical elements (DOEs), where a computer-generated hologram was etched into a material to form a holographic plate of negligible thickness (i.e., no refraction). Here, the underlying concept stems from Denis Gabor's Nobel award-winning development of holographic imaging, allowing for the recording of light's amplitude and phase information. Geometric phase has been exploited for complex beam shaping [100]—which by definition is polarisation sensitive—allowing for the creation of vectorial light fields. A more recent move to sub-wavelength structures in the visible has allowed for polarisation dependent propagation phase control using metasurfaces [101], paving the way for all phases to be exploited.

The aforementioned approaches are all static, which greatly limits their versatility. The introduction of liquid crystal spatial light modulators (SLMs) [102] ignited a plethora of investigations into novel light shaping techniques and their corresponding applications. These devices are void of customer-specific production requirements, span the visible and near-infrared wavelength ranges, and offer instantaneous and rewritable amplitude, phase, and polarization control. Intriguingly, recent times has seen a return to amplitude-only devices in the form of digital micromirror devices (DMDs), which are fast, cheap, and versatile. Direct light shaping at the source, within the laser, allows for high-efficiency and high-purity modes as the output, and can be achieved through a variety of intra-cavity and cavity geometry approaches [103], with complex light possible from very simple laser cavities [104]. For example, a combination of internal and external path control has resulted in eight-dimensional structured light across multiple DoFs (see Figure 4 and Ref. [104]).

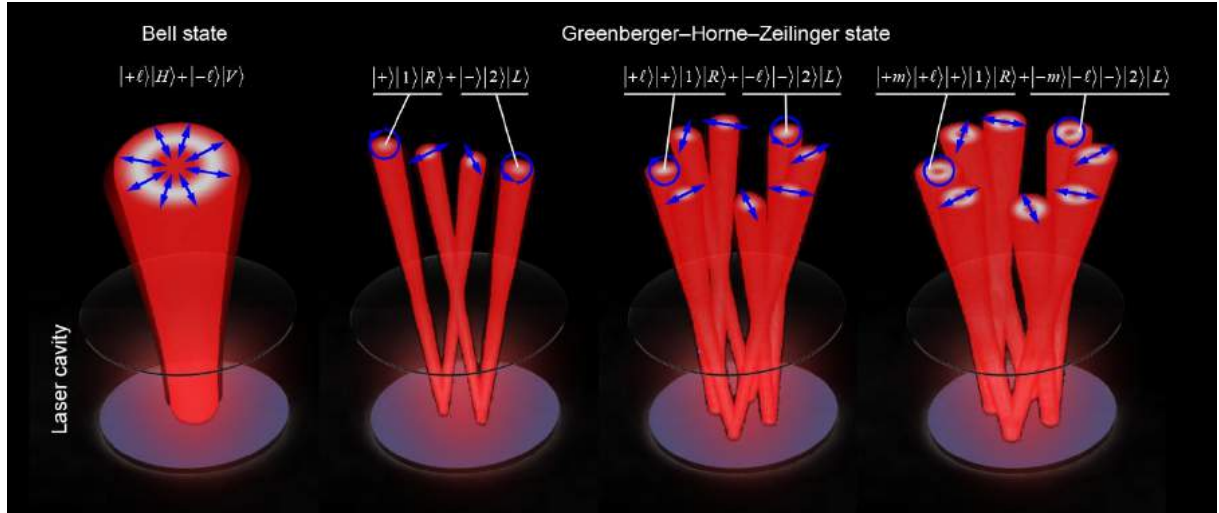

**Figure 1.** A simple laser cavity can shape light that appears ray-like but with wave-like properties, across multiple DoFs. Eight-dimensional structured light has been shown when internal and external light shaping are combined, producing the classical analogue to the famous GHZ states of quantum light. Reproduced from [97] with permission.

**Challenges and opportunities**

The algorithmic approaches to shaping light (the recipes) are very well established [105], dating back to the early work on pattern recognition, and improved recently with the aid of machine learning. The challenge lies mostly in the hardware. Although huge advancements have been made in the light-shaping toolkit and its employment in a diverse range of applications, there remain some challenges. Foremost amongst these are the need for the miniaturization of the modulation technology, increased power thresholds, and a broader range in wavelength control, particularly at shorter ultra-violet (UV) or extreme UV wavelengths. For example, in order to integrate and interface existing (miniature) electronic components with fast photonic technologies that exploit the many photonic DoFs, light-shaping approaches have to be miniaturised. The development of miniature, integrated on-chip devices will bolster the optical computing, imaging, and communications industries. Miniaturising the light-shaping toolbox has been restrained by the material of the beam-shaping device. Liquid crystals typically allow for micro-meter-sized pixels, while refractive elements are limited to feature sizes much greater than the operating wavelength. Recent developments in metamaterials and nanostructured materials are offering promising avenues to engineer structured light with subwavelength thick optics. Gradient metasurfaces are of particular interest in that they possess a spatially varying phase response allowing for arbitrary wavefront control with subwavelength resolution. These novel and minute components allow for integration into existing photonic technologies, such as photonic circuits and optical fibres. Another promising opportunity is based on two-photon polymerization (2PP) which is a direct laser writing technique fabricating complex 3D micro-optic structures. Here wafer-thin optics with feature sizes of approximately 160 nm can be fabricated [106].

Another major challenge is to extend the current light-shaping techniques to high-power levels, where levels exceeding kilowatts are needed for laser machining and laser-enabled manufacturing.

Commercially available single and multimode fibre lasers used for welding, cutting, and additive manufacturing operate at the multi-kilowatt level, but to date, very few beam-shaping technologies can tolerate such extreme power levels, although efforts to remedy this are underway. Optics used in high-power kilowatt systems are large and bulky due to the finite absorption limitation—the complete opposite of miniature-sized components. More recently, SLMs are being supplied with thermal control units, where water-controlled heat sinks allow for a 10-fold increase in incident power. There is promise in revisiting well-known and well-established adaptive optics solutions such as deformable mirrors. These mirrors can tolerate and efficiently reflect powers on the order of kilowatts. However, efforts will need to be made in order to enhance their response rate and stroke to achieve fast, high-purity spatial mode creation while reducing their cost. In attempting to reduce the size of high-power beam-shaping devices, the previously discussed metasurfaces and nanostructures could provide promise; however, their power-handling capabilities have yet to be tested.

The aforementioned approaches are all based on linear optics. An exciting avenue that is very much in its infancy is to shape light by nonlinear approaches. One possible solution is to shape the light at low power and amplify it post-shaping, while another is to transfer shapes from one wavelength to another through parametric processes. These approaches may overcome both the power and wavelength challenges, while on-chip nonlinear optics has a long history, holding promise for compact solutions too. Traditionally, amplification and nonlinear processes have focused on "how much light do I have?", and less so on "what does the light look like?" To explore these possible solution pathways will require a paradigm shift in our thinking on such topics.

Other challenges lie in spatially controlling short wavelengths, as the light shaping toolbox for these wavelengths are rare (neither SLMs nor DMDs work), while hardcoded DOEs and metasurfaces likewise struggle with material choice and feature size. All of the previously mentioned tools are only valid and applicable to coherent beams, but how can we achieve and similarly control the DoFs of incoherent sources? Already, the advancement of "Li-Fi" may benefit from further DoF control if such beam-shaping approaches are feasible for incoherent light.

**Concluding Remarks**

Techniques to achieve light shaping have advanced tremendously during recent years, offering a diverse set of tools, and opening an abundance of exciting applications. Although a broad scope in light shaping functionality has been achieved, photonic control in optical circuits is still primitive in comparison to electronic control, with practical and commercial realisation still far off. Further advances in efficiency, compactness, broader wavelength control, and power-handling capabilities are sorely needed for shaping light to advance from science to application.

## 5. Spatiotemporal optical vortices and OAM

*Sina Zahedpour, Scott W. Hancock, and Howard M. Milchberg*

University of Maryland

**Status**

It is well established that monochromatic beams of light can support vortices where electromagnetic energy density circulates around a local axis. Examples of such orbital angular momentum (OAM)-carrying beams are the Laguerre-Gauss (LG) or Bessel-Gauss (BG) modes in free space, where the OAM axis coincides with the direction of propagation [8], or spatial optical solitons with vortex rings [107]. In all cases, local vortical axes are fixed in space and energy density flow is described in purely spatial coordinates. However, because electromagnetic vortices and OAM are fundamentally associated with energy density circulation, there is no prohibition, in principle, for a local vortical axis to deviate from the propagation direction while embedded in a propagating pulse. This would require a polychromatic beam [19, 21].

Such polychromatic vortices embedded in spacetime were first measured as a naturally emergent and universal structure in the collapse arrest dynamics and self-guiding of intense laser pulses in nonlinear media [21]. Were it not for 'collapse arrest'—a response to high intensity such as ionization—the beam would collapse to a singularity. Collapse arrest enables a quasi-stable accumulation of phase shear—a very sharp phase gradient in the transverse direction--between the inner and outer parts of the beam, triggering a phase defect (and field null) that wraps around the propagation axis and spawns toroidal spacetime vortices (Fig. 1(a)). These toroidal spatiotemporal optical vortices were observed for the first time in [21] and dubbed STOVs; they direct energy density flow in a self-guided pulse, and are robust and topologically protected.

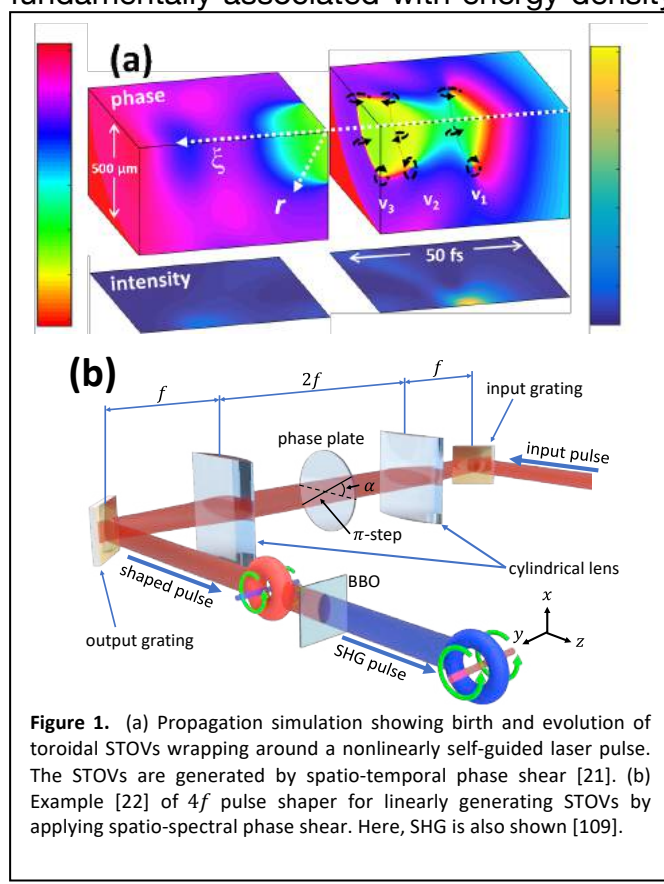

**Figure 1.** (a) Propagation simulation showing birth and evolution of toroidal STOVs wrapping around a nonlinearly self-guided laser pulse. The STOVs are generated by spatio-temporal phase shear [21]. (b) Example [22] of $4f$ pulse shaper for linearly generating STOVs by applying spatio-spectral phase shear. Here, SHG is also shown [109].

The realization that STOVs were generated by phase shear in spacetime led to a method to generate them linearly and controllably, using a pulse shaper to apply shear in the spatiospectral domain [22,23] and then return the pulse to the spatiotemporal domain. The STOV pulse thus generated resembles an 'edge first flying donut' (see Fig. 1(b) and Fig. 2), with phase circulation in spacetime and OAM orthogonal to the propagation direction. Experiments [22] and theory [24,110] have shown STOV propagation is governed by diffraction in both space and time. Later simulations have shown that STOV-carrying pulses can be formed by compact nanostructures that impose a specified field null axis in $(\omega/c, \mathbf{k})$ space, leading to arbitrarily oriented STOV axes in spacetime $(ct, \mathbf{r})$ [111]. That spatiotemporal OAM is a property of single photons was recently confirmed in experiments showing OAM conservation under second harmonic generation (SHG) with STOV pulses [109]. Recent work has investigated extreme ultraviolet STOV photons produced by high harmonic generation [112].

## Current and Future Challenges

Merely measuring STOVs presents new challenges. Unlike spatial optical vortex-carrying beams, whose intensity and phase profiles can be measured by CW beam imaging and interferometry, more complex methods are needed to visualize STOVs. Narrow bandwidth, low resolution multi-shot pump-probe cross-correlation measurements [23] can be used; these are limited to long duration STOV pulses with high pulse-to-pulse stability. For examining ultrafast STOV evolution in nonlinear propagation experiments, where sensitivity to small fluctuations is expected, a broadband single-shot technique has been demonstrated that measures pulses with spacetime phase defects [22]. As STOV-structured light increases in complexity, such as with short wavelength attosecond pulses [112], there will be a need for much higher space and time resolution diagnostics extending across the spectrum.

As the study of STOVs is in its infancy, they could be viewed as a solution in search of a problem. Because they are integral to energy density flow in both nonlinear self-guided propagation [21] and in linear propagation [22,110], an unusual robustness may apply to these beams owing to conservation of topological charge and angular momentum. As with space-defined OAM, the excitation and probing of STOV-OAM states in materials and structures will open completely new research directions. Any system involving spacetime phase shear, such as transient current densities in nanostructures, could be interrogated by STOV pulses with a prepared OAM content. Robust methods for filtering or sorting the OAM components of STOV pulses will therefore be needed. In analogy with super-resolution microscopy using monochromatic OAM beams, STOVs may even provide a super-resolution capability in spacetime.

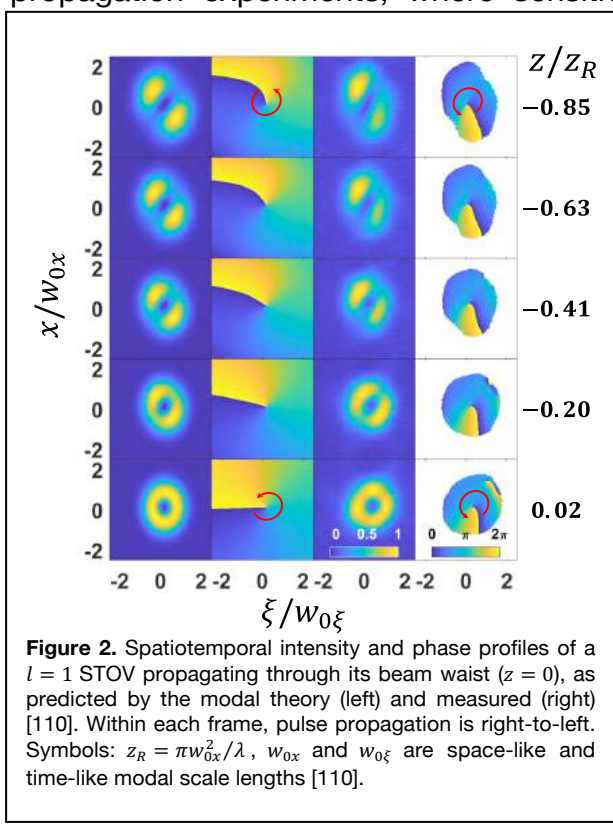

**Figure 2.** Spatiotemporal intensity and phase profiles of a $l = 1$ STOV propagating through its beam waist ($z = 0$), as predicted by the modal theory (left) and measured (right) [110]. Within each frame, pulse propagation is right-to-left. Symbols: $z_R = \pi w_{0x}^2/\lambda$, $w_{0x}$ and $w_{0\xi}$ are space-like and time-like modal scale lengths [110].

At a fundamental level, only very recently has there been theoretical work on angular momentum of STOV pulses. One approach calculates spin angular momentum and OAM of STOV pulses in vacuum [24] and the other derives the mode structure and OAM for paraxial STOV pulses in dispersive media [110]. These papers differ in their choice of OAM operator, with differing results for the OAM per photon $\mathcal{L}$ in a STOV-carrying pulse of topological charge $l$: $\mathcal{L} = l\,\hbar(\alpha + \alpha^{-1})/2$ [24] vs. $\mathcal{L} = l\,\hbar(\alpha - \beta_2\alpha^{-1})/2$ [110], where $\alpha = w_{0\xi}/w_{0x}$ is the ratio of the STOV's time-like to space-like spot sizes and $\beta_2$ is the medium's normalized group velocity dispersion. The latter operator is conserved, and predicts half-integer OAM in vacuum (owing to lack of energy density flux in the local time domain) and the existence of "STOV polaritons" in dispersive media. Just as the pioneering work connecting LG modes to photon OAM [8] was followed by quantum field theories, there are likely to be similar follow-ups to [24,110].

## Concluding Remarks
Experimental realization of electromagnetic pulses with spatiotemporal OAM has opened a promising new avenue for studying OAM and its applications.

## Acknowledgements
The authors thank Konstantin Bliokh for fruitful discussions, and acknowledge the support of the Air Force Office of Scientific Research, the Office of Naval Research, and the National Science Foundation.

## 6. Structured waves in non-Hermitian systems

*Stefan Rotter[1], Franco Nori[2,3], and Şahin K. Özdemir[4]*

[1]Vienna University of Technology (TU Wien)
[2]RIKEN
[3]The University of Michigan
[4]The Pennsylvania State University

**Status**

The structuring of waves typically involves the propagation of an incoming wave field through a device that shapes the transmission in a desired fashion. Engineered structures like gratings and waveplates can equivalently be operated in reflection mode. In both cases, the desired wavefront of the outgoing wave is achieved through appropriate interferences induced by the wave-shaping device that tunes the wave's spatial and spectral (or temporal) degrees of freedom. In a very recent line of research, one tries to extend the possibilities to structure waves of different kinds (electromagnetic, acoustic, etc.) by working with non-Hermitian wave-shaping tools. The term "non-Hermitian" refers here to the fact that the time evolution of the wave passing through the wave shaper is governed by a non-Hermitian Hamiltonian.

While, formally, a simple wall that absorbs all of the incoming waves would already constitute such a non-Hermitian system, one typically refers to non-Hermitian systems only when they contain a non-trivial combination of gain (amplification) and loss (dissipation). Typical examples include non-Hermitian meta-surfaces for steering transmitted or reflected waves in desired ways [113], and waveguides that are designed such as to guide incoming waves around non-Hermitian degeneracies known as exceptional points (EPs) [114]. Such EP-encircling protocols build on the remarkable property that the output state on either side of the waveguide depends only on the direction in which the EP is encircled—a property determined solely by the input port through which the wave is injected into the waveguide.

Going beyond such asymmetric mode-switching protocols, spatially tailored gain-loss landscapes can also be used to guide incoming waves [115–117], even through disordered scattering regions [118, 119]. When the patterning of gain and loss is done right, waves can not only be perfectly transmitted, but can also maintain a well-behaved intensity profile (without any interference fringes) even inside strongly disordered media (see Fig. 1). Under certain circumstances, the details of the system that is patterned with a certain gain-loss distribution do not even have to be known to apply a non-Hermitian structuring. In cases such as random lasers that have an unknown and typically inaccessible internal structure, the spatial pattern of the pump beam that delivers the optical gain to the laser can be optimised through appropriate algorithms to engineer the laser's spectral and spatial emission properties [120, 121]. In this way, single-mode operation and a directional far-field pattern of a random laser can be achieved.

Recently, concepts from topology have also been employed to create robust unidirectional propagation of light and to design lasers that operate on topologically protected edge modes (see [122] for a recent review of this emerging field of research) or on modes that encircle an EP [123].

**Current and Future Challenges**

On the conceptual level, the major lines of research currently concentrate on the question: which new functionalities, advantages, or unconventional features may the engineering of a system's non-Hermitian degrees of freedom bring along? Is it possible to use non-Hermitian engineering to, for example, overcome the stringent material requirements for meta-materials used for optical cloaking? Can one exploit topological concepts to build photonic structures that are robust against fabrication imperfections? Can such progress be achieved without having to work with very exotic materials or cumbersome setups?

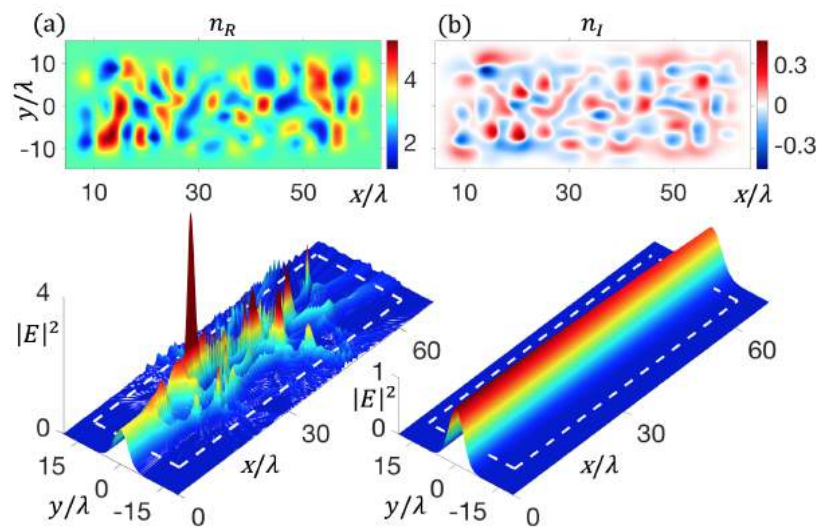

**Figure 1.** (a) Two-dimensional distribution of a real refractive index $n_R$ (top panel). A Gaussian beam entering this disordered scattering landscape and becomes severely distorted (bottom panel). (b) Adding a tailored distribution of gain and loss through an imaginary refractive index $n_I$ (top panel), the Gaussian beam propagates as through a homogeneous medium (bottom panel). Image adapted from [119].

On the technical level, one of the major challenges is the accurate positioning and control of the non-Hermitian (gain and loss) components. While lossy media are ubiquitous and comparatively easy to pattern, media with gain are more challenging to handle. This is because gain necessarily requires an external source of energy to provide the amplification to the wave—think here of an active medium such as a laser cavity that requires an external pump (optical or electrical) to operate the laser. For optical devices in one or two dimensions, one typically works with optically active materials that are pumped through an external beam with a spatial pattern that is controlled by a suitable mask or a spatial light modulator (the size and costs of the latter, of course, also impose limitations). A three-dimensional structuring of the pump profile is naturally more challenging to implement. Moreover, non-linear effects, such as those induced by gain saturation and spatial hole burning, further complicate the situation. For other types of waves, like sound or matter waves, suitable gain mechanisms may not even be available. We emphasise that both the spatial and the spectral tunability of a gain medium have certain restrictions: while the spatial pattern of the pump cannot be arbitrarily fine in its structure, the spectral profile of the gain is determined by the fundamental constituents of the active medium as well as by other restrictions like the Kramers-Kronig relations that follow from the principle of causality.

A viable strategy to overcome these limitations is to consider discrete instead of continuous systems, where the discreteness of the former may refer to their spatial structure or to their time-evolution. A corresponding mapping onto arrays of waveguides or coupled fibre loops has the advantage that each spatial or temporal element involved in such a discrete system can be individually controlled. Recently, this advantage has been used in various circumstances to implement theoretical concepts on non-Hermitian wave engineering—from the creation of constant-pressure sound waves to the realisation of system designs based on the concept of non-Hermitian topology (for a review see [124]).

**Advances in Science and Technology to Meet Challenges**

On the theoretical level, there are still many questions left open for exploration. When considering smaller and smaller constituents of non-Hermitian media, where quantum effects start playing a role, the influence of the noise induced by both gain and loss has to be taken into account. As it turns out, such noise processes are not just an annoying side effect, but they may constitute the major bottleneck for the performance of a non-Hermitian device, such as for sensors that operate at an EP. Dealing with such noise processes remains an outstanding theoretical challenge.

Also in the domain of non-Hermitian topology, numerous puzzles are waiting to be resolved. Going beyond the question of how to correctly transfer concepts from Hermitian topology to the domain of non-Hermitian physics, entirely new topics emerge; consider here, for example, nonlinear effects that lead to strongly correlated states of light and interesting collective processes. It also still needs to be clarified which features non-Hermitian materials can provide that were so far only associated with other types of materials, such as anisotropic media and those with a vanishing or negative index of refraction. Specifically, it will be interesting to clarify if and in which way non-Hermitian media can be used for cloaking an object [125], for breaking Lorentz-reciprocity (in combination with nonlinearity or spatio-temporal modulation), and for imaging beyond the diffraction-limit.

For the experimental implementation of flexible non-Hermitian media, progress on many different sides will be beneficial. Advances in the speed, precision, and cost of spatial light modulators will help to accurately control spatially tailored pump beams. Advances in the development of active materials will help to address the challenging requirements in the gain values required for implementing certain theoretical concepts. Non-Hermitian metasurfaces with well-controlled distribution of discrete gain and loss ingredients may exhibit interference effects originating from gain- and loss-induced phase responses, leading to phase singularities and vortices which can help to control and shape the light transport [126].

**Concluding Remarks**

To conclude, the research on structured waves in non-Hermitian media is still in its infancy. While a whole range of interesting theoretical concepts and experimental platforms for their implementation have recently emerged, the field is still growing in several directions such as towards the inclusion of topological, quantum, and nonlinear effects. Another appealing prospect is to apply non-Hermitian design concepts not just on externally generated light fields, but to integrate them directly into the design of the laser light source itself. Important challenges are the flexible and precise control of the non-Hermitian gain and loss components in both their spatial and spectral degrees of freedom. With rapid technological progress in this direction, we expect non-Hermitian tailoring of light fields to become a standard tool in wave engineering. Ultimately, non-Hermitian elements may achieve comparable relevance to spatial light modulators and conventional diffractive metasurfaces made from dielectric or plasmonic structures without gain inclusion, paving the way to a new era of wavefront shaping.

**Acknowledgements**

S.R. acknowledges support by the Austrian Science Fund (FWF, grant P32300 WAVELAND) and by the European Commission (grant MSCA-RISE 691209 NHQWAVE). F.N. is supported in part by NTT Research, and Ş.K.Ö. by the Air Force Office of Scientific Research (AFOSR) Multidisciplinary University Research Initiative (MURI) Award No. FA9550-21-1-0202.

# 7. Tailoring Random Light for Imaging Applications

*Nicholas Bender and Hui Cao*

Yale University

### Status

Spatially random light has the hallmark appearance of an irregular mosaic of diffraction-limited speckle grains. Speckle formation is a phenomenon inherent to both classical and quantum waves, occurring when a coherent wave undergoes a disorder-inducing scattering process. The speckle patterns are described by a statistically stationary and ergodic random process. Stationarity requires the statistical properties of an ensemble of speckle patterns to be the same as those of an individual speckle pattern within the ensemble. Ergodicity requires the statistical properties of two spatial positions—separated by more than one speckle grain size—to be independent and identical to those of the ensemble. The speckle patterns are categorized by the joint probability-density function (PDF) of their complex-valued field. Rayleigh speckles—the most common family of speckles—obey a circular-Gaussian field PDF which results in a negative exponential intensity PDF [127,128]. The phase PDF is independent of the amplitude PDF, and constant over a $2\pi$ range. The circular invariance of the field-PDF results in a "fully developed" speckle pattern. Typically, non-Rayleigh speckles are classified as either under-developed (the sum of a small number of scattered waves, or the sum of not fully randomized waves) or partially coherent (the sum of incoherent partial waves). Furthermore, fully developed speckles typically possess only short-ranged spatial intensity correlations which are determined by the average speckle grain shape, which is dictated by the diffraction limit.

Because of their pervasiveness, speckle patterns have been adapted for use in a wide range of optical applications ranging from imaging [129, 130] to optical manipulation [131]. While Rayleigh speckles are the most common family of speckled light, their statistical properties and spatial correlations are not necessarily ideal in different applications. There has been a plethora of interest in creating speckle patterns with tailored statistics and spatial correlations [132-136], due to their potential applications in structured-illumination imaging. Specific examples include dynamic speckle illumination microscopy, super-resolution imaging, and pseudo-thermal light sources for high-order ghost imaging. Furthermore, a general method for customizing both the statistics and topology of laser speckle patterns would be a valuable tool for synthesizing optical potentials for cold atoms, microparticles, and active media.

### Current and Future Challenges

Because the conditions required to generate Rayleigh speckles are general, creating fully developed non-Rayleigh speckles is challenging. Specifically, the difficulty lies in altering the intensity PDF without changing other statistical properties, e.g., phase PDF, stationarity, ergodicity. Recently, a simple method for creating non-Rayleigh speckle patterns with a phase-only spatial light modulator (SLM) was developed [132]. High-order correlations were encoded into the field reflected by the SLM, resulting in a redistribution of the light intensity among the speckle grains in the far-field (Fourier-plane). The resulting speckle pattern possesses an intensity PDF with a tail decaying either slower or faster than a negative-exponential function. Subsequently, a general method for tailoring the intensity statistics of speckle patterns was developed based on the same principle of modulating the phase front of a laser beam with a SLM [133]. Experimentally, speckle patterns governed by arbitrary intensity PDFs were created. The speckle patterns exhibit distinct topologies from Rayleigh speckles, without introducing spatial correlations beyond the diffraction-limited speckle grain size.

A foundational principle of statistical physics is the Siegert equation. Specifically for the case of spatial correlations in random light, the Siegert equation proportionally relates the intensity correlation function with the squared magnitude of the field correlation function. As such, the spatial intensity correlation function of a typical speckle pattern does not possess additional structure beyond what is present in the field correlation function, which is determined by the diffraction-limited average speckle-grain shape. In [132,133], the spatial intensity correlations of the customized speckle patterns adhered to the Siegert equation. It was experimentally shown in [134] that a speckle pattern can dramatically break the Siegert relation when non-local correlations are controllably encoded into a speckle field by a SLM, specifically by tailoring the $4^{th}$ order correlations in the Fourier plane.

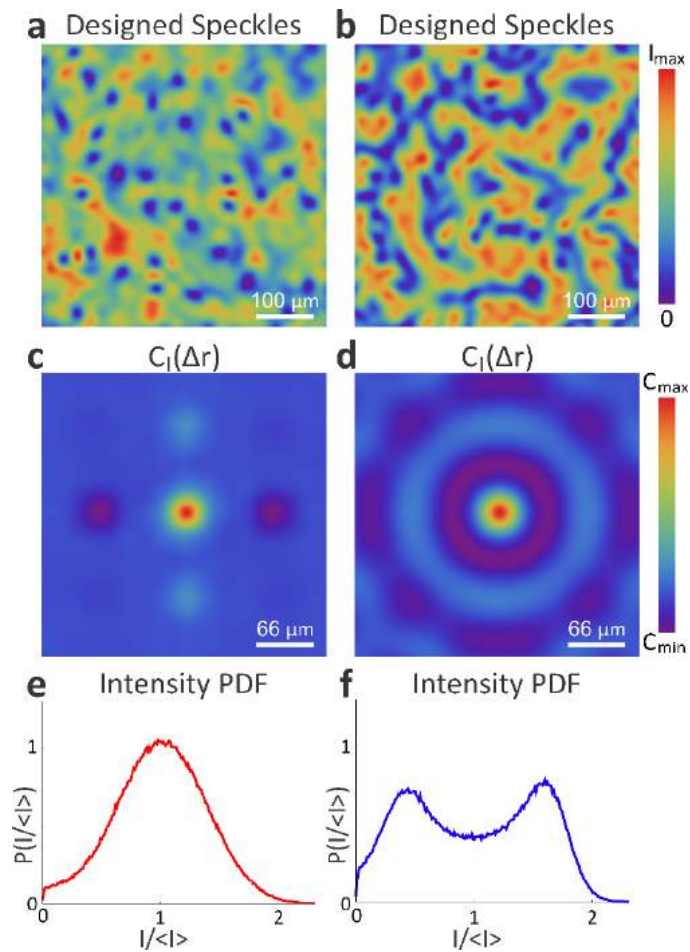

**Figure 1.** Two example speckle patterns (a), (b) from stationary and ergodic ensembles of 100 tailored speckle patterns, with customised spatial intensity-corelations (c), (d) and intensity PDFs (e), (f). The example speckle pattern in (a) possesses non-local intensity-correlations (c), and a unimodal intensity PDF (e). The speckle pattern represented by (**b**) is tailored to have ring-shaped, long-range intensity-correlations (d), and adhere to a bimodal intensity PDF (f).

While the techniques presented in [133, 134] independently modify different properties of speckle patterns, [135] combined these methods to arbitrarily tailor the PDF and spatial intensity-correlations of speckles. Figure 1 presents two examples of fully developed speckle patterns (a), (b) with different tailored spatial correlations (c), (d) and customized statistics (e), (f).

**Advances in Science and Technology to Meet Challenges**

The scientific advances made in speckle customization have begun to translate into speckle-based applications. For example, a proof of principle demonstration [136] has shown that 2D customized speckles can significantly out-perform Rayleigh speckles in nonlinear pattern illumination imaging techniques [130]. Figure 1(a) presents a specially tailored speckle pattern to photoconvert a uniform fluorescent sample. Within the central square, the speckle pattern was designed to consist of a random array of circular vortices embedded in an approximately constant-intensity background; outside the square are Rayleigh speckles. Figure 1(b) is the fluorescence image of the sample after being photoconverted by the speckle pattern. Outside the central square, the fluorescent pattern consists of a sprawling anisotropic web, which reflects the topology of the low-intensity regions surrounding the optical vortices in Rayleigh speckles. In stark contrast, the fluorescence pattern within the central square **c** features isolated isotropic fluorescent spots. The isotropy exhibited by the fluorescent spots originates from the high degree of rotational symmetry of the vortices in the customized speckles. Apart from these spots, the fluorescent intensity is uniformly low due to the homogeneity of the customized speckles' intensity away from optical vortices. Qualitative comparison between the two distinct fluorescent patterns illustrates the degree to which customizing the speckle intensity statistics can enhance the performance of speckled illumination. Quantitatively, the customized speckles were able to create fluorescent spots three times smaller than the diffraction limit set by the illumination optics.

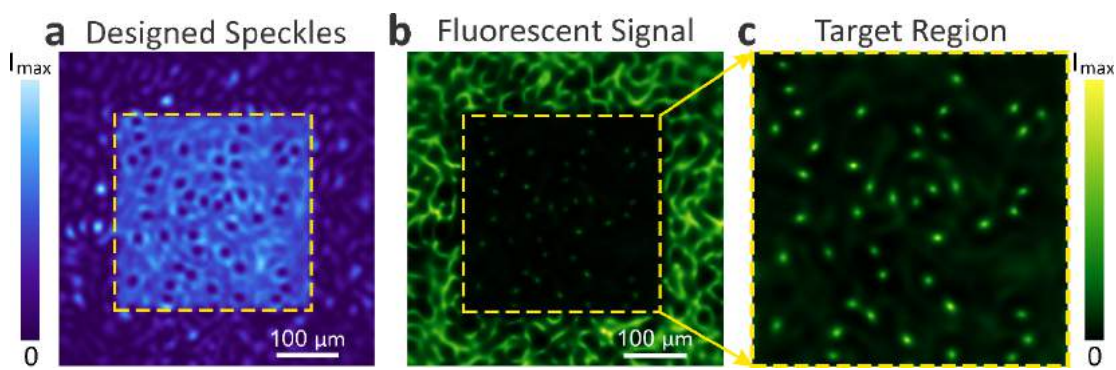

**Figure 2**. A speckle pattern (a) is designed to photoconvert a uniform protein sample. Within the yellow square in (a), optical vortices are randomly embedded in a bright background, and outside are Rayleigh speckles. Fluorescence image of unconverted protein inside the square (b) shows isometric and isotropic spots produced by the vortices in the tailored speckles (c). Unconverted protein outside the square in (b) features large, irregular, and interconnected fluorescent grains.

Nevertheless, significant scientific challenges remain to be solved in order to further speckle-based applications: namely, developing a technique to customize speckled light inside a random scattering medium, tailoring 3D volumetric speckles, and creating vector speckle fields with customized statistics. Technical advances in wavefront shaping devices will facilitate addressing these challenges. For example, increasing the effective number of independent phase modulating pixels on a SLM can provide more degrees of freedom for creating customised 3D speckle patterns, or vector light field for high-NA imaging applications. Future improvements to the SLM operation speed will facilitate tailoring random light inside dynamic scattering media, potentially even allowing random light tailoring inside live biological systems.

**Concluding Remarks**

Recent developments in customizing laser speckles [132-135] have resulted in simple, yet versatile, techniques for creating and controlling random light, which can easily be adapted for use in a diverse range of optical experiments and applications. For example, the ability to arbitrarily control the non-local correlations and intensity PDFs of speckle patterns can be used to create complex optical-potentials for studies on the transport of cold atoms, active media, and microparticles [131]. Potentially, it can also enhance many structured-illumination applications like speckle illumination microscopy [345], super-resolution imaging [129], and high-order ghost imaging [346]. A proof-of-principle experimental demonstration [136] has demonstrated that intelligently tailored speckles can significantly outperform commonly used Rayleigh speckles in nonlinear pattern illumination microscopy. In the future, new techniques for customizing speckle patterns can potentially provide drastic improvements to the myriad of speckle-based applications currently in existence, and potentially lead to the development of new applications.

**Acknowledgements**

The authors thank their co-workers Yaron Bromberg, Hasan Yılmaz, and collaborators Joerg Bewersdorf and Mengyuan Sun for their contributions to the works presented here. They also acknowledge financial support from the Office of Naval Research (N00014-20-1-2197) and the National Science Foundation (DMR-1905465).

## 8. Ultrafast structured beams and intense magnetic fields

*P. B. Corkum[1] and Carlos Hernández-García[2]*

[1]University of Ottawa and the National Research Council of Canada
[2]Universidad de Salamanca

**Status**

With spatial light modulators, Q- or S-plates, it is possible to impose any retardation on any spatial element of an optical beam in the image plane of the retardation plate, enabling the generation of vector or vortex beams. This ability is limited in space by pixel size and the resolution of the image system, in intensity by damage to the phase plates, and in time by the spectral bandwidth of the spatial element that forms the waveplate. The latter restricts the generation of ultrafast structured beams in the femtosecond or even attosecond regimes, where high frequencies and broad spectral bandwidths are required. At the femtosecond timescales, this restriction can be relaxed through femtosecond pulse compression (with fiber compression or in thin dielectric windows), reaching the few-cycle limit of light [137,138].

**Current and Future Challenges**

Femtosecond cylindrical vector beams have been demonstrated recently to allow for coherent control over currents, transferring the light's topology to a material in an ultrafast fashion [139,140]. Electrons created by structured light beams can re-radiate, thereby transferring the original properties of light to a new frequency—higher frequency for generating high harmonic radiation, lower frequency for THz magnetic fields. Information about ultrafast dynamics in the medium is encoded in the generated radiation.

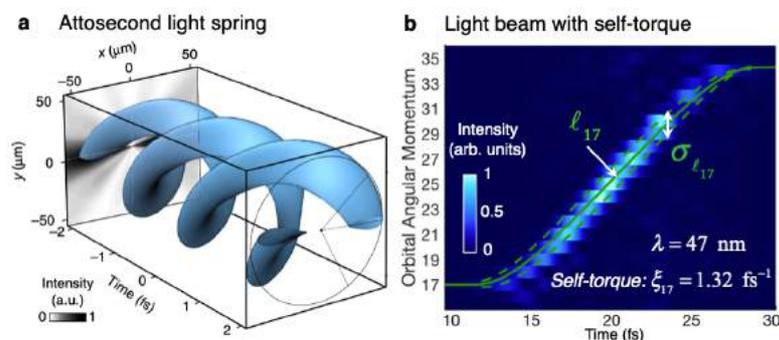

**Figure 1.** (a) Attosecond light spring [129], [130] and (b) XUV beam with self-torque or time-dependent OAM [131].

High-harmonic generation is one of the most robust mechanisms to transfer structured light to the ultrafast, high-frequency regime. In gas-phase high-harmonic generation, the phase-matched emission of all dipole emitters, together with angular momentum conservation rules, allow for the vectorial [141,144] and orbital angular momentum [141,142,145,146] properties from the infrared to be transferred to the extreme-ultraviolet (XUV). Thus, highly nonlinear up-conversion can produce high-harmonic radiation with almost any orbital angular momentum and polarization, and more generally, with any vectorial structure. Interestingly, high-harmonic generation allows for the topological properties of harmonic light to be controlled at the attosecond time scale. If several harmonics with different orbital angular momenta are composed, ultrafast light is arranged like a coil spring—a "light spring" [141,142]. In addition, high-harmonic generation allows for the orbital angular momentum of light beams to be varied in the sub-femtosecond scale, thus creating light beams with time-dependent orbital angular momentum or self-torque [143]. Such ultrafast structured beams, "light springs", or self-torque beams are not found in other spectral regimes.

**a** Scheme to induce intense ultrafast B fields

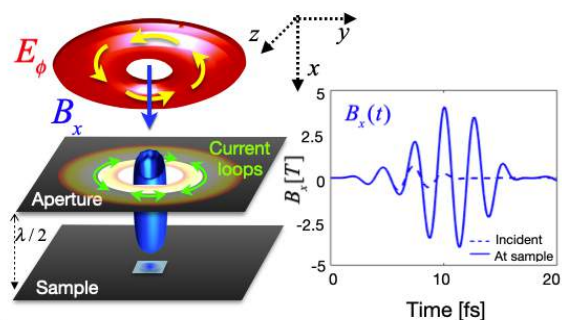

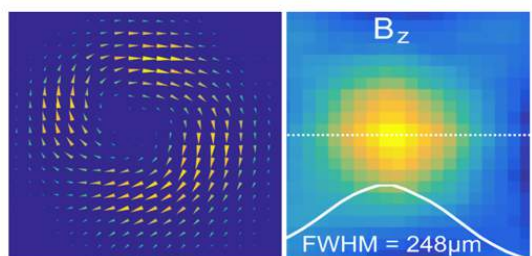

**Figure 2.** (a) Enhanced magnetic field (blue) at a metal sample irradiated by an infrared azimuthally polarized beam [136], [137]. (b) Ring current measured with a 25x25 μm² detector (left) and the magnetic field (right) calculated [127], [128] using the Biot-Savart law.

In gases, where the electron moves primarily in the vacuum, the fundamental field is almost solely responsible for the electron's motion and for the high harmonics that the gas emits. Phase matching forces the conservation of orbital angular momentum [141,142,145,146]. In solids, the electron's interaction with the solid cannot be ignored, but still, angular momentum is conserved. Turning this around, intense infrared pulses can serve as a flexible probe of solids wherein the electron's motion is partially controlled by the infrared light and partially by the electron's interaction with the material, but the high harmonics that emerge still must satisfy phase matching. For example, as high harmonics are developed in metals, metals will produce vector beams and be probed by vector beams [147].

A different way to modify a light pulse is to use a structured material. For example, when a metal containing a hole interacts with azimuthally polarized light [148,149], a large ring current can oscillate at the laser frequency. This current generates ultrafast, intense longitudinal magnetic fields isolated from the electric field (see Fig. 2(a)).

An alternative method is to drive ring currents using quantum control. A quantum-controlled current is not limited by the angular momentum of the incident beam because electrons and holes are created together, and each gain equal but opposite momentum and angular momentum. In semiconductors, quantum control allows for any current structure to be engineered on any pixel [150].

In the 1990s, there was a lot of work on coherent control of semiconductor currents, using linear or circular polarization. For this work, a detector was developed in cold-grown GaAs. That detector has been adapted to measure current driven by azimuthally polarized fundamental and second harmonic light. Fig 2(b) shows the ring currents that are needed to launch a THz "flying torus" [151, 152]. Neither linear nor nonlinear spectroscopy with flying tori have been studied.

Although quantum control is feasible whenever there are interfering pathways to a final state (in this case, the direction of the current), the work in GaAs is in the perturbative limit. In contrast, breakdown in gases or dielectrics will allow high intensity control and high magnetic fields. In gases, the combination of fundamental and second harmonic light transforms a gas into a plasma that has initial conditions imposed by the transformation process.

If we were to ask for the highest magnetic fields that humans can controllably generate, these fields are created within intensely irradiated plasmas. However, magnetic fields within plasmas are difficult to use since they are internal to the plasma. High-intensity quantum control will allow us to generate large, controllable, magnetic fields that are as isolated as possible from the plasma, making these fields useful.

**Concluding Remarks**

Producing and applying XUV/soft X-ray structured beams and ultrafast intense magnetic field pulses will become important forefronts of research with structured light. The study of magnetic helicoidal dichroism or magnetization switching with structured laser pulses [153, 154] are examples of applications of ultrafast structured beams. While it is early days for soft X-ray structured beams, there are two frontiers. First, high-harmonic generation is a robust source of structured beams in the XUV that can be used to observe electrons and, through them, the response of matter to extreme radiation. This matter can be any type of new material—chiral, magnetic, or topological materials. XUV or soft X-ray radiation can also be integrated with attosecond science, allowing for unusual pulses such as "light springs" or light beams with self-torque.

Further advances in the technology of midinfrared structured driving beams may enable the generation of attosecond structured beams deeper in the soft X-rays. In addition, the use of solid targets may hinder new scenarios for structured high-harmonic generation. Secondly, the development of the quantum optics of soft X-ray beams is required for the application of such structured beams in fields such as long-distance space communications which can benefit from their very low beam divergence and high photon energy.

## 9. Metaphotonics with Structured Light

*Haoran Ren[1] and Yuri Kivshar[2]*

[1]Monash University
[2]Australian National University

**Status**

The wave-particle nature of light leads to multiple degrees of freedom such as wavelength, amplitude, phase, polarisation, and angular momentum, which can be controlled in spatial, temporal, and spatial-temporal domains. Structured light patterns were first observed in the double-slit experiment of Thomas Young, where the amplitude and phase interference created bright and dark fringes. Today we understand that a light beam can be structured into millions of transverse modes (e.g., Hermite–Gaussian, Laguerre–Gaussian, etc.) in a square millimetre [96], an extraordinary resource for boosting information capacity. In structured light, singular photonics exhibits topological properties possessing dark singularity centres in a phase vortex with the orbital angular momentum (OAM) of $l\hbar$ per photon ($l$ can take any integer value in $[-\infty,\infty]$, and $\hbar$ is the reduced Planck constant) (Fig. 1(a)); a polarisation vortex manifested by a tensor product of the polarisation and OAM degrees of freedom (defined as $|\psi\rangle = \cos\left(\frac{\theta}{2}\right)|l\rangle|R\rangle + e^{i\alpha}\sin\left(\frac{\theta}{2}\right)|-l\rangle|L\rangle$, where $\theta$ and $\alpha$ denote the weighted contribution of and relative phase between left- ($L$) and right-handed ($R$) circular polarisations) (Fig. 1(b)); and a plasmonic vortex carrying the total angular momentum resulting from spin-orbit coupling (Fig. 1(c)).

Metaphotonics has recently transformed the photonic design for the control of multi-dimensional photonic vortices. To implement phase vortices in real space, different dielectric and plasmonic metasurfaces were designed. A high-index dielectric nanopillar with strong mode confinement was developed as a truncated waveguide with an effective mode index and phase response (Fig. 1(d)). The high-index nanopillar can be designed as a subwavelength waveplate with strong birefringence, exhibiting different phase accumulations for the polarisation along the long and short axe (Fig. 1(e)). Geometric metasurfaces based on the Pancharatnam–Berry phase have been used to create a phase vortex through the in-plane rotation of asymmetric nanopillars [155]. Meanwhile, each anisotropic nanopillar can function as a subwavelength waveplate for implementing polarisation vortices in real space [156].

Huygens' metasurfaces offer an alternative platform to realize phase vortices through spectrally overlapping electric and magnetic resonances (Fig. 1(f)) [157]. Ultrathin plasmonic metasurfaces based on the near-field mode hybridization have been used to create phase vortices (Fig. 1(g)) [158]. Additionally, metal–insulator–metal meta-atoms that support the gap plasmon resonance in a magnetic field could enable highly efficient generation of phase vortices in reflection (Fig. 1(h)). Photonic crystal slabs possess an inherent polarisation vortex in momentum space around bound states in the continuum (BIC) of the periodic structures, featuring the in-plane winding of a vector field and thereby a polarisation vortex (Fig. 1(i)) [159]. In the nonparaxial limit, the space and polarisation degrees of freedom are non-separable, giving rise to the total angular momentum, a measurable quantity through spin-orbit coupling. Surface plasmon polaritons (SPPs)—a tightly confined surface wave beyond the diffraction limit—open the possibility of producing subwavelength plasmonic vortices in the near-field region. For instance, Fig. 1(j) presents a plasmonic nanoring aperture used for the multiplexing generation and detection of different plasmonic vortices [160]. Fig. 1(k) shows the use of plasmonic grooves for the excitation and ultrafast imaging of optical skyrmions in SPPs [161].

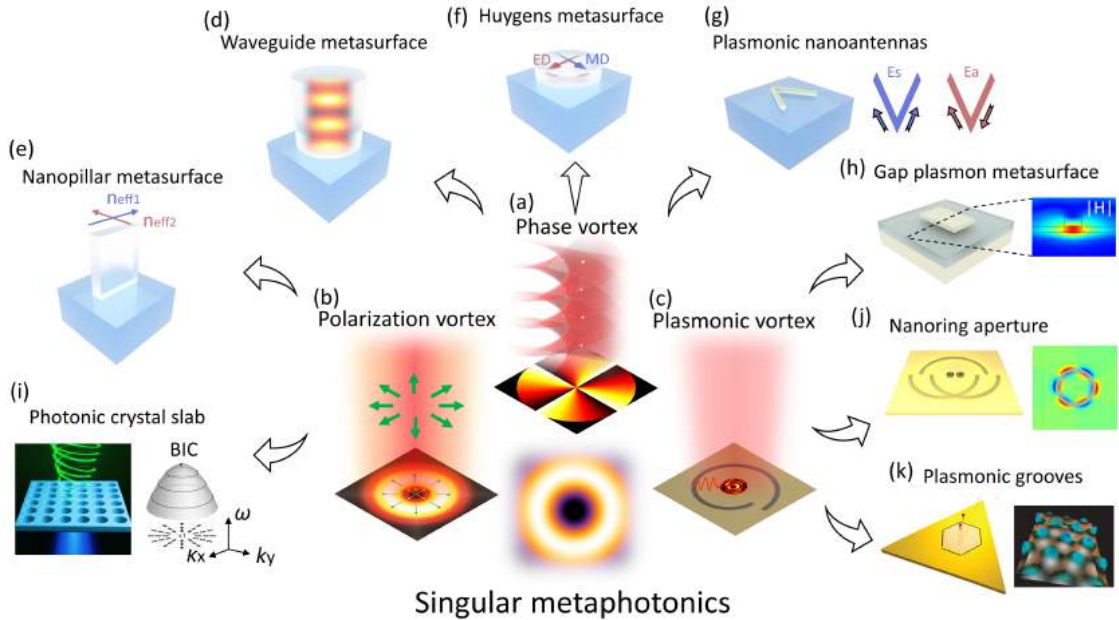

**Figure 1.** Illustration of singular metaphotonics based on the manipulation of phase vortex (a), polarisation vortex (b), and plasmonic vortex (c). (d-h) Design principles of different meta-atoms used for the generation of phase and polarisation vortices, including (d) a high-index dielectric waveguide, (e) a nanopillar waveguide, (f) an ultrathin dielectric cylinder in a Huygens' metasurface, (g) plasmonic nanoantennas with hybridized plasmon modes, (h) a metal-insulator-metal structure supporting gap plasmon resonance. (i) A photonic crystal slab designed for creating polarisation vortex lasing modes in momentum space around the BIC resonance. (j) A plasmonic nanoring aperture used for on-chip OAM multiplexing through the total angular momentum mode-sorting nanoring slits. (k) Plasmonic grooves used for ultrafast imaging of optical skyrmions carried by propagating SPPs. (i-k) Reprinted by permission from American Association for the Advancement of Science in References [159–161].

## Current and Future Challenges

We have provided a review summary on singular metaphotonics and highlighted some nanophotonic structures designed based on different physics for the control of phase, polarisation, and plasmonic vortices. For the phase vortex generation in real space, high efficiency metasurfaces can be designed in both reflection and transmission. For reflection metasurfaces, plasmonic materials featuring a metal-insulator-metal configuration could offer superior efficiency. For transmission metasurfaces, low-loss and high-index dielectric metasurfaces featuring an ultrathin thickness (<λ, where λ is the wavelength of incident light) or a relatively large thickness (~λ) can be designed for Huygens' and waveguide metasurfaces, respectively. However, to achieve the most accurate phase digitalization when considering the fabrication error, geometric metasurfaces exploiting the rotation angle-controlled phase response are perhaps more desired. Meanwhile, to achieve high efficiency polarisation vortex generation in real space, anisotropic meta-atoms based on the metal-insulator-metal configuration and all-dielectrics can be designed for the subwavelength polarisation control in reflection and transmission, respectively. Even though metasurface generation of polarisation vortices faces the same challenges as for the phase vortices, it is generally more difficult to use a metasurface device to distinguish and sort structured polarisation singularities.

In addition to the fabrication challenges, metaphotonics devices are most usually passive and unlikely to be able to dynamically switch vortex modes. Besides, structured optical fields are typically limited to a 2D transverse plane without the wavefront control in the propagation direction. The ability to tailor light beyond 2D structured light, towards 3D control (in all spatial coordinates and field components), and even 4D control with spatiotemporal control of structured light, is of fundamental and practical interest for future research. On the other hand, even though BIC-induced polarisation vortices in momentum space feature robustness, alleviation of coaxial beam alignment, and unrestricted choice of materials, symmetry-protected photonic crystal slabs designed near the BIC at the Γ point are extremely sensitive to the change of refractive index, incident wavelength, and incident angle that may easily break the system symmetry. To create plasmonic vortices, nanogrooves engraved in a metal film were generally employed, but they usually have a low coupling efficiency of SPPs. Traditional noble metals such as gold and silver also suffer from high metal losses due to interband transitions in the ultraviolet and visible frequency ranges. More critically, strong dissipation of the highly localized plasmonic vortex fields in the near-field region hinders the SPPs applications for on-chip vortex transmission and processing.

**Advances in Science and Technology to Meet Challenges**

Nowadays, nanofabrication techniques are available to create functional metaphotonics with the resolution down to a few nanometres. Systems such as EBL, focused-ion beam (FIB), and 3LN can enable ultrahigh resolution (EBL, FIB) or 3D manufacturing, but are limited by a low throughput due to a sequential writing process. Masked techniques, such as photolithography, soft lithography, NIL, and colloidal lithography enable high throughput by replicating the entire mask simultaneously but are often limited in resolution or flexibility. The NIL technique is scalable, and large-area roll-to-roll or roll-to-plate techniques have already been developed to enable high-throughput metasurface production towards industrial applications. Alternatively, self-assembly techniques have emerged and could circumvent the need for clean-room facilities and expensive equipment, though this technique lacks the flexibility to create different-shaped nanostructures.

Recently, digital vectorial holography has enabled the generation of advanced vortex beams, in which the phase and polarisation singularity centres can spatially vary either in 3D polarisations [163] or along the propagation direction [164]. Moreover, a 3D wave packet that carries a spatiotemporal optical vortex with a controllable purely transverse OAM has been realized [23]. Multi-dimensional structured light could offer extra degrees of freedom for versatile light-matter interactions, quantum entanglement, optical trappings, harmonic generation, and optical sensing, holding great potential for novel applications that may not be possible otherwise. Even though strong dissipation of the highly localized plasmonic vortex fields hinders on-chip plasmonic vortex transmission and processing, superior transmission efficiency can be offered by low-loss semiconductor nanowires sustaining highly confined optical modes. Recently, an OAM-controlled hybrid nanowire plasmonic circuit was introduced, demonstrating OAM-controlled optical logic operations including AND and OR gates [165]. OAM beams with different topological charges exhibit selective excitation of single-crystalline cadmium sulfide nanowires through coupling OAM-distinct plasmonic fields into nanowire waveguides for long-distance transportation on-a-chip.

**Conclusion**

We believe that metaphotonics provides a great playground for structured light manipulation, and it will lead to a diverse range of ultracompact, ultrahigh-capacity, and ultrahigh-speed devices harnessing multi-dimensional structured light. We believe it is of paramount importance to integrate developed metaphotonics devices with established optical systems for advanced optical imaging, holographic displays, optical and quantum communications, nonlinear and ultrafast light shaping, and turbulence- and scattering-resilient communications and imaging.

**Acknowledgements**

H.R. acknowledges a support from the Australian Research Council DECRA Fellowship DE220101085. Y. K. acknowledges a support from the Australian Research Council (grant DP210101292).

## 10. Structuring Light with Near-Zero-Index Platforms

*Mário G. Silveirinha[1] and Nader Engheta[2]*

[1]University of Lisbon and Instituto de Telecomunicações
[2]University of Pennsylvania

**Status**

Materials provide the means to structure light. Judiciously engineered material platforms, known as metamaterials and metasurfaces, have provided scientists and engineers with versatile tools to control, manipulate, and sculpt electromagnetic waves and fields. In particular, materials whose real part of relative permittivity and/or permeability attain near-zero values at given operating frequencies offer specially interesting platforms for structuring electromagnetic and optical waves [166-168]. At such frequencies, these materials exhibit (a) refractive index near zero, and consequently (b) the wave phase velocity attains very high values (theoretically infinite values) which leads to (c) a "stretched" wavelength and (d) uniform phase distributions. When both relative permittivity and permeability are zero, the electric and magnetic phenomena are effectively decoupled in such materials [166] yielding static-like spatial distributions of electric and magnetic fields, while at the same time they are dynamically time varying. This special feature makes epsilon-near-zero (ENZ), mu-near-zero (MNZ), and epsilon-and-mu-near-zero (EMNZ), which form the general class of near-zero-index (NZI) materials, particularly interesting for wave manipulations, beam shaping and lensing [167-171], for example the wavefronts emerging from an ENZ material block typically inherit the shape of the ENZ-material surface [170].

Since the wavelength of waves in such media can be long even for high frequencies, one can effectively think of this effect as "loosening" the connection between the frequency and the wavelength. Moreover, a block of material in which the wavelength is very long can be viewed electromagnetically as a "point", even though it can be large compared to the free-space wavelength. This phenomenon has enabled numerous exciting features in wave interaction with such NZI media. The "supercoupling" effect [167-169] is an interesting example of such features; when two metallic waveguides are joined together, the connecting segment (i.e., transition region) between the two waveguides can be of any shape and size, if that region is filled with an NZI medium [167,168]. If an ENZ (or MNZ) medium fills the transition region, then the connecting segment needs to be narrow (or wide), while it can be bent and can have arbitrary shapes [167-169] (see Fig. 1(a) and 1(b)). This "electromagnetically point-like" region may provide an unusual coupling between two emitters, effectively causing "near-field coupling" even when the emitters are far apart in space [172] (see Fig. 2(a) and 2(b)). This is an interesting way to structure light in dipole-dipole coupling among quantum emitters.

Another intriguing by-product of such NZI-enabled stretched wavelength can be considered in "sampling and squeezing" waves through narrow channels [167,168], in which one can effectively transfer an "image" through a subwavelength opening (see Fig. 1(c) and 1(d)). The electromagnetic ENZ phenomena have also inspired efforts on other physical, non-electromagnetic, platforms in which other parameters can attain near-zero values. For example, we theoretically studied how one can conceive electronic metamaterials in which effective mass of electrons can be engineered to be near zero, exploring the topic of "transformation electronics" [173].

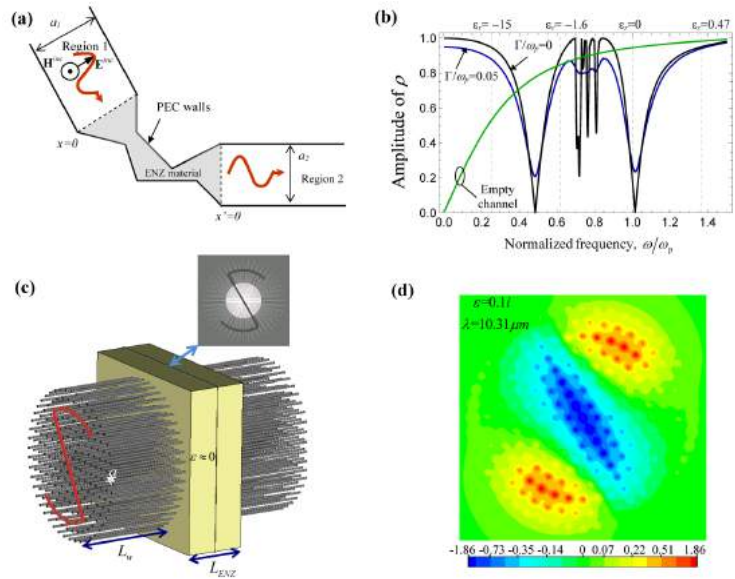

**Figure 1.** Illustration of the ENZ supercoupling effect. (a) Two parallel-plate waveguides are connected through a narrow ENZ channel. As shown in b) for a 180º bend with a narrow channel, the reflection level at the ENZ frequency (ω=ω$_p$) can approach zero in the limit of vanishing material loss and is typically much weaker than for an empty channel. (c) The supercoupling effect may enable the transmission of a complex image through a narrow aperture in a metallic screen embedded in an ENZ material. The image is sampled by an array of metallic wires, which are then "squeezed" through the narrow hole to the other side of the opaque screen. (d) Image transported by the array of metallic wires at the ENZ frequency. Adapted from Refs. [167,168] with permission, copyright (c) American Physical Society.

## Current and Future Challenges

Passive NZI materials must inherently be dispersive, and therefore have a finite bandwidth of interest. However, depending on the feature of interest and the frequency domain of interest, the bandwidth may be sufficient to achieve a desired functionality. For example, consider the phenomenon of "photonic doping" [169], in which a dielectric rod is inserted in an ENZ medium with an arbitrary cross-sectional geometry. As viewed by an outside observer, according to the effective medium theory, for two-dimensional scenarios, this "single-inclusion metamaterial" can be treated as an effective medium such that the effective permittivity is still zero (even though the dielectric rod is inserted in the ENZ host) but the effective permeability can be different from unity (notwithstanding both the dielectric rod and the ENZ host are non-magnetic with unity relative permeability). Such engineered effective permeability exhibits a resonant dispersion as a function of frequency, but this resonance is mainly due to the size and the permittivity of the dielectric rod, which exhibits narrower bandwidth than the bandwidth over which the ENZ host behaves as a material with near-zero permittivity. So in this example, the bandwidth of the ENZ host can be sufficient for the resonant behaviour of the relative permeability.

The idea of photonic doping combined with the fact that in the ENZ media the wavelength is "long" has led to the idea of ENZ-based cavity resonators in which the resonance frequency does not depend on the external shape and geometry of the cavity, but instead it is pinned to the ENZ frequency of the materials [174]. Such geometry-independent cavity resonators have several salient features: (a) they can be the basis for the notion of "flexible photonics", a paradigm in which changing, bending and morphing the external shape of cavities would not affect its resonance frequency; and (b) while changing the external shape and size does not change the resonance frequency, it does affect the quality factor (Q) of the cavity, if a small amount of loss is considered [174]. So here is another unusual situation for structured light in which the cavity's resonance frequency and $Q$ are effectively decoupled from each other, whereas in conventional cavities they are intimately connected.

Furthermore, remarkably, in the limit of vanishing material loss, core-shell ENZ resonators can support embedded eigenstates in the continuum, i.e., non-radiative bound states which despite being coupled to the radiation continuum do not decay in time [175,176]. Geometry-independent ENZ nanoresonators can be exploited in the field of quantum optics [108] where the coupling between an excited atom with a resonant cavity is usually considered. In conventional situations of atom-cavity coupling the resonance frequency of the cavity should match the transition frequency of the excited atoms. This is a delicate balance because a slight change in the shape of the cavity can shift its resonance frequency (since such cavities are usually high-$Q$ cavities) and therefore the cavity would be detuned. Moreover, the vacuum Rabi oscillation depends on the cavity $Q$, which can also be affected by the slight change in the cavity shape. The ENZ-based cavities can provide an interesting solution in this scenario in that one can change the $Q$ of the cavity (and thus engineer the vacuum Rabi oscillation) while the cavity resonance frequency stays tuned [108]. (See Fig. 2c)

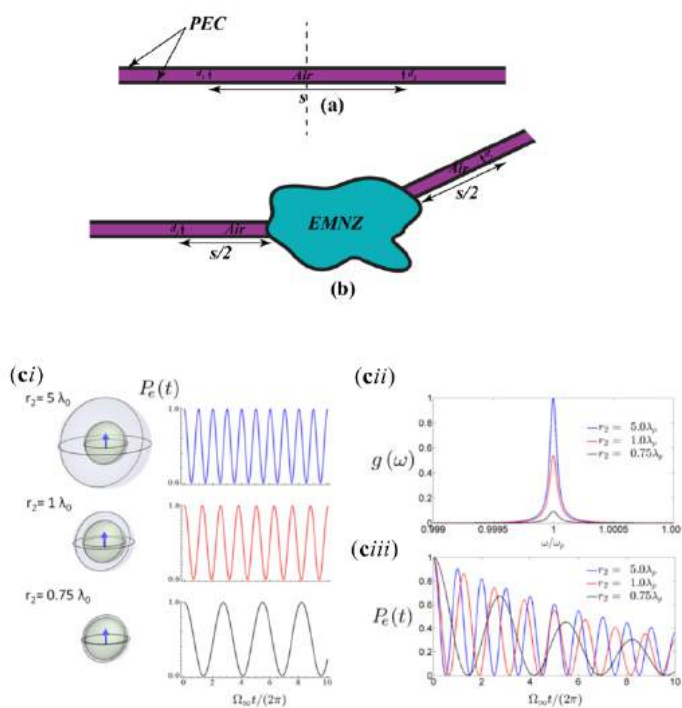

**Figure 2.** (a), (b) An ideal epsilon-and-mu-near-zero (EMNZ) 2D material block with arbitrary cross-sectional shape does not influence the field distribution in the unfilled 2D parallel-plate waveguide sections, and thus behaves electromagnetically as a "single point". In particular, the interaction between two quantum emitters placed in the unfilled waveguide sections is the same as for a straight waveguide regardless of the relative orientation of the unfilled waveguide sections connected to the EMNZ block. (c) Illustration of the decay of an excited quantum emitter enclosed in an ENZ cavity. The cavity resonance is independent of the shell thickness, and thereby the transition frequency of the emitter is always matched to the cavity. However, the coupling strength is sensitive to the shell radius, and hence the Rabi frequency also is. i) Time evolution of the probability of the excited state for different radii of an ideally lossless ENZ shell. ii) Normalized spectral density for a slightly lossy ENZ shell and different shell radii. iii) Time evolution of the excited state in the presence of the lossy ENZ shells. Adapted from Refs. [172,108] with permission. Panels (a) and (b), Copyright is Open access from OPTICA, panels (ci), (cii) and (ciii) Copyright by National Academy of Sciences of the United States.

## Advances in Science and Technology to Meet Challenges

Material loss in some ENZ platforms can be a limiting factor. Since one of the important features of NZI media is the wavelength stretching, it is important to note how material loss can affect this phenomenon. As the refractive index is given by $n = \sqrt{\varepsilon_r + i\varepsilon_i}$, it follows that at the ENZ frequency $\varepsilon_{\mathrm{r}} = 0$, resulting in $n = \sqrt{i\varepsilon_{\mathrm{i}}} = \sqrt{\varepsilon_{\mathrm{i}}}\left(1+i\right)/\sqrt{2}$. Thus the wavelength stretching is approximately limited by $\lambda_{\mathrm{ENZ}} = \lambda_0/\sqrt{\varepsilon_{\mathrm{i}}/2}$. Therefore, the lower the $\varepsilon_{\mathrm{i}}$, the longer the wavelength stretched in ENZ media, and the more uniform the phase distribution in the structure. As a result, for any NZI-based application of interest that depends on the phase uniformity, one needs to ask the following questions: How large is the structure and how much phase variation can be tolerated with the effect still observable? The answer to this question determines how large $\varepsilon_{\mathrm{i}}$ can be. For example, some of the transparent conducting oxides (TCO), such as indium tin oxide (ITO), can exhibit ENZ behavior with $\varepsilon_{\mathrm{i}}$ below unity in the near-IR regime, while silicon carbide (SiC) behaves as ENZ with smaller $\varepsilon_{\mathrm{i}}$ in mid IR wavelength [193]. So each of them can be suitable for a different set of applications, as they have different levels of material loss. It is important to note that one can also engineer metastructures that mimic some of the NZI properties, while the loss can reduced. For example, metallic rectangular waveguides operating at the TE$_{10}$ cut-off frequency possess wave properties resembling some of the ENZ features, so they can be suitable for microwave frequencies. Furthermore, photonic crystals with the Dirac dispersion with accidental degeneracy exhibit effective refractive index near zero, thus providing a platform for NZI properties at optical frequencies [219].

**Concluding Remarks**

Near-zero-index (NZI) photonics is an exciting field of optics and electromagnetics. It encompasses unconventional ways of structuring light due to wavelength stretching, with the material bulk behaving as an electromagnetic point, exhibiting unique features in light-matter interaction, and offering exciting potential applications. Numerous phenomena resulting from the ENZ, MNZ, and NZI features, such as flexible photonics, supercoupling, photonic doping, directive thermal radiation, engineering vacuum fluctuations, ENZ-based quantum optics, super-radiance, emitter-emitter long-range coupling, ENZ electric-based levitation, giant nonlinearity, embedded eigenstates, and more have been explored. Moreover, electromagnetic NZI concepts have also inspired wave phenomena in other physical domains such as acoustics, electronics, amongst others. We are hopeful that this field continues to grow, expand, and reveal other exciting wave phenomena.

**Acknowledgements**

M. G. S. acknowledges partial support from Simons Foundation/Collaboration on Extreme Wave Phenomena Based on Symmetries, from the Institution of Engineering and Technology (IET) under the A F Harvey Research Prize 2018, and from Instituto de Telecomunicações under project UIDB/50008/2020. N.E. acknowledges partial support from Simons Foundation/Collaboration on Extreme Wave Phenomena Based on Symmetries, and from the US Air Force Office of Scientific Research (AFOSR) Multidisciplinary University Research Initiative (MURI) grant number FA9550-21-1-0312.

## 11. Strong Coupling Between Atoms and Guided Light

*Arno Rauschenbeutel, Philipp Schneeweiss, and Jürgen Volz*

Humboldt-Universität zu Berlin

**Status**

The past decade has seen remarkable advances in the field of quantum nonlinear optics, where a strong interaction between individual photons is mediated by quantum emitters [177]. Such strong photon-photon interactions are of both fundamental and technological interest: they are the prerequisite for implementing deterministic quantum logic gate operations for processing optical quantum information [178]. Moreover, photons that strongly interact via a quantum nonlinear medium exhibit complex out-of-equilibrium dynamics that, e.g., enable one to tailor and control the photon statistics of light [179].

Using free-space light fields, photon-photon interactions have been successfully demonstrated in a number of experimental settings. The most established method is to couple atoms with photons that are confined inside a high-finesse optical resonator [180], see Figure 1(b). This allows one to increase the coupling of such a so-called resonator-enhanced atom to the input and output mode of the resonator. In this way, the inherently nonlinear response of the atom mediates strong photon-photon interactions. Alternatively, strong photon-photon interactions have also been demonstrated using the collectively enhanced coupling between propagating light fields and ensembles of strongly interacting Rydberg atoms, so-called Rydberg superatoms [171], see Figure 1(c). Also here the aim is to enhance the coupling of the effective atoms with the input and output light mode to the point where coupling to other modes becomes negligible. In these scenarios, a key figure of merit is given by the so-called $\beta$-factor,

$$\beta = \frac{\Gamma_{trgt}}{\Gamma_{tot}}, \qquad (1)$$

which is the ratio of the emission rate of the initially excited (effective) atom into the target mode, $\Gamma_{trgt}$, and the total emission rate into all possible modes, $\Gamma_{tot}$, see Figure 1(a).

With regards to employing strong photon-photon interactions in future research and technology, it is, however, essential to couple quantum emitters to guided fields in integrated optical platforms. This so-called waveguide quantum electrodynamics (QED) setting has been realised with a variety of emitters and waveguide types [182]. They include semiconductor quantum dots coupled to nanophotonic waveguides, silicon vacancies coupled to diamond waveguides, organic dye molecules coupled to waveguides consisting of an organic crystal-filled glass capillary or to sub-wavelength-diameter silica fibres, so-called optical nanofibres, as well as cold, laser-trapped atoms coupled to nanofibres or one-dimensional photonic crystal waveguides [183, 184]. Finally, resonator-enhanced atoms, realised by single atoms trapped in the evanescent field of whispering-gallery-mode (WGM) microresonators, have been coupled to nanofibres [185], see Figure 2(a).

Waveguide QED systems lend themselves to distributing and processing optical quantum information, to deterministically preparing non-classical states of light, and to realising an almost ideal model-system for strongly correlated, open many-body quantum physics. However, in addition to the nonlinear response at the single-photon level, many of the corresponding experimental protocols require high $\beta$-factors, which should ideally reach 100%. A $\beta$-factor that falls short of this value will, at best, lead to a reduced success probability, e.g., in the case of photon-photon quantum gates [178]. In the worst case, a too small $\beta$-value impedes the implementation of the protocol altogether, e.g., in the case of a photon number-dependent delay line based on quantum nonlinearities [186].

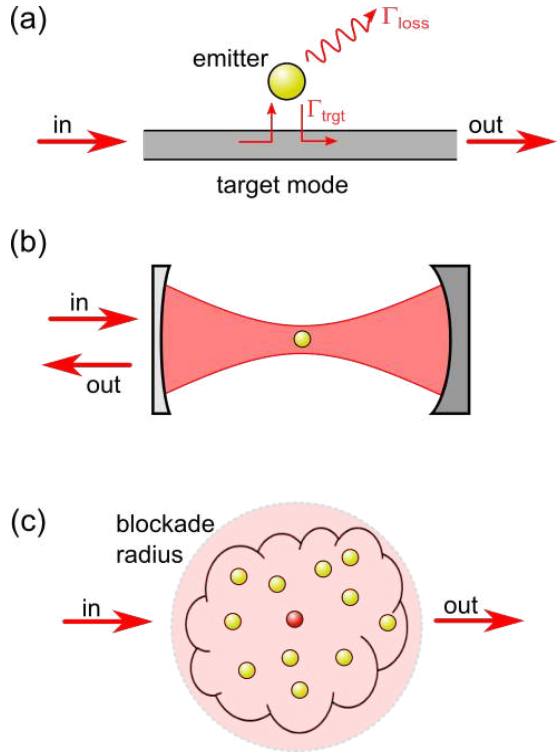

**Figure 1.** (a) Quantum emitter (yellow sphere) coupled to a target mode (grey). The $\beta$-factor is defined according to Eq. (1) with $\Gamma_{tot} = \Gamma_{loss} + \Gamma_{trgt}$. (b) Resonator-enhanced atom with optical input and output modes (arrows). (c) Rydberg superatom composed of a cloud of laser-cooled atoms in the ground state (yellow) and one optically excited Rydberg atom (red). The presence of the latter prevents other atoms inside the so-called blockade radius to be excited.

## Current and Future Challenges

It is important to note that, for the above-mentioned applications, $\Gamma_{trgt}$ in Eq. (1) refers to the emission into a single spatiotemporal target mode, such that all emitted photons exhibit the same lifetime-limited spectrum. Moreover, for many quantum applications, the photons have to be indistinguishable when they are emitted at different times or by different waveguide-coupled emitters. In this context, cold atoms stand out in terms of their superior coherence properties and their negligible spread of resonance frequencies, or inhomogeneous broadening. As a consequence, large ensembles of waveguide-coupled atoms can interact collectively with the guided light. This makes them a prime candidate for scaling up waveguide QED systems. Nonetheless, perfect coupling, with $\beta \sim 100\%$, of a single optical mode to a large number of identical and fully coherent quantum emitters remains an important challenge for existing implementations. For example, when coupling laser-cooled atoms to the evanescent field surrounding optical nanofibres, $\beta$-factors in the few-percent regime are expected and have been realised experimentally.

For a single atom that is coupled to the evanescent field of a nanophotonic waveguide, maximizing the $\beta$-factor for a given decay rate into the free space modes amounts to maximizing the single photon Rabi frequency,

$$\Omega_0 = \vec{d} \cdot \vec{E}_0(\vec{r})/\hbar \,, \qquad (2)$$

where $\Omega_0$ is chosen to be real and positive, $\vec{d}$ is the dipole moment of the atomic transition, and $\vec{E}_0(\vec{r})$ the field per guided photon at the position of the atom, $\vec{r}$. Now, $|\vec{E}_0(\vec{r})|$ increases approximately exponentially when approaching the waveguide. From this perspective, it is thus advantageous to trap the atoms at the smallest possible distance from the waveguide surface. However, for distances smaller than ∼ 200 nm, the van der Waals force becomes so large that it can no longer be straightforwardly counteracted by optical dipole forces. Furthermore, we have $|\vec{E}_0(\vec{r})| \propto 1/\sqrt{A}$, where A is the cross-sectional area of the guided mode, which can in principle be decreased by increasing the refractive index of the waveguide material and concomitantly reducing the transverse waveguide dimensions. However, for the refractive indices accessible with low-loss dielectric materials, even for an atom placed inside a (possibly slotted) waveguide, near-unity $\beta$-values are still out of reach.

**Advances in Science and Technology to Meet Challenges**

We now elaborate on three possible strategies that lend themselves to meeting the grand challenge of reaching near-unity $\beta$-factors for single (effective) atoms. First, $\beta$-factors of about 47%, that have been experimentally demonstrated for the resonator-enhanced atom, can in principle be further increased by improving the Purcell factor of the WGM resonator, $\eta \propto Q/V$. Here, $Q$ is the quality factor of the resonator, and $V$ is the effective resonator mode volume. Currently, for fused silica-based WGM resonators, $\eta$ is limited by scattering-induced losses due to surface roughness and pollution. Here, we thus expect a major step forward by employing advanced resonator production and post-processing techniques.

Along the same lines, the single-atom $\beta$ can be increased by reducing the group index, $n_{gr} = c_0/v_{gr}$, of the guided light using photonic crystal waveguides, see Figure 2(b), where $c_0$ is the speed of light in the unstructured waveguide, and $v_{gr}$ is the group velocity. The $\beta$-factor of the photonic crystal waveguide is then given by,

$$\beta_{pc} = \frac{n_{gr}\beta_0}{n_{gr}\beta_0 + (1-\beta_0)}, \tag{3}$$

where $\beta_0$ is the $\beta$-factor of a single atom coupled to the unstructured waveguide. Here, the major advancement with respect to previous work will be to realise stable trapping of atoms in a region of the waveguide where $\beta_0$ is large while realising a photonic crystal waveguide with small group velocity and small propagation losses. While the latter is mostly a technical challenge, atom trapping in high-$\beta_0$ regions of a photonic crystal waveguide is still subject to current research.

Finally, the coupling of the input and output mode to an ensemble of atoms can be collectively enhanced. Indeed, the collective $\beta$-factor scales with the atom number, $N_{at}$, as,

$$\beta_{coll} = \frac{N_{at}\beta_0}{N_{at}\beta_0 + (1-\beta_0)}. \tag{4}$$

Thus, using only a few hundred atoms, $\beta_{coll} \sim 1$ is reached. However, the response of an ensemble of independent atoms differs from that of a single quantum emitter when the ensemble interacts with more than one photon. Specifically, the inherent nonlinearity featured by each of the atoms is "diluted" because two consecutive photons can interact with different atoms. Thus, in order to profit from a large collective $\beta$-factor for implementing quantum nonlinearities, one needs to introduce atom-atom interactions, e.g., in the form of a dipole blockade in Rydberg superatoms, see Figure 2(c). Here, the major advancement will be to control the detrimental influence of the nearby waveguide on the coherence properties of the atomic Rydberg levels [187], e.g., by working with particularly thin nanofibres that feature super-extended evanescent fields [188].

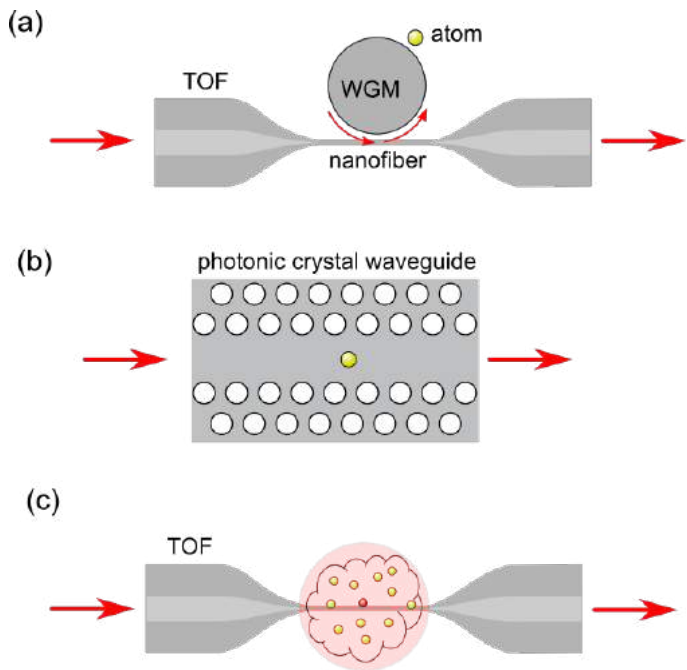

**Figure 2.** (a) Resonator-enhanced atom realised by an atom coupled to the evanescent field of a WGM resonator, interfaced using an optical nanofibre. The latter is realised as the subwavelength-diameter waist of a tapered optical fibre (TOF). (b) Atom coupled to a photonic-crystal-based waveguide with large group index. (c) Rydberg superatom coupled to an optical nanofibre.

**Concluding Remarks**

All three approaches towards reaching near-unity $\beta$-factors laid out above, resonator-enhanced atoms, large group index, and waveguide-coupled Rydberg superatoms, see Figure 2(a)–(c), come with considerable technical and conceptual challenges. Meeting these challenges will mark important advances in science and technology in their own right, ranging from next-generation ultra-high $Q$ factor WGM microresonators to slow-light waveguides with record-low propagation loss to passivation and charge control of dielectrics at the level of single elementary charges. Most of all, the corresponding research and development effort is justified by the exciting applications that are enabled by atomic waveguide QED in the high-$\beta$ range. In particular, it opens the route towards the implementation of near-ideal fibre-coupled nonlinear quantum devices, which will mark a major breakthrough in quantum optics and constitute a key resource in quantum sensing, quantum metrology, quantum communication, as well as quantum simulations.

**Acknowledgements**

We acknowledge funding by the Alexander von Humboldt Foundation in the framework of the Alexander von Humboldt Professorship endowed by the Federal Ministry of Education and Research. Moreover, financial support from the European Union's Horizon 2020 research and innovation program under grant agreement No. 899275 (DAALI) is gratefully acknowledged.

## 12. Surface Waves

*Daniel Leykam[1] and Daria A. Smirnova[2]*

[1]National University of Singapore
[2]Australian National University

**Status**

Surface physics is messy; surface waves are no exception. Analytical solutions are scarce, numerical calculations are resource-intensive, and there is a myriad of possible interface configurations to consider. Emerging from these challenges are elegant theories and numerous applications.

Surface waves are important because they exhibit remarkable properties unattainable using isolated bulk wave media. Long-studied examples include gravity waves at liquid surfaces, elastic waves at the surfaces of solids, subgap Tamm or Shockley electronic states at terminated semiconductors, and electromagnetic plasmon-polaritons at metal-dielectric interfaces, illustrated in Figure 1.

Initial interest in surface waves stemmed from their ability to guide and strongly confine energy, observed most strikingly in the destructive power of seismic Rayleigh waves predicted in 1885. At smaller scales, electromagnetic surface waves guide and localise light below the diffraction limit [189]. More recently, the non-trivial spatial structure of surface waves such as their transverse spin enables chiral coupling between localised sources and guided modes.

A long-standing challenge has been ab-initio prediction of the existence and properties of surface waves. Even in the simplest case of homogeneous media described by a few material parameters, novel solutions continue to be discovered, such as Dyakonov surface waves of anisotropic electromagnetic media [190].

Since the 2000s, studies of surface waves have been reinvigorated thanks to the development of topological band theory [191]. Topological band theory enables prediction of novel types of surface waves of periodic media such as photonic crystals, i.e. systems with wavelength-scale variations in their material parameters.

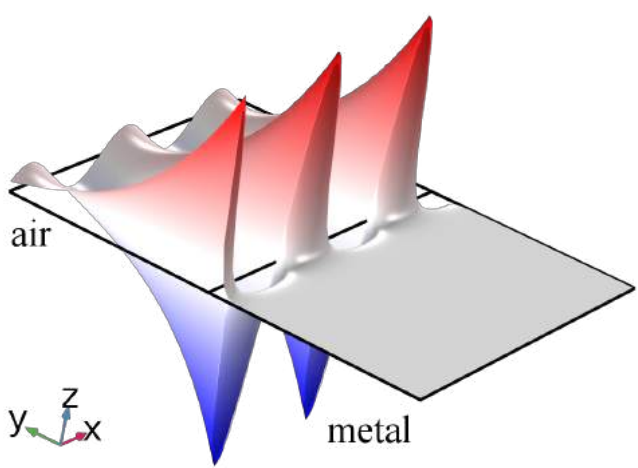

**Figure 1.** Magnetic field distribution (the out-of-plane component $H_z$) in the TM-polarised surface plasmon-polariton wave, showing different transverse localization scales in the two media (air and metal) brought into contact.

Topological band theory shows promise as a systematic approach for designing surface waves and optimising their properties for applications including precision sensing, compact waveguides, and signal processing. At the same time, techniques from the structured light community are being fruitfully applied to study topological bands [192].

Ongoing research aims to better understand connections between surface waves emerging for different classes of waves. This will not only allow us to design and optimise surface waves in a variety of wave systems, but also observe novel physics. For example, Weyl semimetals are topological materials supporting coexisting surface and bulk modes exhibiting Weyl quasiparticles originally hypothesised in 1929. In 2015, Weyl semimetals were observed for the first time using an electronic system (TaAs) and an analogous microwave photonic crystal [122].

**Current and Future Challenges**

While there has been great success in emulating edge modes of 1D and 2D condensed matter systems, studies of protected surface waves of 3D systems remain in their infancy [194]. Such surface waves exhibit linear Dirac-like dispersion, with locking between their spin and momentum, illustrated in Figure 2 and elaborated on further in Section 14.

One challenge is that many models of topological surface waves were originally formulated for electronic condensed matter systems with fermionic spin-orbit coupling. To implement similar surface waves for classical wave systems requires other effects such as bi-anisotropy, use of orbital angular momentum modes as a spin-like degree of freedom, or additional crystalline symmetries. Limits to the strength of these effects may lead to non-ideal dispersion relations, such as surface waves co-existing with bulk bands, resulting in bulk scattering losses.

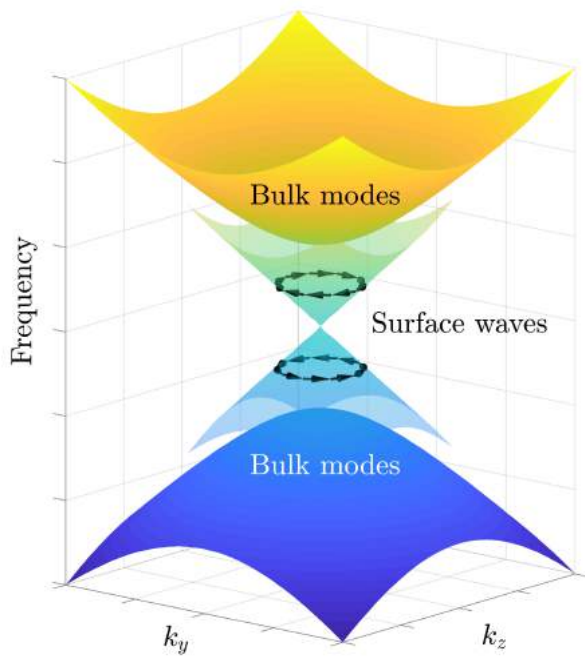

**Figure 2.** Massless Dirac-like dispersion of topological surface waves with spin–momentum locking within the bulk gap in momentum space. The (pseudo-)spin texture is illustrated by black arrows.

Not all classes of surface waves may be accessible in a given material platform. For example, sound-based phononic crystals are based on manipulating scalar waves and can therefore be well-described by simple tight binding models [195]. By contrast, 3D photonic crystals typically have band structures complicated by orbital and polarisation degrees of freedom, making space group arguments insufficient to guarantee the existence of protected surface waves.

The holy grail is the ability to identify the best possible surface wave for a given application subject to material or design constraints. At first glance, this seems like a hopeless task given the explosion in degrees of freedom compared to uniform media. Topology allows us to make concrete statements about some properties of surface waves, such as the difference between the number of forward and backward propagating waves, independent of details such as precise material parameters or the interface shape. However, continuous parameters such as the wave speed or degree of localization are not topologically protected and must be optimised using conventional methods.

Another challenge is to understand the robustness of surface waves against effects including scattering losses, fabrication imperfections, imperfect symmetries, and incomplete band gaps. Topological band theory as originally developed for electronic condensed matter materials was limited to lossless, non-interacting wave systems described by the Schrödinger equation, analogous to the paraxial wave equation. Generalising beyond these constraints will allow us to identify robust surface states for new kinds of wave media.

**Advances in Science and Technology to Meet Challenges**

Advances in fabrication technologies will expand the platforms available for implementing surface waves and give us new tools to tune their properties. Areas of active research include heterogeneous multilayer metamaterials, novel two-dimensional materials such as graphene, hexagonal boron nitride and twisted monolayers [196], nanostructured metasurfaces, hybrid polariton systems, surface magnetoplasmons in gyrotropic materials, and even electronic topological materials [197].

3D printing is maturing as a fast and flexible approach towards prototyping topological surface waves, with many recent high-profile works in acoustics [195] covered in Section 17. For electromagnetic waves, 3D printing is practical for microwave photonic crystals, but scaling up to optical frequencies remains challenging due to the need for 3D nanofabrication. State-of-the-art methods such as direct laser writing have been used to realise topological edge and surface waves in the terahertz and near-infrared [122]. Advances in 3D printing including finer resolution and the ability to incorporate more combinations of materials will open up new possibilities for surface waves in the visible frequency range.

Methods from the field of structured light may offer an easier route towards generating and finding useful applications of topological surface waves. Specifically, one can use internal degrees of freedom such as orbital angular momentum as synthetic dimensions, with hopping along the synthetic dimension mediated by periodic spatial or temporal modulation [198]. Surface waves that mix spatial and internal degrees of freedom show promise for applications such as robust, high-efficiency mode conversion and non-reciprocity in planar integrated photonic circuits. Realising this goal will, however, require the integration of high-efficiency optical modulators.

Generalisations of topological band theory to broader classes of wave media are being actively pursued. These include active or lossy (non-Hermitian) systems (see also Section 6), effective medium theories describing metamaterials [199], non-periodic, dispersive, and nonlinear media [200]. Advances in these directions will allow us to identify novel combinations of materials supporting surface waves and better understand their robustness to losses and other perturbations. For example, for high power device applications, it is essential to understand the conditions under which nonlinear surface waves remain stable.

There is growing interest in applying machine learning techniques to physics problems [201]. Potential applications to surface waves include discovery of interface designs with superior surface wave properties via generative modelling, identification of new classes of topological wave media supporting robust surface modes, and mesh-free neural network-based beam propagation methods for numerical simulation of complex interface geometries [202]. Applications of machine learning to the broader field of structured waves are discussed further in Section 25..

## Concluding Remarks

Topological band theory has led to a resurgence of interest in surface waves in quantum and classical systems in a similar vein to how analogies with bulk electronic band structures gave rise to the fields of photonic and phononic crystals. While researchers are still attempting to understand all the subtleties of topological phases, there is no doubt that these discoveries will require band structure textbooks to be rewritten.

So far, the flow of ideas has largely been unidirectional, from electronics to photonics and acoustics. Given the greater appreciation of the universality of bulk and surface waves occurring in various fields, there is great potential for recent advances in structured light to be applied to electronic and acoustic systems [44].

Engineered interfaces and surface waves will serve as a flexible testbed for probing relativistic physics (fundamental science, quasiparticles) and strong light-matter interactions (due to the field confinement). Advances in their basic science will lead to technological breakthroughs in more applied areas, such as ultra-thin, resilient, and flexible surface wave-based devices. We have only scratched the surface of potential applications.

## Acknowledgements
D. A. S. acknowledges support from the Australian Research Council (DE190100430).

## 13. Photonic spin-orbit interactions at metasurfaces: stochastic, Rashba and quantum effects

*Kexiu Rong[1], Bo Wang[2,1] and Erez Hasman[1]*

[1]Technion – Israel Institute of Technology
[2]Shanghai Jiao Tong University

**Status**

Light possesses both spin and orbital angular momentum (OAM); the former is associated with circular polarisation states, and the latter arises from azimuthal phase gradients of the light field. A coupling between spin and OAM (or linear momentum) occurs when light interacts with anisotropic or inhomogeneous structures, giving rise to optical phenomena in which the spin of light affects and controls the spatial degrees of freedom of light, such as the vectorial field distribution and propagation path [203,204]. These spin-orbit interactions (SOIs) bring forth novel spin-optical effects (e.g., photonic spin Hall effect (PSHE) and photonic Aharonov-Bohm effect) and enable efficient spin-dependent light manipulations [205-208].

Metamaterials are artificial structures assembled from multiple elements smaller in scale than the wavelength of external stimuli, endowing a medium with unique electromagnetic responses and functionalities. Metasurfaces [209-215], metamaterials of reduced dimensionality, are phased arrays composed of resonant optical nanoantennas, which facilitate substantial control of local light scattering properties. Controlling the electromagnetic response of metasurfaces can be achieved by a geometric phase (e.g., Pancharatnam-Berry phase) mechanism [88,209], enabling an excellent platform to investigate new types of SOI effects from the classical to quantum regime. In this roadmap, we would like to introduce novel types of SOIs utilising geometric phase metasurfaces (GPMs): (i) Stochastic PSHE [211], (ii) Photonic Rashba effect [212], (iii) Quantum entanglement between the spin and the OAM of photons [213].

(i) The study of SOIs in disordered systems offers a wealth of interesting effects and numerous potential applications, such as suppressing undesired optical scatterings and achieving ultra-sensitive optical metrologies. An optical metrology that can detect extremely weak disorders in a deep-subwavelength resolution is critical for nanotechnology. Recently, we reported on a stochastic PSHE arising from space-variant Berry-Zak phases, which are generated by disordered magneto-optical effects. This effect is observed from a spatially bounded lattice of ferromagnetic meta-atoms displaying nanoscale disorders. Our approach may be used for sensing deep-subwavelength disorders by actively breaking the photonic spin symmetry and may enable investigations of fluctuation effects in magnetic nano-systems.

(ii) Heterostructures combining a thin layer of quantum emitters and planar nanostructures enable custom-tailored photoluminescence in an integrated fashion. Recently, there has been a surge of interest in selectively manipulating quantum emitters—that is, valley excitons—in transition metal dichalcogenide (TMD) monolayers due to their potential as an alternative information carrier in valleytronics. GPMs constructed of anisotropic nanoantennas with space-variant orientations allow the manipulation of light by the spin degree of freedom [209,210]. This is enabled by the polarisation evolution of light on the Poincaré sphere, thus generating spin-dependent Pancharatnam-Berry phases for the spin-flipped components. Hence, the GPMs represent an attractive candidate to perform the desired valley separation required by valleytronics, inspired by spin-dependent phenomena such as the PSHE and photonic Rashba effect underpinned by SOIs [216-218].

(iii) Quantum information provides a route to solve problems in reduced time and complexity by exploiting fundamental quantum principles such as superposition and entanglement. Moreover, due to a relatively easy manipulation and long quantum coherence time, single photons encoded with quantum states are an appealing candidate to implement quantum algorithms. Hence, generating and manipulating entangled photon states using SOIs mediated by metamaterials is at the heart of the field of photonic quantum information.

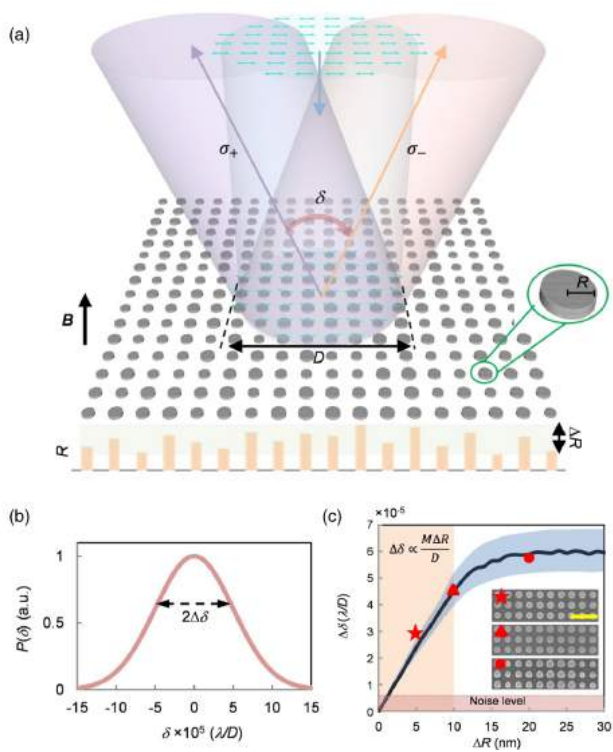

**Figure 1.** Optical metrology with stochastic PSHE. (a) Sketched PSHE from a magnetized disorder metasurface. A polarised incident beam (polarisations indicated by the cyan arrows) is reflected and split into spin-up ($\sigma_+$) and spin-down ($\sigma_-$) components with a subdiffraction-limited angle $\delta$, due to disordered magneto-optical Kerr rotations (the disordered cyan arrows). $R$ indicates the radius of a circular nanoantenna, $D$ is the beam's diameter, and $B$ is the magnetic field. The bottom panel exemplifies a radius distribution of disordered nanoantennas, with $\Delta R$ being the fluctuation. (b) Probability distribution of stochastic PSHEs $P(\delta)$, with $\Delta\delta$ being the standard deviation of the Gaussian distribution. $\lambda$ is the wavelength of light. (c) Experimental and calculated $\Delta\delta$ vs. radius fluctuation $\Delta R$. $M$ is magnetization. Insets: scanning electron microscopy images. Scale bar, 1 $\mu$m. Reprinted with permission from [211], copyright 2020 The Authors.

## Current and Future Challenges

A challenge of dealing with weakly disordered nanostructures is the limited opportunities to extract the information from subtle light-matter interactions. The disorder and stochastic nature hinder light from detecting any useful information other than that from a homogeneous medium. The previous strategies to overcome this limitation involve the implementation of special optical conditions including critical angle, symmetry broken, and resonant enhancement. Specifically, the emerging photonic spin-dependent effects due to symmetry broken in arrays of anisotropic nanoantennas provide a SOI mechanism to achieve a sensitive optical metrology [214]. However, the spin-dependent effects in these architectures naturally disappear when the anisotropic nanoantennas are replaced by isotropic ones. More importantly, how to accurately quantify the structure fluctuations as a function of the measurable spin-split effects remains largely unexplored.

In the pursuit of new spin-optical devices possessing large information capacity and high processing speed, a long-thought goal is to interface spinoptics and spintronics for an interchange of spin information between photons and electrons. This requires the miniaturisation of spin-polarised sources down to a nanometric scale and beyond. Currently, a family of atomic-thin materials has triggered intense research due to their exotic electrical, optical, and thermal properties. Particularly, direct bandgap TMD monolayers show opposite electronic spins at ±K valleys. Consequently, the valley information can be selectively encoded and retrieved by the photonic spin according to the valley-dependent optical selection rules. Although great efforts have been devoted into this field, previous strategies using metallic structures inherited intrinsic losses and limited functionalities. Alternatively, versatile GPMs represent an attractive candidate to perform such a task, inspired by, for example, the photonic Rashba effect that describes a momentum-space spin-split dispersion from inversion-asymmetric structures [215]. However, conventional GPMs are generally designed for plane waves, preventing an efficient interaction between nanoantennas and integrated valley excitons behaving as in-plane circular dipole emitters. A similar issue also arises when single quantum emitters are integrated with GPMs to investigate SOIs in the quantum regime, where a further requirement of long quantum coherence times (or low losses) should be fulfilled.

**Advances in Science and Technology to Meet Challenges**

*(i) Stochastic PSHE*: In a recent work [211], we approached the metrology challenge in weakly disordered systems by exploiting magnetised disordered metasurfaces (Figure 1). Nanoscale size fluctuations were revealed by the probability distribution of a stochastic PSHE, which was induced by disordered magneto-optical Kerr rotations. Here, the metasurfaces are consisted of circular nickel nanoantennas with radii randomly fluctuated in several nanometers. The random variations in sizes of nanoantennas give rise to disordered geometric phases from magneto-optical Kerr rotations. This leads to a spin-dependent beam shift being several orders of magnitude smaller than the diffraction limit of light, i.e., a PSHE. By evaluating the PSHEs via weak measurements from many disordered metasurfaces with different randomisations, we observe a Gaussian probability distribution for the spin shifts. Notably, the standard deviation of the Gaussian distribution is proportional to the size fluctuation of the nanoantennas. This result enabled us to detect a five-nanometer size fluctuation of nanoantennas.

*(ii) Photonic Rashba effect from quantum emitters:* On the other hand, we tackled the weak interaction between integrated valley excitons and nanoantennas by exploiting a novel platform of Berry phase defective photonic crystals (BP-PhCs) [212,213]. The BP-PhCs are composed of a PhC slab with isotropic nanopillars and a GPM with space-variant anisotropic nanoantennas that serve as defects (Figure 2). By utilising the bandgap of the PhC slab, the insertion of the GPM into the PhC slab gives rise to a near-field geometric phase defect mode, which couples the defects for an effective interaction with the integrated valley excitons, resulting in site-controlled excitation, photoluminescence enhancement, and spin-dependent manipulation of individual valley excitons. Consequently, a spin-split dispersion from valley excitons is observed in momentum space, manifesting as the photonic Rashba effect. Particularly, the spin-up and -down branches correspond to emission from ±K valley excitons, respectively, indicating a valley separation in momentum space at room temperature. Moreover, this basic interaction mechanism between circular dipole emitters and nanostructures can be generalised to quantum emitters with arbitrary in-plane polarisations and PhC structures with distinct symmetries.

*(iii) Quantum photonic metasurfaces:* In the preparation of entangled photons [213], we used lossless dielectric metasurfaces, the near unity efficiency of which enables the manipulation of single photons under a sufficient long quantum coherence time. To entangle single photons' spin and OAM, a GPM embedded with a spin-dependent helical phase was designed. Depending on the sign of the photon spin, the GPM performs a unitary transformation that adds or subtracts one quanta of OAM.

In our case, the spatial wavefunction of the photon is paraxial; therefore, the spin and the OAM are independent and have Hilbert spaces of different dimensions. Consequently, single photons in entangled spin and OAM states, and photon pair with nonlocal correlations between the spin of one photon and the OAM of another photon are generated.

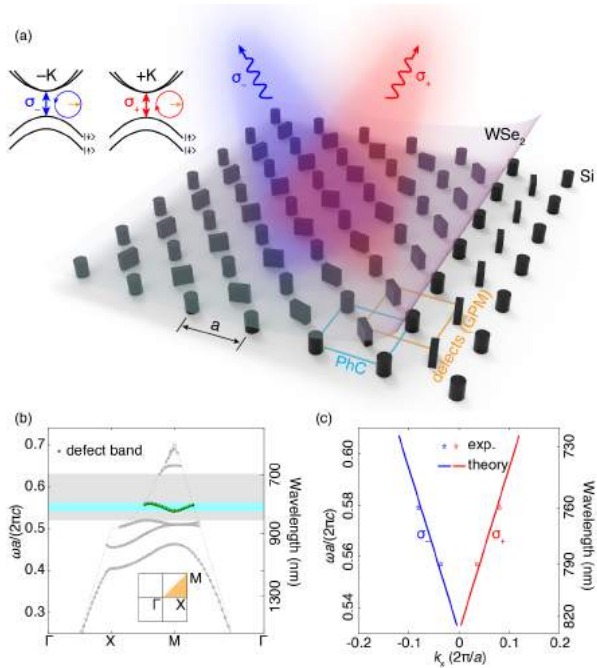

**Figure 2.** Photonic Rashba effect from valley excitons. (a) Illustration of a heterostructure combining a BP-PhC and a WSe₂ monolayer. The PhC slab composed of isotropic nanopillars is arranged in one lattice, and the GPM composed of anisotropic nanoantennas is arranged in another lattice, serving as defects to the PhC slab. By exploiting the emerging Berry-phase defect mode, valley excitons effectively interact with the defects for coherent geometric phase pickups, leading to a photonic Rashba effect in momentum space. Inset: Schematics of valley-dependent optical selection rules for ±K valley excitons in WSe₂ monolayers. (b) Simulated defect band of a BP-PhC. (c) Measured spin-split dispersion in momentum space (i.e., photonic Rashba effect) from valley excitons. Reprinted with permission from [212], copyright 2020 The Authors.

## Concluding Remarks

Optical SOIs are ubiquitous in nano- and atomic-scale systems. An in-depth understanding of them contributes to both fundamental physics and advanced applications. Our results suggest that Pancharatnam-Berry phase optical elements are suitable for discovering new types of SOIs, which show promising applications in novel photon transport controls, such as entangled photons, spin-polarised light sources, and ultra-sensitive optical metrologies utilising splits of non-degenerated spin modes distinguished by quantum weak measurements. To investigate nanophotonics under unprecedented extreme conditions, the spin-controlled generation, manipulation, and detection of atomic-scale light sources of various statistical properties—e.g., spontaneous emission (super-Poissonian), stimulated emission (Poissonian), and quantum emission (sub-Poissonian)—are promising fields, which we foresee many possibilities in the coming future. In general, introducing spin-orbit coupling of electromagnetic waves into contemporary photonics and atomic-scale optics may result in the development of a new area of research, that is, atomic-scale spinoptics.

## Acknowledgements

The authors gratefully acknowledge financial support from the Israel Science Foundation (ISF), the U.S. Air Force Office of Scientific Research (FA9550-18-1-0208) through their program on Photonic Metamaterials, the Israel Ministry of Science, Technology and Space. The fabrication was performed at the Micro-Nano Fabrication & Printing Unit (MNF&PU), Technion.

## 14. Spin, momenta, and forces in evanescent waves – towards spatial and temporal structuring.

*Michela F. Picardi[1,2], Anatoly V. Zayats[1] and Francisco J. Rodríguez-Fortuño[1]*

[1]King's College London and London Centre for Nanotechnology
[2]ICFO – Institut de Ciencies Fotoniques

### Status

Coupling between spin and orbital angular momenta is a fundamental property of electromagnetic (EM) waves, but it is especially pronounced in the evanescent fields. When modes are spatially confined along at least one dimension, as is the case of dielectric or plasmonic waveguides or fibres, their wavevector becomes complex. The wavevector component along the direction of confinement at the interface between two media is imaginary, resulting in the field exponentially decaying away from the interface. Many interesting phenomena originate from the topological features of evanescent waves related to spin-momentum locking [207]. The transversality condition ($\mathbf{k} \cdot \mathbf{E} = 0$) requires the evanescent fields to be elliptically polarised with a field component parallel to the propagation direction, in sharp contrast to free-space fields. This locks the handedness of the waveguided fields with their propagation direction [207], meaning that when either of the two is reversed, the other must be reversed too. This behaviour arises from basic laws of electromagnetism and exists even in unpolarised light [220]. Such topological protection of the propagation direction of guided waves is valid until the spin flips, e.g., due to scattering.

Because of spin-momentum locking, elliptically polarised EM sources excite guided modes *unidirectionally* via evanescent coupling [221]. Selective mode excitation was also achieved considering the reactive power of evanescent waves $\mathrm{Im}(\mathbf{E}^* \times \mathbf{H})$. Being perpendicular to the interface, the reactive power is locked with the direction of evanescent decay so that multipolar EM sources may excite a guided mode, without any directionality, only if their reactive power is parallel (not antiparallel) to that of the evanescent wave, as in the case of Janus dipoles [222]. Selectivity and directionality of the guided mode launching may also be understood as near-field interference, with the same evanescent modes being excited and interfering with their different symmetries inherited from different EM source multipoles [221,359].

Reciprocal effects in controlling the polarisation of far-field directional scattering with the direction of guiding modes were also demonstrated [223]. The spinning fields of guided modes were widely exploited to achieve polarisation-dependent optical forces on achiral objects [360]. All these effects stem from topological properties of evanescent fields of the guided modes and have been observed in an extremely broad spectral range, from optical to radio frequencies, with various types of waveguides and EM sources, such as multipolar emitters and scatterers, atoms and quantum dots, and predicted for other kinds of evanescent waves, such as acoustic and gravitational ones.

### Current and Future Challenges

The phenomena described above follow from classical photonics but can also be applied in the quantum realm. The coupling between quantum dots and complex light, such as evanescent waves, leads to chiral directional single photon routing [361]. Single-photon evanescent waves and their entanglement open up new possibilities in the development of light-matter interfaces and quantum technology.

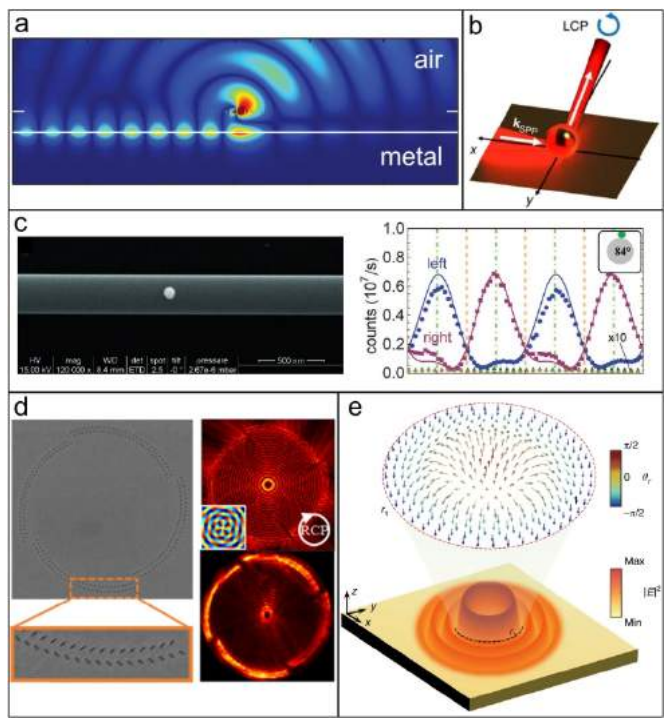

**Figure 1.** (a) Magnetic field amplitude of a surface plasmon-polariton unidirectionally excited by a circularly polarised electric dipole, adapted from reference [221]. (b) Reciprocal control of far-field scattering polarization via unidirectional near field excitation, adapted from reference [223]. (c) SEM image of the nanosphere placed on the waveguide and light intensity at the two ends of a fibre measured by varying the polarisation of the light incident on the nanosphere. The nanosphere acts as a point dipole, and when it is circularly polarised, it will couple unidirectionally to the mode guided by the fibre, adapted from reference [224]. (d) Structured surface plasmon-polariton vortex excited via a plasmonic vortex lens, reprinted from [225]. (e) Skyrmion-like structure of the spin angular momentum vector (top) and the intensity distribution (bottom) of a surface wave vortex, adapted from reference [31].

Structured evanescent waves are obtained by relaxing the condition of a single-wavevector, such as in surface plasmon vortices [225,362]. Structured evanescent waves are already proving a fertile ground for physical phenomena such as dynamic field-skyrmionic lattices. Exploiting spin to orbital angular momentum coupling, spin-skyrmions in the evanescent field have been demonstrated (Fig. 1(e)) [31]. Structured evanescent waves are at their infancy, and inspiration can be taken from the myriad of structured beams considered in free space to find and exploit analogues in the near-field.

As well as structuring in space, time-domain structuring can be introduced. For propagating waves, time-dependence leads to remarkable polarization patterns in multicoloured light [38]. It is therefore to be expected that the routing and metrological capabilities of single-coloured evanescent waves—based on their unique polarization patterns—will acquire additional features and versatility when the time-harmonic assumption is dropped, in favour of multi-coloured and pulsed evanescent waves.

The combination of both spatial structuring and time-dependence, e.g., in spatio-temporal vortices [19-24], will greatly extend the evanescent wave playground, providing time-dependent photonic topologies in the presence of a complex wavevector. Guided spatio-temporally structured waves could be used for encoding classical or quantum information into topological invariants such as vortex winding numbers as a novel means of guided wave information transfer.

Optical forces near surfaces is another vast field enabled by the properties of guided waves. Counter-intuitive effects such as levitation, polarisation-controlled forces, or chiral sorting for particles and molecules is an active field of study [363-365] based on the peculiar spin and momentum properties of simple non-structured time-harmonic surface evanescent waves. The use of structured and multi-coloured evanescent wave illumination can lead to precisely engineered optical force fields for optically tuneable manipulation of nanoparticles, atoms, and molecules near surfaces.

Even more layers of complexity will be introduced if the assumption of polarised and coherent light is dropped. Transverse spin of the evanescent field persists even when light is fully unpolarised [220]. Incoherent light (with a fluctuating phase) will show similar behaviour, opening a vast playground of statistical optics with evanescent waves. This could enable new applications, for example, chiral optical force separation of enantiomers using non-laser light sources, and applications using thermal emission or even sunlight.

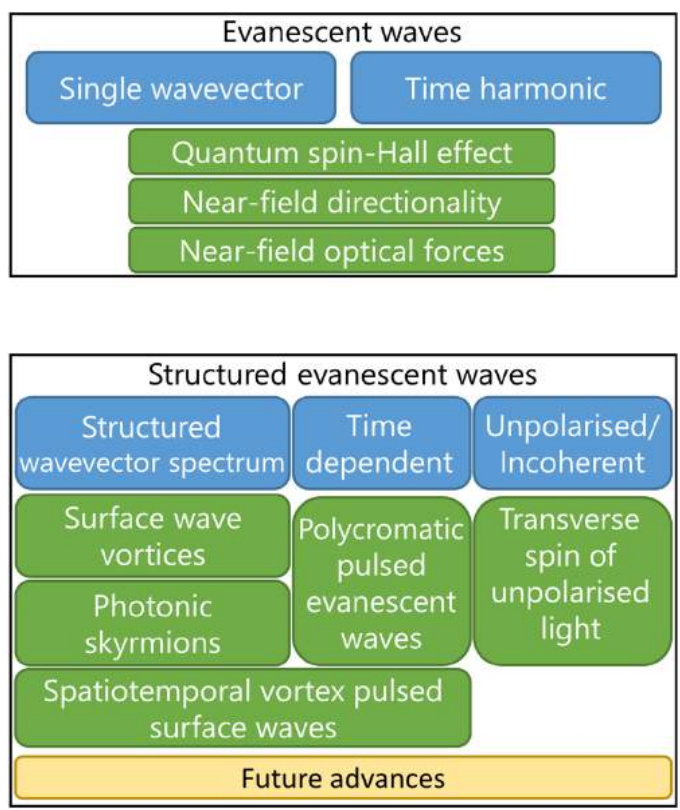

**Figure 2.** Schematic representation of the main interesting phenomena observed in non-structured time-harmonic surface waves (top), and the possible direction of expansion into new domains which can be accessed via space and time structuring of surface waves (bottom).

## Advances in Science and Technology to Meet Challenges

Despite significant progress in describing spin-orbit properties of light fields, further advances in the understanding of dynamical properties of near-fields are needed. Light properties are typically defined via their effect on material probes. For instance, the spin angular momentum density of an evanescent wave is proportional to the mechanical torque exerted by the wavefield on a dipolar particle. Novel dynamical properties are uncovered when considering more complex particles. The optical torque on a quadrupolar particle is not proportional to the spin, but instead to another quantity: the spin of the field gradient [227]. Given potential applications of optical spin effects in evanescent fields, such as transverse spin and skyrmions, higher-order field properties—such as the above-described field gradient spin and other related properties that arise in the evanescent field interactions with higher-order multipoles—call out for an in-depth analysis and deeper understanding of light properties in the near field, especially in structured, dynamic, and multi-colour settings.

More immediate important technological challenges are related to reducing the effects of losses if surface plasmons are chosen as a platform for evanescent waves, and to improving the coupling efficiencies from emitters and scatterers to evanescent wave platforms such as rib and slot nanophotonic waveguides and nanofibres. The temporal structuring of surface waves will need advances in attosecond physics and the control of the intra-pulse polarization evolution.

Advances in material science, nanophotonic platforms, and optical force instrumentation will all be needed to exploit the full potential of spatially and temporally structured surface evanescent waves and their effects. The miniaturisation of EM emitters and improvement in their coupling efficiencies as well as the successful integration of quantum dots, molecules, and atoms into nanophotonic devices are requirements for many of the proposed and potential applications: from nanoscale quantum technologies, requiring single-photon spin routing based on evanescent wave spin-momentum locking, which could form part of optical quantum computing or secure quantum communication platforms, to the use of evanescent fields for on-chip chiral optical forces that might enable important applications like all-optical on-chip separation of enantiomers.

**Concluding Remarks**

The applications in quantum technologies, nanophotonics, sensing, metrology and nano-opto-mechanics drive requirements on making use and manipulation of all degrees of freedom of guided light. In addition to polarisation-controlled routing and coupling of electromagnetic waves, which have already resulted in applications in position sensing with nanometric displacement sensitivity, integrated miniaturised polarimeters and coherent-optical receivers as well as unusual lateral and repulsive optical forces, the use of structured evanescent waves with new dynamical and topological properties above and beyond spin-orbit interaction offer innovative solutions towards miniaturisation, energy efficiency, and ultrafast operation. These future advances will be made possible in close interaction of theory, developing new understanding of angular momenta properties in structured and dynamic evanescent fields, nanoscience, and photonic instrumentation for realising and measuring these properties, and development of photonic and nanophotonic platforms tailored for specific applications.

**Acknowledgements**

This work was supported by the European Research Council projects iCOMM (789340) and Starting Grant ERC-2016-STG-714151-PSINFONI.

## 15. Momentum and spin of electromagnetic, sound, and water waves

*Konstantin Y. Bliokh*

RIKEN

**Status**

Wave momentum has a long (and probably never-ending) history starting from 18th century studies by Euler, with the landmarks of the electromagnetic momentum by Poynting and quantum-mechanical momentum by de Broglie. It includes numerous controversies: the momentum of sound waves [228], the Abraham-Minkowski dilemma for the momentum of light in a medium [229], the Belinfante-Rosenfeld problem for the energy-momentum tensor in field theories [230], etc. Angular momentum (AM) is closely related to the momentum, and is also involved in these problems. Recent great interest in structured waves, materials, and wave-matter interactions prompted thorough revision of local momentum and AM properties of generic inhomogeneous (although often restricted by monochromaticity) wave fields.

Starting with the general field-theory approach, there are two types of momentum and AM densities: (i) the *kinetic* momentum density $\mathbf{\Pi}$ (e.g., the Poynting momentum in electromagnetism) and kinetic AM density $\mathbf{M} = \mathbf{r} \times \mathbf{\Pi}$; (ii) the *canonical* momentum density $\mathbf{P}$ and the canonical AM density $\mathbf{J} = \mathbf{r} \times \mathbf{P} + \mathbf{S}$, where $\mathbf{r} \times \mathbf{P} = \mathbf{L}$ and $\mathbf{S}$ are the *orbital* and *spin* (intrinsic AM) densities. The kinetic and canonical quantities are related by the Belinfante-Rosenfeld equation [230]:

$$\mathbf{\Pi} = \mathbf{P} + \frac{1}{2}\mathbf{\nabla} \times \mathbf{S} \,. \tag{1}$$

One can regard the canonical momentum and spin densities as two fundamental independent quantities; other momentum/AM characteristics are their derivatives. Both the kinetic and canonical quantities are important and have their own advantages. For example, in relativistic field theory, the kinetic quantities are expressed via fields and are explicitly gauge-invariant, while the canonical ones follow directly from Noether's theorem, i.e., provide generators of translations and rotations in the quantum-mechanical formalism. Notably, the canonical momentum and spin densities allow meaningful gauge-invariant expressions only for monochromatic fields, but these are directly observable via forces and torques exerted by monochromatic fields on dipole particles [231].

For monochromatic electromagnetic fields in free space, the canonical momentum and spin densities read [231–234],

$$\mathbf{P} = \frac{1}{4\omega}\text{Im}[\varepsilon\,\mathbf{E}^* \cdot (\mathbf{\nabla})\mathbf{E} + \mu\,\mathbf{H}^* \cdot (\mathbf{\nabla})\mathbf{H}] \,, \tag{2}$$

$$\mathbf{S} = \frac{1}{4\omega}\text{Im}(\varepsilon\,\mathbf{E}^* \times \mathbf{E} + \mu\,\mathbf{H}^* \times \mathbf{H}) \,. \tag{3}$$

Here, $\mathbf{E}(\mathbf{r})$ and $\mathbf{H}(\mathbf{r})$ are the complex electric and magnetic field amplitudes, whereas $\varepsilon$ and $\mu$ are the vacuum permittivity and permeability. Substituting Eqs. (2) and (3) into Eq. (1) and using Maxwell equations yields the kinetic Poynting momentum $\mathbf{\Pi} = \frac{1}{2c^2}\text{Re}(\mathbf{E}^* \times \mathbf{H})$, where $c = 1/\sqrt{\varepsilon\mu}$ is the speed of light.

Importantly, there is a freedom in defining the canonical momentum and spin densities (2) and (3) even when the Poynting momentum is fixed. While Eqs. (2) and (3) represent symmetric 'arithmetic mean' of the electric and magnetic contributions, one can use either purely electric or purely magnetic quantities instead. The reason for choosing the dual (electric-magnetic) symmetric expressions is mostly aesthetic: to preserve the dual symmetry inherent in free-space Maxwell equations [231–234]. However, the presence of charges breaks this symmetry because only electric charges exist in nature. In addition, electric-dipole and magnetic-dipole particles effectively interact with the electric and magnetic parts of the canonical densities (2) and (3) so that their electric/magnetic/symmetric form is not fixed fundamentally but rather chosen in each particular problem using auxiliary arguments.

Sound waves in a fluid or gas have similar momentum and spin properties. Although sound waves are often regarded as 'scalar' or 'spinless', these are actually *vector* waves; the local displacement (or velocity) of the medium particles provides the vector wavefield. Akin to electromagnetic waves, one can describe monochromatic sound waves via two complex fields: velocity $\mathbf{v}(\mathbf{r})$ and pressure $p(\mathbf{r})$. The canonical momentum and spin densities of monochromatic sound waves are [41–44]:

$$\mathbf{P} = \frac{\rho}{2\omega}\text{Im}[\mathbf{v}^* \cdot (\boldsymbol{\nabla})\mathbf{v}],  \tag{4}$$

$$\mathbf{S} = \frac{\rho}{2\omega}\text{Im}(\mathbf{v}^* \times \mathbf{v}),  \tag{5}$$

where $\rho$ is the mass density of the medium. Substituting Eqs. (4) and (5) into Eq. (1), supplied with the acoustic wave equations for $\mathbf{v}$ and $p$, yields the acoustic analogue of the Poynting vector: $\boldsymbol{\Pi} = \frac{1}{2c_s^2}\text{Re}(p^*\mathbf{v})$, where $c_s = 1/\sqrt{\rho\beta}$ is the speed of sound with $\beta$ being the compressibility of the medium.

From the field-theory viewpoint, the canonical quantities (4) and (5) also allow different forms compatible with the same kinetic momentum. Namely, instead of the velocity-related quantities (4) and (5), one can use the pressure-related quantities ($\mathbf{P} = \frac{\beta}{2\omega}\text{Im}(p^*\nabla p) = \boldsymbol{\Pi}$ and the spin density vanishes in this case), or an 'arithmetic mean' of the velocity-related and pressure-related contributions [43,44]. Furthermore, monopole and dipole acoustic particles are effectively coupled to the pressure-related and velocity-related forms of the canonical quantities [43]. Then why do we prefer the 'asymmetric' velocity-related definitions (4) and (5)? This is because, in contrast to electromagnetism, sound waves exist only in a medium, and one can associate their dynamical properties with *microscopic mechanical properties of the medium particles*. Indeed, local rotation of the medium particles in a generic sound-wave field with an elliptical polarisation (polarisation of the velocity field $\boldsymbol{v}$ corresponds of the microscopic *real-space trajectory* of the particle) generates exactly the canonical AM density (5) [41,42]. Furthermore, the medium particles slowly drift in the sound-wave field due to the second-order difference between the Euler and Lagrange coordinates. This is the *Stokes drift* with the velocity $\mathbf{u}$ exactly corresponding to the canonical momentum density (4): $\mathbf{P} = \rho\mathbf{u}$ [55,235]. Thus, microscopic mechanical properties of the medium allow one to unambiguously determine the canonical momentum and spin densities in sound waves. Remarkably, Eqs. (4) and (5) are quite general and are also valid for *elastic* waves in isotropic solids [49,50,53] or Langmuir *plasma* waves [236].

The above features make the canonical momentum and spin in acoustic waves directly observable, at least in principle, via microscopic motion of the medium particles. In practice, such observation is challenging with typical sound waves. Larger-scale waves, such as water-surface waves, can serve as a perfect platform for the observation of microscopic medium properties in structured wave fields. Considering monochromatic deep-water gravity waves with the dispersion $\omega^2 = gk$ ($g$ is the gravitational acceleration) as a quasi-2D wave system, we recently derived the canonical momentum density in the unperturbed surface $(x, y)$-plane and the corresponding spin density in the vertical $z$-direction [55]:

$$\mathbf{P} = \frac{\rho}{2\omega} \operatorname{Im}[\mathbf{V}^* \cdot (\boldsymbol{\nabla}_2)\mathbf{V} + W^* \boldsymbol{\nabla}_2 W] \,, \tag{6}$$

$$\mathbf{S} = \frac{\rho}{2\omega} \operatorname{Im}(\mathbf{V}^* \times \mathbf{V}) \,. \tag{7}$$

Here, $\boldsymbol{\nabla}_2 = (\partial_x, \partial_y)$, whereas $\mathbf{V} = (v_x, v_y)$ and $W = v_z$ are the in-plane and vertical velocity components of water particles. Substituting Eqs. (6) and (7) into Eq. (1), and using the equations of motion for gravity waves, yields the kinetic momentum $\boldsymbol{\Pi} = \frac{\rho k}{\omega} \operatorname{Im}(W^* \mathbf{V})$, which is consistent with Ref. [237]. Akin to sound waves, Eqs. (6) and (7) correspond to the mechanical momentum (due to the Stokes drift) and the microscopic mechanical angular momentum (due to the local elliptical trajectories) of water particles, as shown in Figure 1 [55].

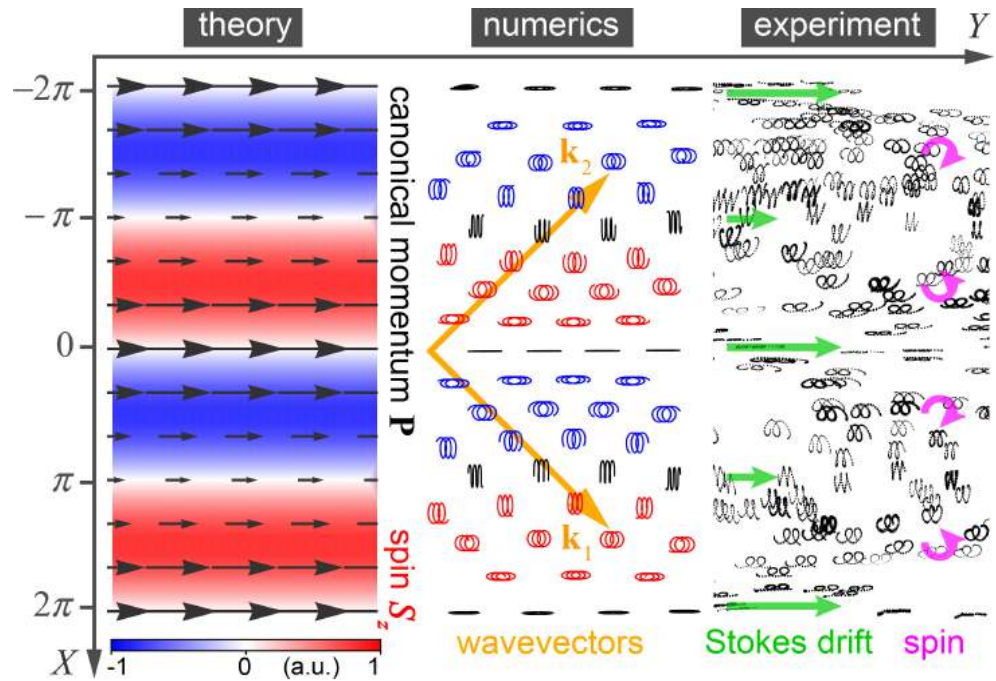

**Figure 1.** Theoretically calculated surface distributions of the canonical momentum $\mathbf{P}$ (black arrows) and spin $S_z$ (blue-red) densities, Eqs. (6) and (7), in the interference of two plane gravity (water-surface) waves with orthogonal wavevectors $\mathbf{k}_{1,2}$. Numerical and experimental plots show trajectories of microscopic water particles for three wave periods $6\pi/\omega$. The Stokes drift of the particles and their elliptical motion correspond to the canonical momentum and spin, respectively. The normalized surface coordinates are $X = \sqrt{2}kx$, $Y = \sqrt{2}ky$. Adapted from Ref. [55].

Thus, the momentum and angular momentum properties of electromagnetic, acoustic, and water waves have profound similarities related to the presence of spin and the fundamental Belinfante-Rosenfeld relation (1). There is also a principal difference in that the free-space electromagnetic quantities are not associated with any medium ('ether') and cannot be derived from microscopic mechanical considerations. However, electromagnetic waves in a *medium* represent *mixed* light-matter waves and do involve microscopic mechanical properties of the medium. The kinetic momentum of an electromagnetic wave in an isotropic dispersive medium with permittivity $\varepsilon = \varepsilon(\omega)$ and permeability $\mu = \mu(\omega)$ is still given by the Poynting (or Abraham) momentum $\boldsymbol{\Pi} =$ $\frac{1}{2c^2}\text{Re}(\mathbf{E}^* \times \mathbf{H})$, while the canonical momentum and spin densities can be described by Eqs. (2) and (3) with the substitution [236,238],

$$\varepsilon \to \tilde{\varepsilon} = \varepsilon + \omega \frac{d\varepsilon}{d\omega}, \quad \mu \to \tilde{\mu} = \mu + \omega \frac{d\mu}{d\omega}. \tag{8}$$

In this case, microscopic contributions from the motion of the medium particles (e.g., electrons or atoms) play a crucial role [235,236,238,239], and the canonical expressions can be regarded as the Minkowski-type momentum and spin densities [238]. Equations (2), (3), and (8) are found within the dual-symmetric formalism, but the electric-field-biased approach is also possible and can be relevant because the medium is usually electric-biased on the microscopic level [236] (i.e., consisting of electric charges and dipoles).

**Current and Future Challenges**

Despite the great progress in the description of the momentum and angular momentum of structured waves, there are still many unsolved questions. To name a few: extension of the above approach (if possible) to anisotropic media, polychromatic fields, and fields with complex frequencies (e.g., Mie quasimodes).

**Advances in Science and Technology to Meet Challenges**

Rapid development of nanophotonics, including metamaterials and plasmonics, provides a perfect platform for theoretical, numerical, and experimental studies of fundamental dynamical properties of complex fields and their interactions with matter.

**Concluding Remarks**

We have briefly described the canonical and kinetic momentum and AM properties of monochromatic electromagnetic, acoustic, and water waves. Despite enormous differences in scales and their nature, the momentum and AM of these waves share profound similarities. This reflects the universality of these concepts, as well as the remarkable role of the spin and field-theory relations even in 'spinless' classical waves.

## 16. Acoustic spin

*Chenwen Yang and Jie Ren*

Tonji University

**Status**

The acoustic waves in fluids were traditionally mis-regarded as spinless fields because of their curl-free nature. Recently, several works about elastic and acoustic waves show that even pure longitudinal waves that can be fully described with a scalar field still have the ability to carry spin angular momentum (SAM). The existence of SAM does not require an extra spatial degree of freedom but needs a locally temporal rotation of the vector field, like displacement or velocity. A clear definition of SAM in acoustic wave systems will pave a new way to understand and realise various structured acoustic wave systems, included but not limited to the spin-momentum locking in acoustic wave systems and the symmetry selective excitation.

The expression of spin angular momentum (SAM) in acoustic/elastic waves could be derived from the definition of angular momentum in acoustic/elastic waves, which is [42,50,53,240,355]:

$$\mathbf{S} = \frac{\rho\omega}{2}\mathrm{Im}[\mathbf{u}^* \times \mathbf{u}] = \frac{\rho}{2\omega}\mathrm{Im}[\mathbf{v}^* \times \mathbf{v}], \qquad (1)$$

where $\mathbf{u}^*$ represents the conjugate of the displacement vector $\mathbf{u}$, $\mathbf{v}$ is the particle velocity which is the time derivative of $\mathbf{u}$, $\omega$ is the angular frequency of wave, and $\rho$ is the density of media. Under linear configuration of small wave amplitude, $\rho$ is regarded as a constant. It is worth noting that the spin angular momentum of phonons is also studied as early as the 1960s [48,241], which are quasiparticles more suitable for describing quantised lattice vibration in the quantum scene. The spin angular momentum of phonons shares a similar expression as Eq. (1), the SAM of acoustic/elastic waves, reflecting the fact of wave-particle duality. However, initial works about phonon spin ignored the spin of curl-free longitudinal mode [355]. The SAM of elastic/acoustic waves describes the locally temporal rotation of displacement/velocity vector, not the spatial curl of the vector field. In other words, if a displacement/velocity field has both longitudinal and transverse components (or could be non-zero decomposed into two directions), it could have the SAM, and this is irrelevant with the curl of the vector field. As such, even longitudinal waves (curl-free wave), e.g., acoustic waves, possess the ability to carry SAM. Next, we will give a brief explanation through several derivations.

The displacement field $\mathbf{u}$ could be written as the combination of the gradient of a scalar potential and curl of a vector potential [50,51]:

$$\mathbf{u} = \boldsymbol{\nabla}\phi + \boldsymbol{\nabla} \times \boldsymbol{\psi}. \qquad (2)$$

In longitudinal wave fields, $\boldsymbol{\nabla} \times \mathbf{u} = 0$ and $\boldsymbol{\psi} = \mathbf{0}$. Note that in the field of acoustics, people usually describe the acoustic wave with the media particle velocity $\mathbf{v}$ and acoustic pressure $P$ instead of displacement $\mathbf{u}$ and the scalar potential $\phi$, which gives [35,242]:

$$-\boldsymbol{\nabla}P = \frac{\partial(\rho\mathbf{v})}{\partial t} = -i\rho\omega\mathbf{v},$$

$$\mathbf{v} = \frac{\partial\mathbf{u}}{\partial t} = -i\omega\mathbf{u}. \qquad (3)$$

Under this configuration, the acoustic SAM could be express as $\mathbf{S} = (\rho/2\omega)\,\mathrm{Im}[\mathbf{v}^* \times \mathbf{v}]$ as shown in Eq. (1) [42,43,45,240,243]. These configuration differences do not affect our discussion below.

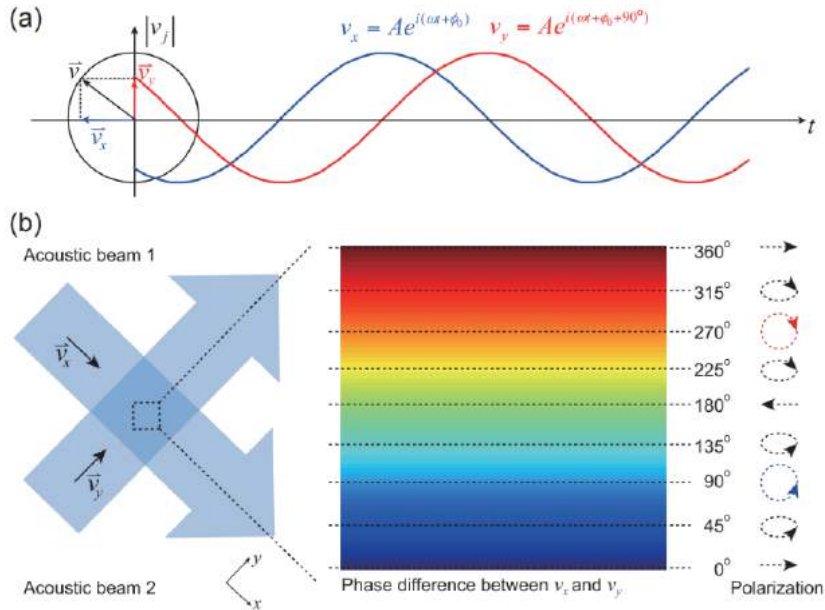

**Figure 1.** Acoustic spin as a rotating particle velocity field (8). (a) A rotating particle velocity (black arrow) can be decomposed into two components $v_x$ (blue arrow) and $v_y$ (red arrow) along the $x$ and $y$ directions. (b) Acoustic spin in the interference of two acoustic beams. Two beams with equal amplitudes propagating along the $x$ and $y$ directions contribute $v_x$ and $v_y$ components of the particle velocity field, respectively.

To clarify the existence of SAM in longitudinal waves, we could start with a common form of the scalar potential in $x$-$y$ plane:

$$\phi = \phi_0 e^{ik_y y} e^{ik_x x} e^{-i\omega t}, \tag{4}$$

Where $k_x$ ($k_y$) represents the wave number along $x$ ($y$) axis. According to Eq. (2), the displacement field of a longitudinal wave can be written as:

$$u_x = \frac{\partial \phi}{\partial x} = ik_x \phi,$$
$$u_y = \frac{\partial \phi}{\partial y} = ik_y \phi. \tag{5}$$

Clearly, the displacement vector $u_y$, or the transverse component perpendicular to the transport direction, is non-zero in this evanescent wave. Although the material which only supports longitudinal waves has no shear modulus, the transverse component $u_x$ still exists. As such, the spin angular momentum is:

$$\frac{\rho\omega}{2}\mathrm{Im}[\mathbf{u}^* \times \mathbf{u}] = \hat{\mathbf{z}}\frac{\rho\omega}{2}\mathrm{Im}[u_x^* u_y - u_y^* u_x]$$
$$= \hat{\mathbf{z}}\frac{\rho\omega}{2}\mathrm{Im}[(k_x^* k_y - k_y^* k_x)\phi_0^2], \tag{6}$$

where $\hat{\mathbf{z}}$ is the unit vector along the $z$-axis. If both $k_x$ and $k_y$ are real constants, e.g., a plane wave transported in the $x$-$y$ plane, nothing interesting happens. As $k_x^* k_y - k_y^* k_x = 0$, the acoustic SAM is zero. However, things change when we consider a more complex situation, e.g., the evanescent acoustic wave transported along the $x$-axis and decaying along $+y$-axis from $y = 0$. As such, $k_y$ could be rewritten as $i\tau$, where $\tau$ is a real component. Also, the acoustic SAM will be a non-zero quantity: $\hat{\mathbf{z}}\frac{\rho\omega}{2}\text{Im}[(k_x(i\tau) - (-i\tau)k_x)\phi_0^2] = \hat{\mathbf{z}}\rho\omega\tau k_x\phi_0^2$. We may call this acoustic SAM as the transverse spin because it is perpendicular to the wave direction, similar with the transverse spin of an optical wave [231]. Note that the sign of $k_x$ determines the sign of SAM in this evanescent acoustic wave configuration. This indicates the spin-momentum locking effect in evanescent acoustic waves [42,43,45], which is similar with the electromagnetic evanescent waves in optics [207].

Nevertheless, similar things happen to the phase difference of strain components $\epsilon_{xx}$ and $\epsilon_{xy}$. For example, a surface longitudinal wave in fluid, like air or water, can have transverse shear strain, which could be expressed as:

$$\epsilon_{xx} = \frac{\partial u_x}{\partial x} = -k_x^2\phi,$$
$$\epsilon_{xy} = \frac{1}{2}\left(\frac{\partial u_x}{\partial y} + \frac{\partial u_y}{\partial x}\right) = -k_xk_y\phi. \tag{7}$$

Although $\boldsymbol{\nabla}\cdot\mathbf{u} = 0$, the transverse shear strain $\epsilon_{xy} \neq 0$, and $\frac{\epsilon_{xx}}{\epsilon_{xy}} = \frac{k_x}{k_y}$. If we consider a plane wave in both the x and y direction, $\frac{k_x}{k_y}$ is always a real number, and there is no phase difference between $\epsilon_{xx}$ and $\epsilon_{xy}$. However, if we consider a surface wave, $\frac{\epsilon_{xx}}{\epsilon_{xy}} = \frac{k}{\tau}e^{i\frac{\pi}{2}}$, this gives a $\pi/2$ phase difference between $\epsilon_{xx}$ and $\epsilon_{xy}$. Also, one can say that the phase difference between the normal and shear strains shows the existence of acoustic SAM in this scene. It is worth noting that the phase difference between $\epsilon_{xx}$ and $\epsilon_{xy}$ is also important in SAW-induced nonreciprocal ferromagnetic resonance [244]. According to the analysis above, the acoustic SAM may also excite the ferromagnetic resonance.

Actually, the existence of non-zero acoustic SAM relies on the local phase difference between $u_x$ and $u_y$ (or $v_x$ and $v_y$), as indicated by Eq. (1) and (6). Non-zero SAM exists when the phase is different between two different directions [42,240,356], e.g., the acoustic field generated with two acoustic beams with different phases, as shown in Figure 1.

**Current and Future Challenges**

*Acoustic spin induced torque.* The acoustic wave with non-zero SAM will introduce a torque to an absorption particle. This spin-matter interaction can be characterised by the rates of the angular momentum transfer between the acoustic field and the particle [43]. Because the monopole vibration mode is isotropic and does not provide the circularly polarised local states, only the dipole momentum can introduce the torque. With the help of a meta-atom which supports dipole resonance, one can measure the torque induced by the acoustic SAM [42], as shown in Figure 2(a). These acoustic spin and torque also exist in the topological meta structure [245]. Although the present work shows a clear correlation between the pseudo-spin state and SAM in this kind of quantum spin hall effect acoustic metamaterial, due to the challenge in directly measuring the SAM-induced torque in complex acoustic system, the experimental evidence is still missing, and the physical origin of this correlation still needs further clarifying.

*The structured acoustic wave with SAM.* The acoustic SAM also improves the fundamental understanding of the inherent near-field symmetry and directional coupling in acoustics. Along with the time average energy flow and reactive power, acoustic SAM density can be used to characterise the time and space symmetry of the evanescent acoustic wave [45], as shown in Figure 2(b). These evanescent wave modes could be selectively excited by the acoustic source with particular symmetries, which provides a feasible approach for designing functional acoustic devices.

Besides the near-field acoustics, the spin-dependent transportation could also be realised in wave guides with symmetry-breaking boundary conditions [243], as shown in Figure 2(c). The momentum of the acoustic wave in such waveguides are tightly coupled with the acoustic SAM. This spin-momentum locking effect in wave guides will raise the SAM dependent selective transportation and enhance the backscattering suppression.

As a kind of special structure, skyrmions are also proposed in acoustic systems [35,29]. Under the help of a well-designed hexagonal acoustic metasurface, the acoustic velocity fields can raise clear skyrmion lattice patterns, which unveil a fundamental property of acoustic fields and may inspire future research in structured acoustic waves. However, the excitation and controlling method of skyrmion patterns of acoustic SAM is still not presented yet. In the future, acoustic skyrmions may pave a new way for the research in structured acoustic waves and functional acoustic devices.

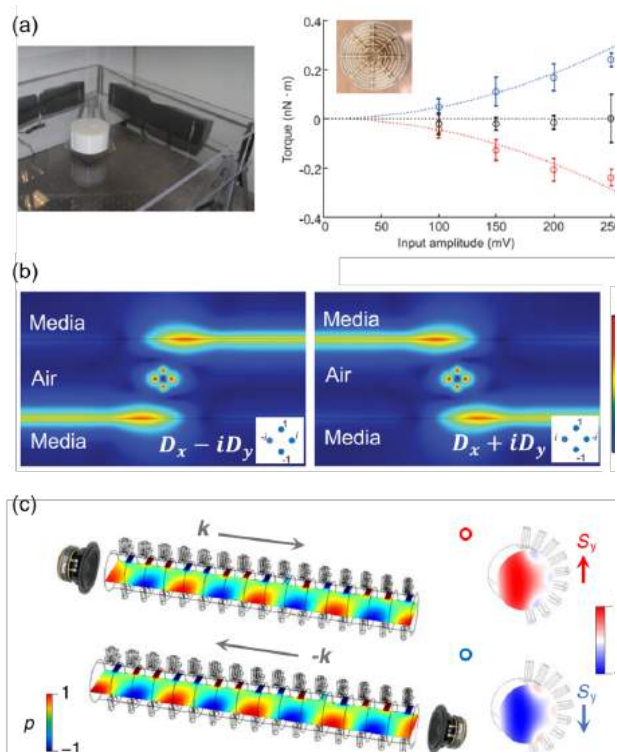

**Figure 2.** (a) Experimental set-up for the measurement of acoustic spin and the measured SAM-induced torque of acoustic waves which possess positive (red line) and negative (blue line) SAM [42]. (b) The near-field selective excitation of spin source [45]. (c) The spin-dependent propagations in waveguide with symmetry-breaking boundary conditions [243]. The excited waveguide modes have different propagation directions and carry opposite SAM texture.

*The SAM interaction between elastic and magnetic/optical systems.* As the acoustic spin we talk about here can carry real SAM, it is natural to assume that angular momentum could transfer between acoustic systems and magnetic/optical systems.

In the field of research about magnetic materials, the interaction of spin waves and ultrasonic waves in ferromagnetic crystals was theoretically demonstrated with the phonon-magnon interaction as early as 1958 [246]. The experimental demonstrations about the interaction between surface acoustic waves (SAW) and magnon systems also attract many attentions, including but not limited to the SAW spin-pumping [247], SAW-driven ferromagnetic resonance [248] and SAW-controlled magnetisation [237]. Meanwhile, the angular momentum interaction transfer between elastic and electromagnetic systems is also proposed with the help of optical fibres [250] or piezoelectric materials [251]. These works may offer a new way to control dynamic states of magnetic, optical, and elastic systems. However, existing research mainly describes this spin transfer with the elastic strains or phonons rather than acoustic/elastic SAM. To give a simpler and more practical theory of the SAM coupling between various physical systems in the micro scene, a definition of the transition between acoustic SAM and magnetic/optical SAM is still needed.

**Concluding Remarks**

Spin angular momentum helps people to give a fundamental explanation of structured waves and practical implications for wave devices. The acoustic SAM proves that the ability of wave to carry SAM is more related with the polarised vibration other than the number of intrinsic spatial degree of freedom. We expect that these concepts of acoustic SAM will assist the development of structured waves in optical, acoustic, and elastic systems, as well as the SAM transfer between different physical systems.

## 17. Acoustic Pseudospins for Wave Control and Topological Protection

*Alexander B. Khanikaev[1] and Andrea Alù[2]*

[1]The City College of New York
[2]Photonics Initiative, CUNY Advanced Science Research Center

**Status**

In recent years, synthetic degrees of freedom deliberately introduced into the design of metamaterials via symmetry engineering have significantly expanded the landscape of classical wave phenomena. In particular, synthetic pseudo-spins spanning Hilbert spaces of desirable dimensions and leveraging effective Hamiltonians with nearly any form and structure have been enabling a wide range of eigenstate and spectral engineering in metamaterial responses. Advances in material engineering and manufacturing on an unprecedented deeply subwavelength scale have been enabling a precise control over the structure of such effective Hamiltonians, leading to the emulation of numerous fascinating physical systems, from field theory with synthetic gauge fields to relativistic and topological phenomena.

In this context, acoustic metamaterials arguably represent the most straightforward and easy-to-work platform for emulating complex wave phenomena. In addition to the relative simplicity of acoustic experiments, recent advances in additive manufacturing have made acoustic systems very appealing. On the other hand, unlike other vector wave-fields, such as electromagnetic and mechanical platforms, which offer natural intrinsic degrees of freedom that can be leveraged to construct effective Hamiltonians, the acoustic pressure field is scalar in nature. Therefore, synthetic degrees of freedom represent an absolute necessity for the emulation of effective Hamiltonians with acoustic pressure-wave systems [252]. Figure 1(a) illustrates how synthetic pseudospin can be produced in an acoustic Kagome lattice of 3D printed trimers and used for directional (valley) excitation of bulk Bloch waves. For mechanical waves of vector nature, engineering the additional degrees of freedom represents a powerful tool to further expand the Hilbert space and entangle synthetic and natural degrees of freedom.

Based on this approach, a broad range of topological phenomena, including higher-order 2D and 3D topological phases, and emulation of Dirac and Weyl physics, have been realised in acoustic metamaterials [253-256]. These advances have led to numerous demonstrations of unprecedented control over spectral features, propagation, and scattering of sound waves. Spectral pinning of resonant higher-order topological states via lattice symmetries, pseudo-spin polarised topological edge transport, and topological resilience represent just some examples of recent demonstrations.

Beyond these demonstrations, artificial acoustic media with synthetic pseudospins are posed to bring even more fascinating advances, both in the demonstration of fundamentally new wave phenomena and in practical applications. These advances will likely stem from new physics enabled by sound-matter interactions and more exotic responses of structured acoustic media, which can radically expand the range of attainable effective Hamiltonians. Indeed, recent demonstrations of multiphysics-enabled phenomena in classical-wave polaritonic systems have shown how structured nature of waves in one physical subsystem can be transferred onto the second one via interactions, to yield a new type of topological excitations, as illustrated by Fig. 1(b) [257]. From the technological point of view, synthetic pseudospins have not been exploited yet, the context of controlling scattering and radiative properties of acoustic metamaterials, which is another promising direction to pursue, e.g., for achieving pseudo-spin-controlled directional emission of sound waves.

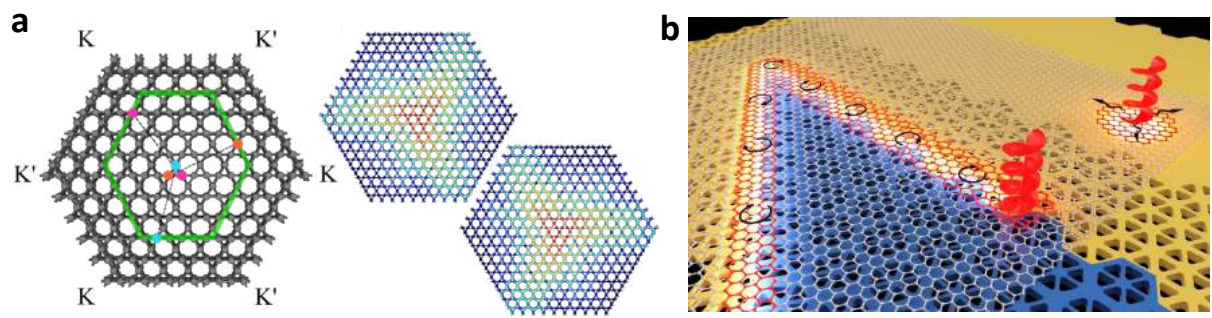

**Figure 1.** (a) Acoustic pseudo-spin in 3D printed kagome lattice of acoustic resonators (left) and control over the propagation directions (K or K') of the spin-full bulk modes via circularly polarized excitation (right). (b) Illustration of multiphysics with synthetic degrees of freedom—polaritonic systems where structured optical field endowed with pseudo-spins induces chiral vibrational modes in van der Waals materials [257].

## Current and Future Challenges

Emergent topological phenomena involving nonlinear and active regimes, where new pseudo-spin-dependent and topological nonlinear responses and nonlinear modes could be observed, are therefore of great interest [258-261]. The traditional approach to enable this challenging vision has been to realise acoustic metamaterials loaded with active elements by incorporating electric circuits with nonlinear elements and amplifiers, which may provide desirable feedback via microphones and transducers. While this approach does enable testing of both nonlinear and non-Hermitian regimes, it is hardly scalable, and therefore, it is not suitable beyond proof-of-principle demonstrations.

Moreover, acoustic excitations carry no charge, spin or magnetic momentum, and thus they do not easily interact with other types of excitations. This in turn makes it difficult to actively control acoustic waves, exploit nonlinearities that can be inherited from such excitations, and enable multi-physics phenomena. Indeed, by mixing and hybridizing acoustic modes with excitations of different a nature, such as mechanical waves, electromagnetic and optical waves, magnons and spin-waves, etc., as illustrated in Fig. 2, one could promote nonlinear effects, drain their energy to induce acoustic gain, or expand the number of available degrees of freedom to explore richer physics, e.g., by engineering synthetic Hamiltonians of a more complex structure and enlarged dimensionality. It is, therefore, a major current challenge in the field of topological acoustics to find material systems that could efficiently and actively interact with acoustic waves and, at the same time, can be suitable for integration into acoustic metamaterials [258-260].

In addition to these materials' related challenges, there is still a great need in better understanding how synthetic pseudospins and spatially varying gauge fields can affect the radiative properties of open acoustic metamaterials. While photonic pseudospins have been shown to enable unprecedented control over radiation, from basic control over the radiative lifetime of topological modes to the generation of vortex beams within topological cavities, these successes have not been translated into the acoustic domain yet. Besides the control over far-field radiative profiles, the possibility to control near-fields with synthetic pseudospins could enable a plethora of applications unique to acoustics, including trapping and moving objects along pathways defined by synthetic gauge fields, and transfer of angular momentum from acoustic fields to trapped objects for controllable rotations.

**Advances in Science and Technology to Meet Challenges**

While basic models—such as tight-binding model and coupled mode theory, which are widely used to describe acoustic metamaterials with synthetic degrees of freedom—can sometimes be sufficient for understanding basic properties of metamaterials, they also tend to oversimplify the physics by largely ignoring the structure of pressure, displacement and velocity fields, and long-range coupling within metamaterials. On the other end, first-principles methods, such as finite element methods which allow to directly solve wave-equations, provide little insight into the fundamental features of topological metamaterials. While these two approaches typically agree well and meet the needs of systems based on discrete resonances, such as discrete lattices of printed or machined acoustic cavities, they diverge when considering systems where the continuous nature of wavefields should be properly treated. One example is long-range interactions that naturally exist in non-discrete systems and can give rise to significant corrections to spatial dispersion of the modes, and even to new types of topological excitations [262]. Similarly, in open systems where acoustic fields may have an evanescent nature, discrete models would not capture the complex structure of the near-fields carrying nonzero angular momentum [242,263]. It is therefore crucial to develop analytical and semi-analytical techniques that can be simple enough to provide an effective Hamiltonian description, yet capable of accounting for the structure of the wavefield. One of the candidates for such description is mode matching, which has been widely used in various contexts, but not for the description of pseudospins and gauge fields in acoustic metamaterials.

A comprehensive theoretical description of the acoustic field structure is also crucial for understanding and modelling the interaction of sound with active matter. For example, overlap of chiral hotspots of the near-field in topological materials can be used for the generation of pseudospin-dependent synthetic gauge fields and spin-selective control of sound waves. Active matter could be mechanically, electrically, or optically driven, and contain internal degrees of freedom which could be engineered to interact selectively with synthetic acoustic pseudospins. In this case, interactions can be leveraged to induce a desirable form of non-Hermitian and nonlinear synthetic gauge fields.

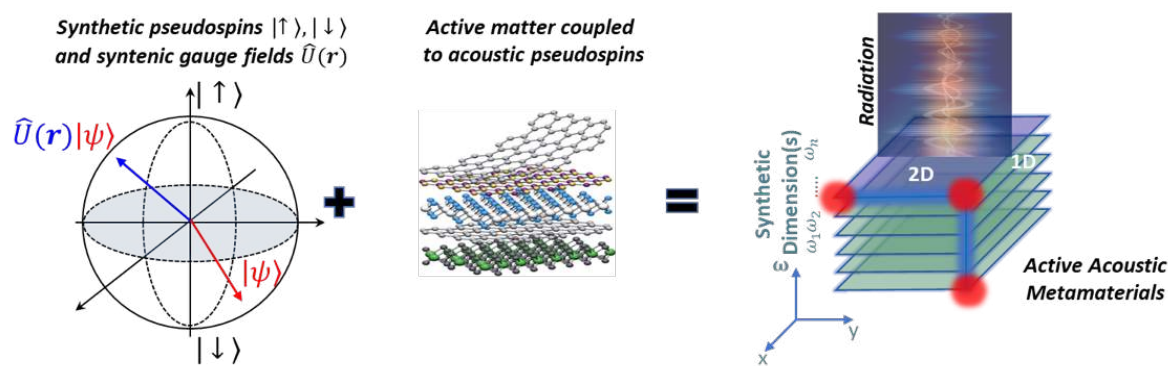

**Figure 2.** Concept of nascent acoustic metamaterials with synthetic pseudo-spins coupled to active and nonlinear materials for unmatched control over propagation and radiation of acoustic waves. Judicious design of coupling of structured sound waves with active, nonlinear, and time-modulated materials envisions emulation of non-Hermitian and interacting topological phases in real and synthetic dimensions.

**Concluding Remarks**

Acoustic metamaterials with symmetry-engineered pseudospins have already become an exciting platform for control of propagation of sound waves via synthetic gauge fields acting in the expanded Hilbert space span by pseudospins. Coupling such metamaterials with active materials interacting with sound waves opens even broader opportunities via realisation of pseudospin-dependent tuneable gauge potentials, including the realisation of phenomena that have so far evaded the field of acoustics, such as novel non-Hermitian and nonlinear topological phases, including real and synthetic dimensions. Hybrid acoustic/active-matter systems also offer a broad range of novel applications, from active control of acoustic radiation as well as structure of the near- and far-fields which can enable new approaches for acoustic communications, imaging and for mechanical trapping—acoustic tweezers, all enhanced with additional degrees of control via synthetic degrees of freedom.

## Acknowledgements

Our work in this area has been funded by the National Science Foundation, the Office of Naval Research, and the Simons Foundation.

## 18. Mechanical effects of structured sound waves

*Etienne Brasselet*

University of Bordeaux, CNRS

**Status**

Acoustic waves are mechanical in nature because their existence requires a medium whose vibratory characteristics allow us to classify the waves into two families: longitudinal (compression) and transverse (shear). Considering the idealized situation of a plane wave propagating in a homogeneous medium, the former refers to a 1D back-and-forth motion along the direction of propagation, while the latter refers to a motion, usually 2D, in a plane orthogonal to the propagation direction.

More than a century ago, physicists discovered that the propagation of acoustic waves is accompanied by mechanical effects on the media in which they propagate. This is illustrated by several pioneering experimental works on the spatial manipulation of microscopic [264] or macroscopic [265] solid objects, and on the deformation of fluid interfaces [266]. These phenomena involve a rich set of physical effects such as heating, fluid dynamics, and radiation stresses. The relative simplicity of the experimental implementation contrasts with the difficulty of a quantitative description which must account for the transfer of energy, linear momentum, and angular momentum between the wave and the material.

Figure 1 illustrates dissipative and non-dissipative wave-matter interactions, which occur jointly at different length scales depending on the nature of the material inhomogeneities and the properties of the media involved. In homogeneous media, the attenuation of waves, which corresponds to the thermo-viscous dissipation inherent to the setting in motion of matter at the microscopic scale, leads to a transfer of energy and momentum. Energy transfer produces a heating while a force per unit volume exerted on the matter results from momentum transfer, which generates flows in usual fluids, deformations in viscoelastic media, and mechanical stresses in solids. The transfer of momentum can also occur at the interface between two homogeneous media where the discontinuous change in material properties results in a force per unit area exerted on the interface, which can then be deformed, even—and perhaps somewhat counterintuitively—if the interface is acoustically transparent [347].

All these phenomena have led to the emergence of numerous applications such as quantitative imaging of inert or living media, metrology of wave or material properties, non-contact manipulation and processing of fluids and objects. Knowing that any real-world field is a superposition of plane waves, and that any system is finite, *structured acoustics meets structured matter* is the norm, for which ongoing conceptual and technical developments aim to fully exploit the advantages of acoustomechanics based on translational and rotational degrees of freedom.

**Current and Future Challenges**

When considering the mechanical effects of (compression) sound waves, it is striking how often theoretical and experimental developments are carried out in the context of their electromagnetic counterparts. This highlights the power of analogies in wave physics as well as the fundamental distinctions and opportunities associated with fields of a distinct nature. To name but a few, we are dealing with: longitudinal (sound) versus transverse (light) waves; coupled scalar and vectorial fields (pressure and velocity) versus coupled vectorial fields (electric and magnetic); a series of material parameters involved in the constitutive relations describing how the fields evolve, which offer more practical flexibility in acoustics (density and compressibility) than in optics (dielectric and magnetic permittivity); and generic wave scattering problems involving multipole expansions in the treatment of the wave-matter interaction.

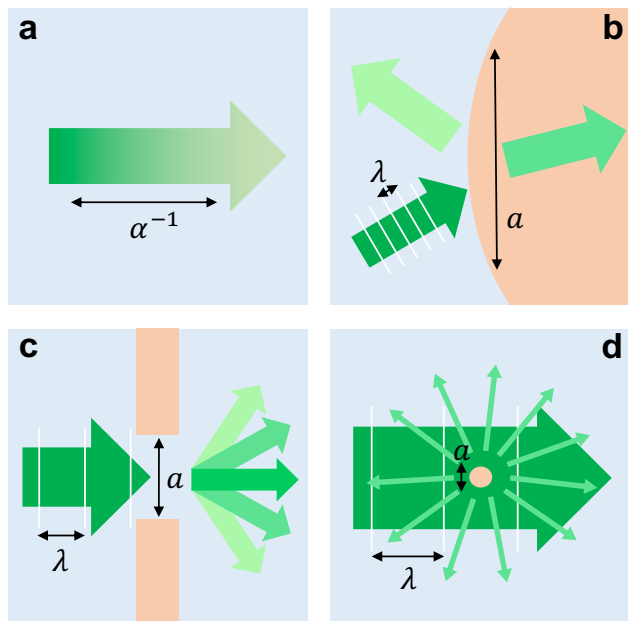

**Figure 1.** Illustration of the main mechanisms driving mechanical effects mediated by wave-matter transfer of energy and momentum. (a) Wave attenuation during propagation, which is associated with the characteristic length $\alpha^{-1}$, where $\alpha$ is the field attenuation coefficient. (b) Geometric reflection and refraction of acoustic beams when the characteristic length $a$ associated with inhomogeneities of the material properties typically satisfies $a/\lambda \gg 1$, where $\lambda$ is the wavelength. (c) Wave diffraction when inhomogeneities are of the order of the wavelength, hence typically $a/\lambda \sim 1$. (d) Scattering for point-like inhomogeneities, hence in the regime $a/\lambda \ll 1$. Note that there is no formal boundaries between the latter regimes; still, the parameter $ka$ is a key parameter to define the relevant framework when describing the wave-matter interaction.

Analytical toolkits are emerging which provide a better fundamental understanding of the generic and specific aspects of the physics involved, e.g. [43,267]. Furthermore, they equip experimentalists with rational design and fabrication strategies of soft metamaterials whose properties go beyond those of their constituents, mediated by spatially extended acoustic force landscapes. Exploiting acoustomechanics opens the way to programmable functionalities [268] provided that the structured sound fields can be adapted to demand. Usually, the soughtafter "meta-atom" architectures correspond to periodic networks, for which standing waves are well suited. However, this prevents the development of meta-devices endowed with spatially distributed functionalities. Non-periodic radiation force networks therefore appear as a natural and time-dependent next step as a way to actively control the local interactions of many-body systems and thus their collective behaviour [269], seeding the development of active matter fuelled by sound.

Although rotational mechanical effects date back to the beginnings of acoustomechanics, as recalled by the implementation of the sound mill by Dvorak and Mayer in the 1870s [265], it is only recently that angular momentum characteristics of structured sound are experimentally exploited. On the one hand, pressure fields endowed with phase singularities make it possible to exert acoustic radiation couples on matter either by dissipative [270] or non-dissipative [271] orbital angular momentum transfer processes. On the other hand, a spin contribution to the angular momentum of sound, which is intimately linked to inhomogeneous fields (two plane waves are sufficient) and is associated with the local elliptical vibration of the medium, has also been demonstrated experimentally recently by using a dissipative and polarisable subwavelength object [42]. While taming the different facets of spin and orbital angular momentum of sound is still in its infancy, the interest it has generated suggests that it will not remain a mere scientific curiosity for long.

## Advances in Science and Technology to Meet Challenges

Now that conceptual frameworks, structural design approaches, and fabrication tools are available for waves and materials, the study and exploitation of the mechanical effects of structured sound has a bright future, which depends in part on the ability of often distinct research communities to open up to each other.

The contactless manipulation of matter highlights how fundamental and technological advances can come together in a simple way. As an example, we mention the prototypical situation of a focused vortex beam interacting with a spherical particle, which refers to the recently introduced single-beam acoustic tweezers [272] and inherently involves mechanical translational and rotational degrees of freedom. This represents the acoustic analogue of their famous optical counterpart celebrated by the 2018 Nobel Prize in Physics, but endowed with force and torque resources that are—Watt per Watt—several orders of magnitude larger. Indeed, the linear and angular acoustic momenta respectively scale as $1/c$ and $1/\omega$, where $c$ is the speed of sound and $\omega$ is the angular frequency of sound.

To date, the spin, orbital, and viscous contributions to the mechanical actions exerted on a particle trapped in vortex tweezers remain little explored. This invites one to address the acoustohydrodynamic problem as a whole towards developing applications such as sound-driven micro-machines from physical, chemical, and biological perspectives. In particular, it appears necessary to go beyond the simple, yet instructive, case study of an isotropic spherical particle immersed in an isotropic standard fluid. Conversely, technologies relying on the mechanical effects of "unstructured" sound, such as radiation-force based ultrasound imaging techniques, are likely to benefit from the advantages deriving from structured waves since other mechanical degrees of freedom are involved, such as acoustic vortex elastography [348].

From a general point of view, the growing interest in acoustic spin and orbital angular momentum opens up questions such as: How can the polarization state of sound be used to manipulate anisotropic media as is done in electromagnetism for many decades with optical spin angular momentum? How can the singularities and topological textures of inhomogeneously polarised acoustic fields be used to shape matter in non-trivial ways? How can spin-orbit interaction mediated by anisotropic or inhomogeneous media be used to enrich the toolbox of acoustomechanics? In this context, topological properties of artificially structured materials and elasticity are some of the new key players at play.

## Concluding Remarks

Although speaking of the mechanical effects of sound is formally a pleonasm—sound being itself a mechanical movement—their surprisingly rich consequences across length scales, both for waves and for matter, make them still attractive. Recalling that, more than a century ago, the first steps of acoustomechanics dealt with the use of spatially textured sound fields owing to wave interferences [264], the exploration of linear and angular mechanical effects exerted on objects [265] and the use of deformable materials allowing nonlinear feedback phenomena [266], it is remarkable how easy it is to find echoes of them in current research themes. Now that acoustics has entered into our everyday life as sensors, transducers, and imaging systems, among other things, the mechanical effects of sound remain a source of inspiration for improving knowledge as well as for designing and implementing new technologies. We may hear about structured sound for a long time to come.

## 19. Transport of surface matter in structured water waves

*Michael Shats*

The Australian National University

### Status

Water surface waves share many similarities with their optical and acoustic counterparts, except for a very different dispersion relation, $\omega^2 = gk + \sigma k^3/\rho$, where $\omega$ and $k$ are the wave frequency and wave number, $g$ is the gravity acceleration, $\sigma$ is the surface tension coefficient, and $\rho$ is the density of the liquid. The restoring force for the surface perturbation at long wavelengths (longer than about 20 mm in water) is the gravity force and $\omega \sim \sqrt{k}$. For shorter waves (less than 10 mm), the restoring force is capillary and $\omega \sim k^{3/2}$. Though water waves have been studied for centuries, there is no universal theory which would describe the motion of fluid particles even in relatively simple structured waves. On the other hand, recent progress in laboratory studies of the particle motion on the water surface perturbed by waves revealed rich phenomenology related to the generation of horizontal vortices and vortex lattices [273], direct and reversed jets [274], and to developed two-dimensional turbulence [275]. Some of these phenomena, e.g., vortex lattices, are related to the wave momentum and spin [55], while others, for example, turbulent motion of fluid particles driven by steep nonlinear waves, require new theoretical approaches. Horizontal motion of fluid particles at the surface is coupled to the wave pattern and to the width of the wave spectra [277]. Chaotic fluid motion at the surface leads to the increased disorder in the wave field as manifested in the broadening of the wave spectrum. This suggests new theoretical approaches which would allow to predict rms velocities of the fluid particles from the wave spectrum width and vice versa. It is interesting to note that, to generate 2D turbulence at the liquid-air interface, it is not necessary to drive turbulence in the wave field; a slightly broadened wave spectrum results in a broad spectrum of the horizontal fluid velocities matching classical Kolmogorov-Kraichnan spectrum $E_k \propto k^{-5/3}$.

The mass transport driven by surface water waves is well recognised in natural applications, for example, in oceanology. However, there is also growing interest in controlled manipulation of particles on a liquid surface for engineering applications, such as mixing, particle sorting and clustering, as well as for controlling properties of tunable 'metafluids' [278]. Better understanding of the wave-driven transport of particles opens opportunities for the development of new biomaterials in liquid media by applying waves to the growing culture [279]. Waves can also be used to promote or discourage the formation of biofilms on solid substrates [279].

### Current and Future Challenges

Recent progress in experimental research advanced our understanding of the mass transport driven by the surface waves. Experiments revealed a different nature of the particle motion in small-amplitude, or weakly nonlinear waves, and in parametrically excited, strongly nonlinear waves, also known as the Faraday waves. Traditionally, the mass transport by propagating waves was described within the framework of the Stokes drift in 2D wave fields. Such drift along the *y*-axis (*z*-axis is in the vertical direction) produces vertically polarised trochoids, or the motion with horizontal spin. Recently, it was shown that, in 3D waves, fluid particles have both horizontal and vertical spin and corresponding generalized Stokes drift [55]. An example of the particle trajectory is illustrated in Fig. 1. Good agreement between theory and experiments gives hope that the field theory approach can be productive in developing theories and models capable of predicting the mass transport for the weakly nonlinear waves.

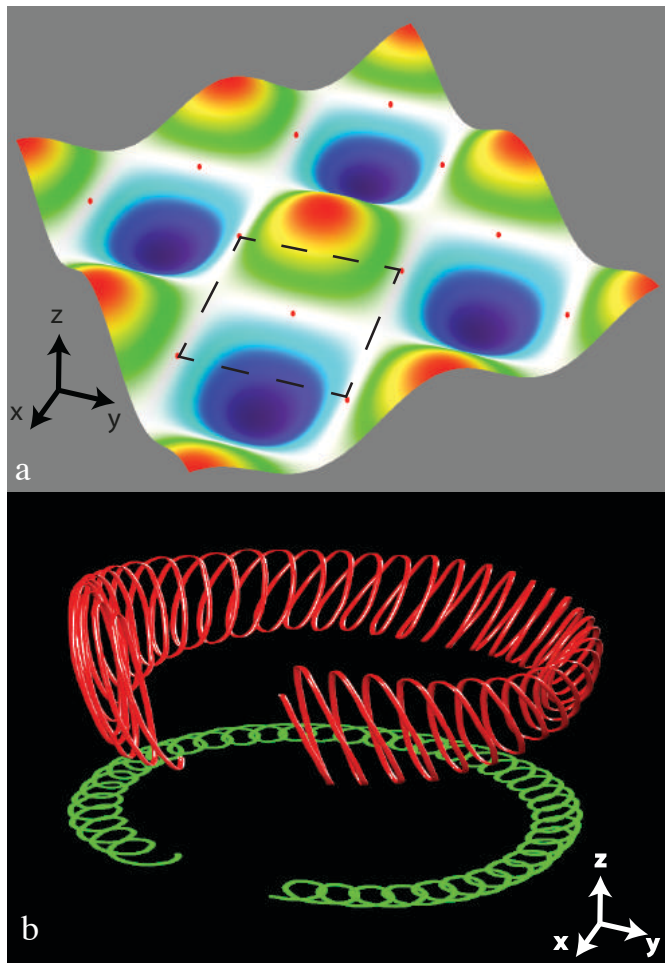

**Figure 1.** (a) Measured surface elevation produced by two orthogonal standing waves. (b) Measured 3D trajectory (red) of a surface particle drifting within a unit cell (dashed line in (a)) and its projection on the horizontal plane (green).

The existence of the vertical and horizontal spin in 3D waves is also important for the motion of larger inertial particles possessing internal spin. The motion of such particles is governed by the interaction between the wave spin and particle spin, and it opens the opportunity to manipulate and sort spinning particles using structured surface waves [280].

The existence of the vertical spin in some wave configurations offers a new conceptual base for the development of the surface wave spintronics [281], where the spin of passive particles can be controlled by imposed surface waves [55]. The challenge here is to account for the return flows in continuous medium as a reaction to the wave-driven drifts.

The situation is more complex for steep nonlinear Faraday waves. Fast motion of the surface fluid particles differs qualitatively from the slow drift in linear waves; particle trajectories in such waves have no resemblance with classical Stokes drift [275], as seen in Fig. 2. The development of theory considering particle inertia is a challenging problem. The importance of the inertia of fluid parcels in Faraday waves is manifested in the extended inertial interval in the spectra of horizontal fluid velocities [275]. Though Faraday waves generate chaotic particle motion, the mass transport is statistically predictable and allows fine control over particle dispersion at the surface [282]. The main challenge is the development of theory based on the Lagrangian description of fluid motion. The progress can be made through the development of theoretical models verified and fine-tuned in experiments.

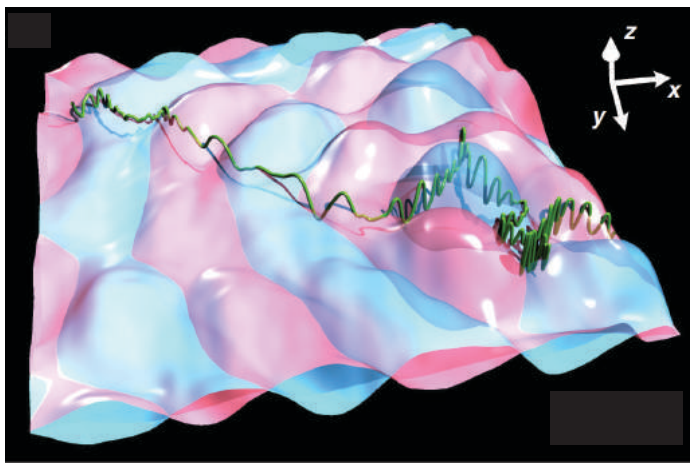

**Figure 2.** Fluid particle motion in Faraday waves. Pink and blue wave fields correspond to two consecutive phase extrema of the waves separated in time by a half-wave period. Green: three-dimensional particle trajectory followed for 100 Faraday wave periods.

## Advances in Science and Technology to Meet Challenges

Recent demonstration of importance of the wave-generated spin of fluid parcels on the surface perturbed by structured waves is an important step towards developing new applications relevant for particle manipulation and sorting. These would require overcoming several problems related to (a) incorporation into theoretical models of the transverse angular momentum–induced transport in continuous media, or (b) balancing in experiments of the return flows caused by the spatially varying Stokes drift. These problems are related to the configuration driven by two orthogonal propagating waves [55], while they are not important in the field produced by two orthogonal standing waves, as in Fig. 1(b), where the Stokes-drift flow closes on itself and thus remains stationary. In this example, waves generate large-scale vortices which form a vortex lattice.

The generation of a vortex lattice is also a feature of the nonlinear Faraday waves [275]. However, in that case, due to the larger fluid velocities, vortices strongly interact with each other causing 2D turbulent motion. Experiments demonstrated that turbulence is dominated by the randomly moving coherent bundles of particles, or meandering 'rivers', whose width is about half the Faraday wavelength [282]. The existence of such 'rivers' is an important feature of the wave-driven turbulence. In particular, the 'rivers' can be guided by solid boundaries within the flow which allows the rectification of the turbulent (mean-zero) velocity fluctuations. It has been shown that this effect can be used to create unidirectionally propagating floaters which tap turbulence energy (self-propelled floating devices) [283]. Similarly, the turbulence-driven rotors powered by turbulence have been demonstrated in laboratory experiments [284]. The latter ability can be used for efficient utilization of the wave energy. This direction requires better theoretical basis and models capable of deriving flow parameters relevant to engineering applications from the properties of the disordered but structured wave fields.

Periodic and quasi-periodic wave-driven flows have a potential to be used in biological flows, such as bacterial suspensions. It has been demonstrated that waves can shape the patterns of the bacterial biofilms developing in the wave fields [55]. This should also be investigated in the wave-driven turbulence. A potentially important theme is the formation of the structure of the bacterial cellulose. This important biomaterial grows near the media-air interface, and it is affected by the surface waves. This would require joint efforts by physicists and microbiologists to understand the formation of extracellular polymeric matrices in moving fluid environment.

**Concluding Remarks**

The structured water wave field is an emerging tool to control mass transport at the gas-liquid interface. Weakly nonlinear surface waves share many similarities with optical and acoustic waves [55] and can drive deterministic transport of the surface matter. They promise new directions, such as liquid interface spintronics based on the transverse angular momentum-induced transport, resembling the spin Hall effect. Due to the modest wave steepness, the prediction of the mass transport in such waves is applicable to oceanic waves, or they can be generated in various industrial flows.

Steep nonlinear waves (Faraday waves) generate intense turbulent motion on the surface due to the strong interaction between wave-driven horizontal eddies. Such turbulence shows statistical properties of 2D turbulence since it is based on the inverse energy cascade, a process of the spectral energy transfer from intermediate to large scales. In a bounded domain, such a transfer can lead to the accumulation of spectral energy at the domain scale, a.k.a. spectral condensation, which is a form of turbulence self-organization into a coherent vortex. Wall-guided self-organization of turbulence can be used to rectify turbulent energy for the development of self-propelled surface vehicles and unidirectional rotors for the wave energy conversion.

**Acknowledgements**

This work was supported by the Australian Research Council Discovery Project DP190100406.

## 20. Structured electron waves

*J. Verbeeck[1] and P. Schattschneider[2]*

[1]EMAT, University of Antwerp
[2]TU Wien

### Status

Electron beams are used in a wide variety of applications ranging from vacuum tubes that formed the basis of electronics more than half a century ago with specific high power and high frequency tubes still being indispensable today, to electron microscopes providing atomic resolution images of materials, and free electron lasers providing intense X-ray beams, radiotherapy and surface treatment, e- beam lithography and chip inspection tools, displays, portable X-ray sources and many more.

Most of these applications rely on a classical picture of a ray of accelerated electrons, providing a current through vacuum. But as with light, the wave nature of electrons limits the spatial resolution to the de Broglie wavelength reaching picometer levels for energies greater than a keV, at least five orders of magnitude smaller than the wavelength of visible light. This unique property of electron beams is the essence of its use in electron microscopy and e-beam lithography, providing spatial resolution that reveals the atomic structure of materials on a routine basis. In order to approach the ultimate resolution limit, wave aberrations induced by magnetic round lenses of electron microscopes had to be corrected. The tremendous evolution of such correctors in the last two decades, based on phase modifying magnetic multipoles, has generated ideas to apply phase control of the electron for even more sophisticated applications.

Quantum mechanically, electrons are described by the Schrödinger/Dirac equation. The relevant paraxial solutions are highly similar to solutions of the Helmholtz equation used to describe wave phenomena in optics and acoustics. This indicates that all wave shaping that is investigated in these areas (see e.g. Sections 7 and 18) can, at least in principle, be carried over to electron beams.

Wave shaping of electrons can be obtained by interaction with electric or magnetic fields provided either by (macroscopic) sources, as realized in aberration correctors, by phase plates, or by the microscopic electric or magnetic potential of matter (Fig. 1).

In this roadmap, we want to look beyond *resolution revolution*, exploiting the quantum nature of the electron, in analogy to the revolution of *adaptive light optics* or phased arrays in radiocommunication/radar and acoustics.

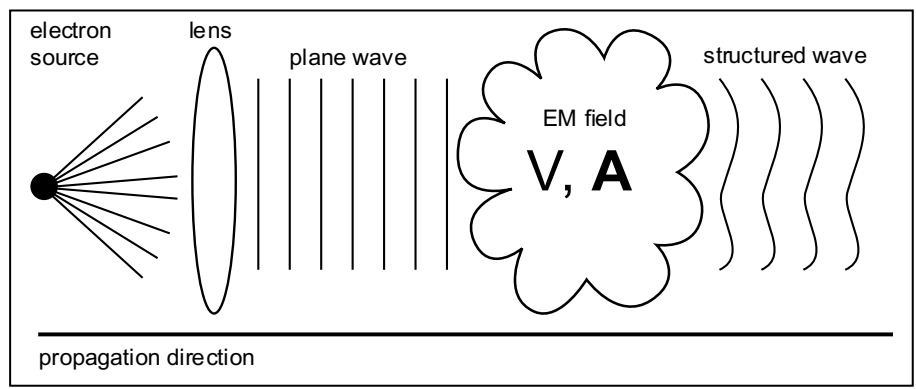

**Figure 1.** Sketch of the interaction of a paraxial electron beam with purposely designed scalar (V) and vector (A) electromagnetic potentials creating a structured electron wave.

**Current and Future Challenges**

So far, *electron vortex beams* (EVB) have received the most attention with a wide range of experiments showing ways to create these topological electron waves, carrying quantized orbital angular momentum (OAM), similar to its optical counterpart. The charge of the electron results in a magnetic moment mµB parallel to the propagation axis, with m a quantum number defining the OAM and µB the Bohr magneton -—a unique property of charged matter waves. So far, EVBs have been created, by and large, using forked amplitude gratings [285], magnetic monopole-like fields, and thin refractive elements with spiral height profiles. These methods are static and absorb a significant fraction of the beam intensity [286,226]. A phase plate in combination with cylinder lenses (magnetic quadrupoles) conserves the intensity and can be tuned so as to achieve EVBs with + or − helicity. It can as well be operated in reverse [287]. Tuneable electrostatic phaseplates have been demonstrated as well [276]. Figure 2 shows some vortex beams and an overview of wave shaping methods. Besides creating EVBs, significant progress has been made in EVB filters to decompose arbitrary electron beams in its OAM components [349].

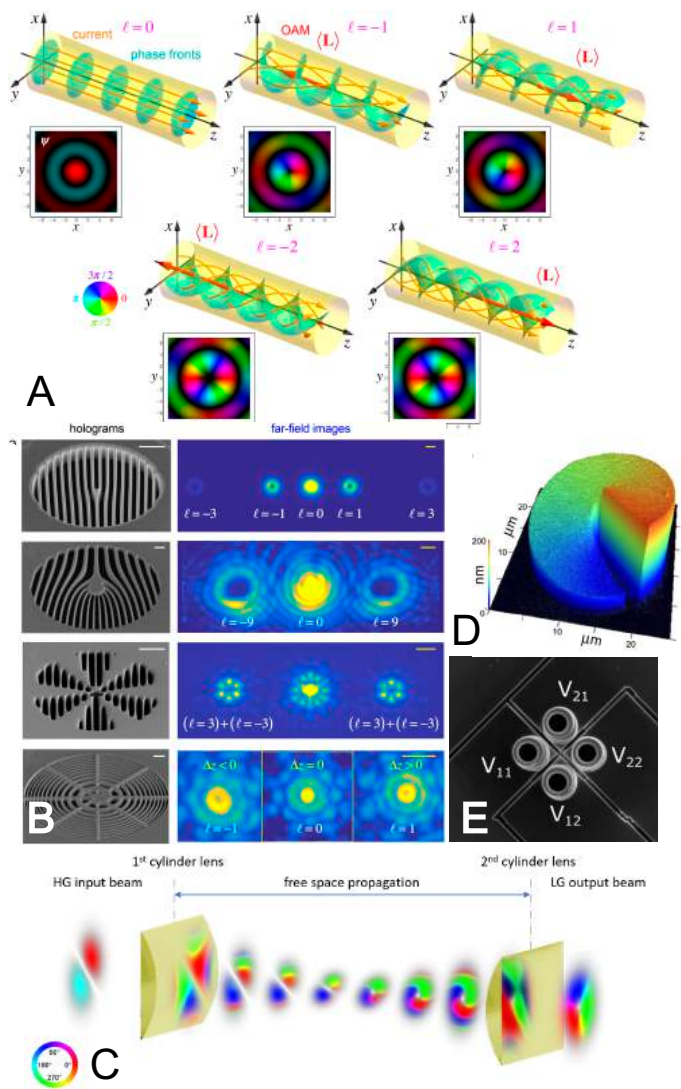

**Figure 2.** A) a series of electron vortex beams carrying topological charge -2 to 2, with their wavefront structure. B) A series of holographic gratings producing a variety of structured waves. C) Mode converter with cylindrical lenses.  D) Spiral phase plate. E) 2x2 electrostatic programmable phase plate for electrons. Reprinted from ref 2 (A, B, D), ref 3 (C), ref 7 (E) with permission from Elsevier.

EVBs can reveal chirality in crystals [288] or apply torque on nanoparticles and atoms with predicted very high rotational speeds. Interaction with optical excitations in materials could reveal local information on optically chiral nano-objects [289]. Structured electron waves carry information equivalent to polarized light, but at far higher resolution [289], a fact upon which EMCD, the electron counterpart of XMCD, is based [290].

Besides EVBs, other non-diffracting beam classes have been studied, like Airy beams seemingly following a parabolic trajectory in field-free space (like a curve ball), or Bessel beams that stay focused over long distances, providing attractive prospects to study thicker samples [350-353].

Challenges in *arbitrary wavefront shaping* relate to pushing experimental trials to technical maturity, such as scaling up an array of electrostatic Einzel lenses [291], or using the ponderomotive effect when electrons travel through regions of intense light, shifting the challenge towards making a programmable light field with extreme intensities [292]. Other attempts use light fields in combination with an electron transparent thin film, increasing the electron-photon coupling at the expense of inserting material in the beam [293]. Yet other experiments have achieved EVBs by pulsed laser illuminating of round apertures with circularly polarised light via interaction with surface plasmon polaritons [294].

**Concluding Remarks**

The emerging ability to freely shape electron wavefronts in practical instruments opens interesting avenues for future research and application.

One is the ability to encode quantum information in electron beams e. g. in a vortex basis. This provides a quantum communication channel, which could open new avenues for quantum computing. The short wavelength combined with strong interaction with electromagnetic fields and the robustness against decoherence, the topological protection and ease of single particle detection provides attractive features.

Another emerging opportunity is the ability to implement adaptive optics in electron beam instruments, where the instrument can optimise its beam conditions in a feedback loop to avoid tedious alignment procedures and to provide the highest possible contrast for specific features of interest [354]. This is especially attractive for life science as contrast is notoriously poor, demanding a high electron dose, often damaging the material before images can be obtained. Dynamic structuring of the wavefronts could maximize the information per dose.

Other applications include imaging through thick objects, coded aperture approaches to directly reveal the phase, dynamic phase scrambling to reduce effects of dynamic scattering in both imaging and diffraction and many more.

**Acknowledgements**

JV acknowledges funding from the eBEAM project supported by the European Union's Horizon 2020 research and innovation programme under grant agreement No 101017720 (FET-Proactive EBEAM), FWO project G042820N 'Exploring adaptive optics in transmission electron microscopy' and European Union's Horizon 2020 Research Infrastructure - Integrating Activities for Advanced Communities grant agreement No 823717 – ESTEEM3. PS acknowledges the support of the Austrian Science Fund under project nr. P29687-N36.

## 21. Structured neutron and atomic waves

*Dusan Sarenac, David G. Cory, and Dmitry Pushin*

University of Waterloo

**Status**

The successful extension of the structured waves toolbox to neutron and atomic beams promises an array of exciting applications in fundamental physics and material characterization techniques. For example, neutrons offer a complimentary probe of nature and materials when compared to photons and electrons, as they possess unique penetrating capabilities and interaction strengths due to the strong force and electroneutrality. However, given the wave-particle duality, the methods first developed for generating and characterizing optical structured waves are typically the backbone of the methods to create structured neutron and atomic waves [366]. Although the methods are theoretically and conceptually analogous, the practical realizations are complicated by the technical challenges associated with controlling and manipulating these de Broglie waves. For example, the extension of holography techniques to neutron waves was accomplished via perfect-crystal silicon interferometer which through Bragg diffraction provided a coherent superposition of an angled reference beam and a beam that had passed through a macroscopic object [367]. This is a direct adaptation of the two-beam wedge optics technique introduced by Leith and Upatnieks [368]. The first experiments with neutron orbital angular momentum (OAM) focused on manipulating the OAM of incoming neutrons, though given that the input beam had a small transverse coherence length (ranging from ≈nm to μm) relative to the beam diameter (≈cm) the value of the OAM was not well defined [369]. Several theoretical studies of incorporating spin correlations to OAM [370-372] led to an experiment that prepared and characterized neutron lattices of spin coupled OAM beams [373]. Shown in Fig. 1 is the experimental setup of Ref. [373] that relied on a sequences of magnetic field gradients produced by spatially oriented triangular coils to achieve programmable spin topologies.

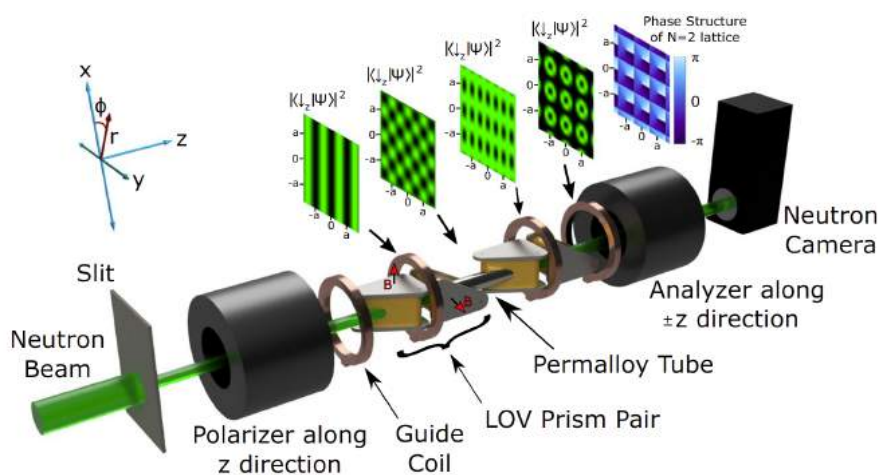

**Figure 1.** The setup used in Ref [8] to prepare and characterize neutron lattices of spin coupled orbital angular momentum. Sequences of triangular magnetic fields are used in conjunction with ³He cells. Here the magnetically polarized ³He cells act as neutron spin polarizers due to their neutron absorption cross section being highly dependent on the neutron spin direction relative to the helium polarization state. Shown are the simulated spin dependant intensity profiles at each stage of the setup. Reprinted with permission from Ref. [373].

In regards to inducing azimuthal phase shifts over the wavepacket coherence lengths, the first experimental achievement of atomic and molecular beams carrying quantized OAM came in 2021 [374]. The demonstration was done with helium atoms and metastable helium dimers. This work opens new avenues in using the OAM to probe particle collisions between atoms and/or molecules. Soon after, the first demonstration for quantized neutron OAM (depicted in Fig. 2) was achieved in 2022 [375]. The convenient integration of this method with material characterization studies at Small Angle Neutron Scattering facilities promises to extend neutrons as probes of topological material's bulk properties [376,377], which cannot be directly probed via photons or electrons.

**Current and Future Challenges**

The grand scientific challenges revolve around incorporating the additional degrees of freedom brought forth by structured waves, such as OAM, into the existing scattering theory and material characterization methods. Spin textures and spin topologies, such as skyrmions and merons, possess a non-trivial coupling between spin and other dynamical degrees of freedom which manifest a rich variety of emergent dynamics and phases of matter. While conventional probes provide indirect transport measurements of topological excitations, neutrons with specific spin-orbit couplings are strongly desired because they may act as direct probes of the target's topology.

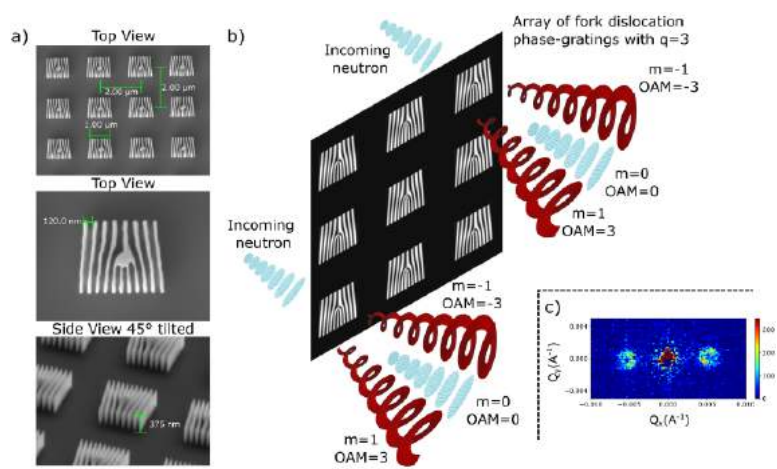

Figure 2. Grating arrays have been the enabler of experimental demonstrations of neutron, atom, and molecular helical waves. a) The SEM images of the grating arrays used in the neutron demonstration. b) The conceptual illustration where each grating of the array coherently acts on individual neutrons. c) Example of the observed intensity in the far field where the OAM signature profiles are observed in the diffraction orders. Reprinted with permission from Ref. [375].

The technical challenges that deter the progression of the field revolve around the difficulties in preparing, manipulating, and detecting neutron and atomic beams. It is expected that as those capabilities evolve, the field follows the progression of optical structured waves. First to note is that there is no device equivalent to a laser which outputs coherent light. Whereas coherent and well-defined Gaussian states are the common inputs to optical experiments with optical structured waves, experiments with neutrons and atoms are limited to working with beams whose transverse coherence lengths are much smaller than the size of the actual beam. Typical methods of beam collimation rely on circular slit pairs to define the beam divergence and thus the transverse coherence. Here we can also note that the low neutron flux ensures that only one neutron at a time is present in the entire setup, and therefore all the experiments in essence are done with a post-selection on there being a neutron. In relation to this, a notable challenge is the access and availability of high intensity neutron sources. Likewise, it is worth mentioning that many other electromagnetic components which are taken for granted in optics setups are also not yet practical. For example, the neutron index of refraction for most materials is around n≈1-10$^{-5}$, which makes the production of a neutron lens impractical. Lastly the position sensitive neutron detectors possess much poorer spatial resolution when compared to optical cameras. For example, the OAM demonstration with neutrons of Ref. [375] used a neutron camera with a pixel size of around 5mm by 5mm.

**Advances in Science and Technology to Meet Challenges**

The holy grail technological advance that would push the field of neutron and atom structured waves is a component analogous to a spatial light modulator (SLM). The SLM's ability to provide arbitrary wavefront shaping has revolutionized the field and enabled the widespread use of optical structured waves. However, a practical active device that enables either phase or intensity modulation is currently out of reach for neutron and atomic beams. For the time being, the main enablers have been the advances in programming and control the full range of neutron and atomic degrees of freedom, for example the nanofabrication methods of passive devices. It is interesting to note that both OAM works of Ref. [374] with helium atoms and molecules as well as Ref. [375] with neutrons relied on nanofabricated arrays of gratings. The former relied on absorption gratings while the latter on phase-gratings. Due to the small transverse coherence length in both cases the gratings need to have periodicities on the nanometre scale in order to induce coherent diffraction. The arrays in Ref. [375] consisting of 6,250,000 individual fork dislocation phase gratings and they are depicted on Fig. 2. The motivation for the large arrays was to increase the observed signal which otherwise would have been too small to detect. Even with such an increase in signal every intensity image took around an hour of acquisition time. In the case of absorption gratings, the thickness of the gratings needs to be sufficient to spatially remove parts of the beam. Whereas in the case of phase-gratings, they need a high aspect ratio to induce an appreciable phase-shift. Given the suitable material options, the available/possible fabrication methods determine whether phase or intensity gratings are more practical.

**Concluding Remarks**

The power of neutrons was most striking in the hallmark fundamental physics experiments such as the first demonstration of gravity on a quantum particle and the observation of the $4\pi$ symmetry of spinor rotation, and the impactful industry applications with the neutron imaging of fuel cells and lithium batteries. This was enabled by the control of the three easily accessible degrees of freedom: spin, path, and energy. The addition of the structured waves toolbox and the OAM degree of freedom is expected to set forth the next generation of fundamental experiments [378] and material characterization techniques [379,380]. The neutron's penetrating abilities are well suited for bulk studies of materials with spin textures and spin topologies such as skyrmions.

**Acknowledgements**

The authors would like to thank their many collaborators including Wangchun Chen, Charles W. Clark, Lisa DeBeer-Schmitt, Huseyin Ekinci, Melissa Henderson, Michael Huber, Connor Kapahi, Ivar Taminiau, and Kirill Zhernenkov. The authors would also like to acknowledge their funding sources: the Canadian Excellence Research Chairs (CERC) program, the Natural Sciences and Engineering Research Council of Canada (NSERC), the Canada First Research Excellence Fund (CFREF).

## 22. Structuring the Quantum State of Light

*Michael Birk, Alexey Gorlach, and Ido Kaminer*

Technion–Israel Institute of Technology

**Status**

The shaping of various degrees of freedom of light has been a key factor in the progress of optical sciences and engineering; many degrees of freedom have been brought under control. Spatial shaping has been used to create nondiffracting beams such as Bessel [298] and Airy [295-297] beams, and to imbue light with orbital angular momentum (OAM) [8], while frequency and temporal shaping lie at the core of optical communications and ultrafast optical sciences. Advances that combine both spatial and temporal shaping now push the boundaries of this field even further [299,300]. The richness of capabilities for structuring light made clear that the more degrees of freedom of light are under control, the more applications become possible.

However, it is pertinent to remember that all degrees of freedom addressed so far in the context of structured light are related to classical light. From a quantum perspective, these degrees form only a small subspace that the photonic state can inhabit. The quantum degrees of freedom can be described as a function of generalized phase-space coordinates. This function is known as a Wigner quasiprobability distribution, describing the quantum state of light for a single or multiple optical modes. Our purpose in this roadmap is to propose the concept of structured quantum light. The impact made by the field of structured light shows the prospects that can emerge from developing capabilities for shaping the quantum properties of light. Below, we discuss some of the many open challenges in quantum optics that remain to be resolved if one wishes to acquire control over the quantum shape of light—*structuring of the photonic Wigner function*.

Current efforts and proposals in this direction can be divided into deterministic schemes [301-303] or schemes based on post-selection [304,305]. Most of these efforts have been focused on creating single- or few-photon quantum states. Even though few-photon quantum states have applications in quantum technologies, the ability to create many-photon quantum light states has seen a rising need for a wider range of applications, both classical and quantum. Such states are needed for ghost imaging, precision measurements, quantum communication protocols, and photonic quantum computation based on the so called "bosonic codes" or continuous-variables quantum information [307]. A prime example is the quest toward the generation of the Gottesman–Kitaev–Preskill (GKP) states [308], which are the much-needed resource for scalable fault-tolerant photonic quantum computation [304, 307], and yet they have never been generated in the optical range. From the perspective of basic science, robust control over the quantum light state can be used to significantly enhance even the most fundamental nonlinear optical processes [301].

The applications of structured quantum light will certainly be greatly expanded when tools for many-photon Wigner function shaping will be developed—as demonstrated in the past by the richness of discoveries brought forth by the invention and commercialization of spatial light modulators (SLMs) for spatial shaping of light.

**Current and Future Challenges**

The basic theoretical toolset for shaping the light Wigner function in the optical domain is provided by the operations of rotations, displacement, squeezing, and amplitude dispersion, which can be performed by phase delays, beam splitters, amplifiers, and nonlinear optical effects such as parametric down conversion and the Kerr nonlinearity. However, what is possible in theory turns out to be difficult in practice. While it has been predicted that Kerr nonlinearity can generate Schrödinger cat states [309], experimental efforts to generate macroscopic quantum states in free space have been thwarted by low interaction strengths and dissipation. More recent attempts to overcome these challenges use optical nonlinearities in microcavities, optomechanical cavities, and various integrated photonic platforms. Such approaches have been successful in generating few-photon quantum states but have not yet led to the generation of many-photon states with useful quantum properties.

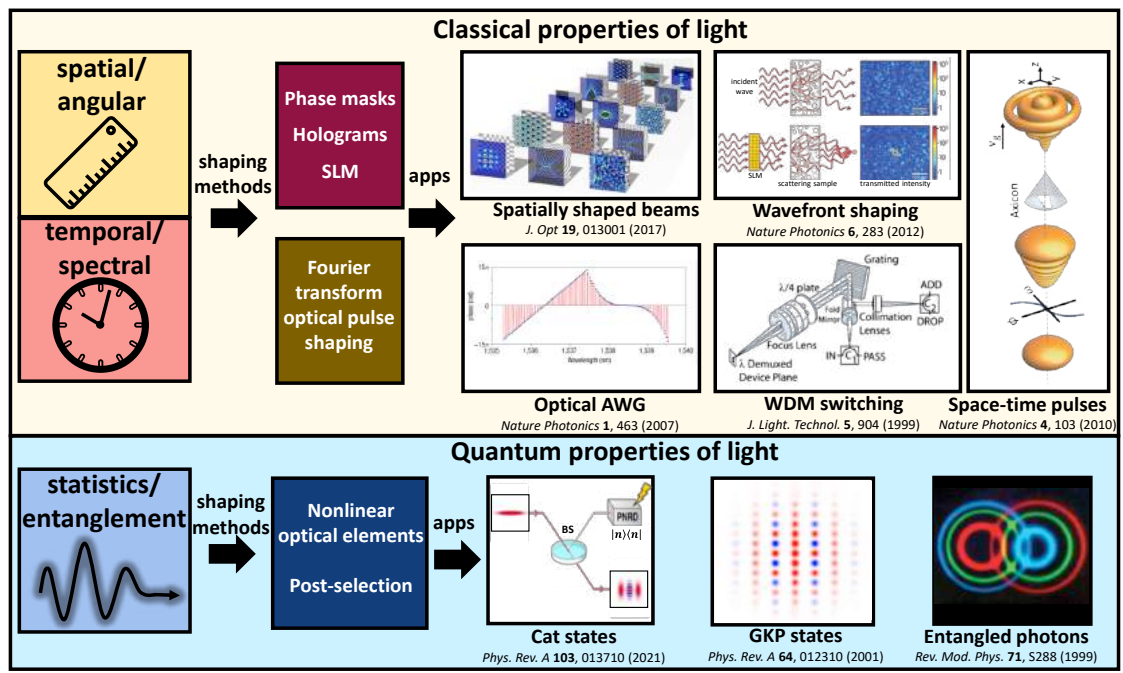

**Figure 1. Structured light across all degrees of freedom.** Current efforts focus on shaping the classical properties of light (top), including spatial/angular/temporal/spectral/polarization degrees of freedom. We envision developments in shaping the quantum properties of light (bottom), including the photon statistics and high-order coherences. While shaping classical properties relies on established technology such as spatial light modulators (SLM), there are currently no standard methods for generic control of quantum properties, but they are expected to rely on nonlinear optics and on post-selection using photon number resolving detectors (PNRD). Techniques for structuring classical properties of light enabled a vast range of capabilities such as imaging and focusing in complex media, optical arbitrary wave generation (AWG), and wavelength division multiplexing (WDM) for applications in optical communication. Spatial and temporal shaping can be combined to gain full 3+1D control over the light field. However, the quantum nature of light contains many more degrees of freedom, such as squeezing and photon entanglement. Finding ways to structure the quantum degrees of freedom could lead to breakthroughs in many areas of quantum technologies. For example, many-photon superpositions of coherent states (cat states) and Gottesman-Kitaev-Preskill (GKP) states are the sought-after building blocks for fault-tolerant optical quantum computation.

The most advanced demonstrations of many-photon squeezed light states to date are using four-wave mixing and Kerr nonlinearities in fibres, on-chip nonlinearities in waveguides and cavities or parametric down-conversion in free-space optical setups. Such experiments enabled manipulation of photon statistics, leading to enhancement of nonlinear processes and generation of exotic indefinite-mean photonic states [302]. These efforts are mostly limited to generation of a limited set of photonic states (called Gaussian states), insufficient for quantum computation [307].

Other methods of quantum state generation involve post-selection methods [306,381] rather than relying on optical nonlinearities. These methods have a different bottleneck—the probability to measure the desired outcome of a heralding observable, and the ultimate sensitivity of the detector. Such schemes are inherently probabilistic, mandating either low generation rates (hence low throughput), or the use of resource-intensive architectures such as parallelizing a large number of generators to raise throughput [304].

A conceptually different approach for the shaping and entanglement of light is utilizing the interaction of light with free electrons [305, 310]. The key to this approach is the nonlinear nature of electron-light interactions and the ability to pre-shape [311] the electron wavepacket before the interaction. A recent experiment demonstrated the effect of quantum photon statistics on the interaction [312], and recent predictions proposed novel methods for utilizing the interaction to create the desired many-photon quantum states [305, 310], including a GKP state [313]. Such schemes may also involve post-selection on the electron energy, since it becomes entangled with the photonic state. The advantages of electron-based approaches are the use of mature techniques for electron wavepacket shaping [311], and a post-selection process that avoids the dependence on photon number resolving detectors, which are a bottleneck in conventional post-selection schemes. The obstacle facing these approaches is the complexity of high-quality electron sources, which are currently mostly studied in expensive state-of-the-art electron microscopes [312].

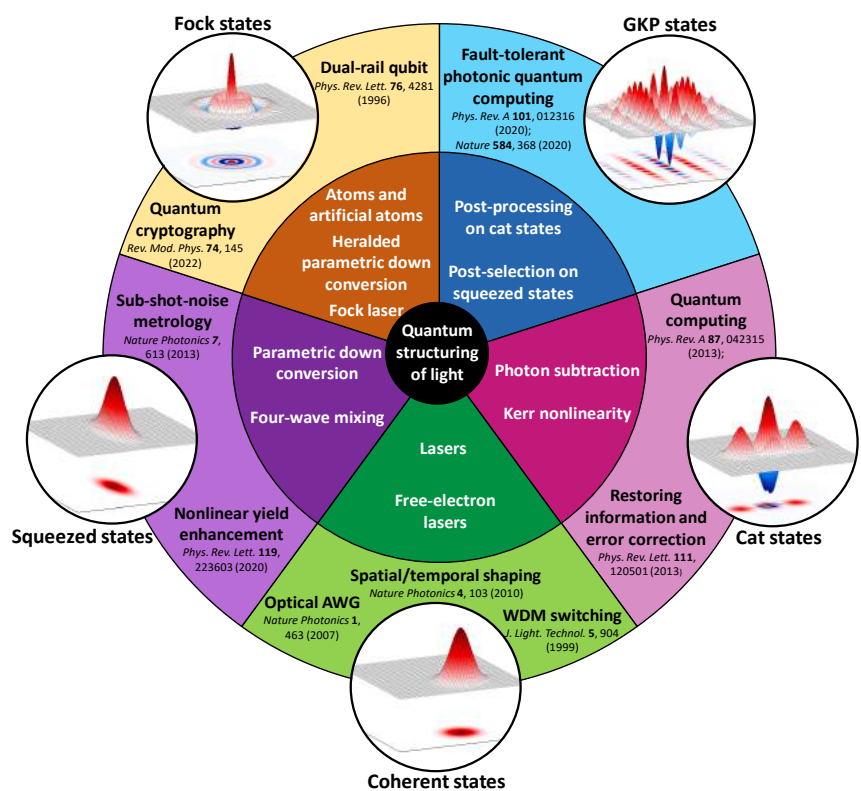

**Figure 2. Exemplary quantum states of light along with selected generation methods (inner circle) and examples of applications (outer circle).** The plots along the outer circle present example single-mode Wigner functions of the corresponding light states. The text in the inner circle describes several known and theorized methods of generation of the quantum light states. The text in the outer circle provides examples of current and potential applications of the states. The special case of classical light – the (Glauber) coherent state – is nested in the green segment. All other segments contain quantum states that cannot be described classically, with example applications in quantum metrology, computing, and communications.

**Advances in Science and Technology to Meet Challenges**

Considering the described challenges, we highlight a few selected paths toward the full quantum structuring of light:

1) The use of photon number resolving detectors is critical for the current post-selection schemes enabling the creation of low-number quantum light states [304]. Efforts for development of better detectors capable of resolving higher photon numbers will propel the frontiers of the field toward the goal of generating many-photon quantum states of light.

2)  The ability to generate quantum light using optical nonlinearities depends on the ratio between the nonlinearity coupling efficiency and losses. This ratio is being gradually improved using better quality nonlinear microcavities, nonlinear photonic crystal fibres and integrated-photonic waveguides [382]. An utterly different method is to utilize mechanisms of extreme nonlinear optics, such as the ones behind high-harmonic generation, which may lead to strong emission of quantum light, bypassing the limitations in efficiency and losses [383].

3)  Light-electron interaction schemes for quantum light shaping can become more practical by miniaturizing high-quality electron sources and integrating them with existing methods for shaping electron wavepackets [313]. Additional advances in nanophotonic-based electron–light couplers are necessary to increase the intrinsic interaction strength, an important requirement for generating many-photon states.

**Concluding Remarks**

The development of generic methods for shaping the quantum state of light, akin to those available for the spatial and temporal degrees of freedom of light, could spawn a great variety of applications in quantum optics and the wider fields of quantum technologies. While such methods may not be immediately in sight, the many advances in quantum optics over the past decade leave us optimistic about the prospects for structured quantum light in the coming decade.

## 23. High-dimensional quantum communication

*Ebrahim Karimi*

University of Ottawa

**Status**

Electromagnetic (EM) waves are widely used in our global communication network to transmit information through free-space, underwater, and fibre networks. EM fields—photons in the quantum regime—are the main 'resource' for classical and quantum communication infrastructures because they do not possess a net electric charge and rest mass. Generation, manipulation, transmission, and detection of the EM field's internal degrees of freedom (DOF) play curtail roles in both classical and quantum communications. For instance, the classical communication bandwidth is directly proportional to the dimension in which the information is encoded and logarithmically depends on the detection signal-to-noise ratio ($S/N$), respectively—according to the Shannon–Hartley theorem, the channel capacity is given by $k \operatorname{Log}_2(1 + S/N)$, where $k$ is the communication bandwidth [314]. This clearly indicates the importance of multiplexing in order to increase the communication bandwidth.

EM fields possess several DOF, including polarisation, frequency, amplitude, phase, and spatiotemporal modes, which can be used to share information. An EM field that is a coherent or incoherent superposition of all of these DOFs is referred to as structured light (or structured photons in the quantum regime) [96]. Polarisation is inherently bi-dimensional, and thus can only be used to share one bit ('0' or '1') of information. Meanwhile, frequency, amplitude, phase, and spatiotemporal modes—although completely different in nature—are unbounded, and thus can be used to increase information beyond '0' and '1'. Frequency (wavelength), amplitude, and phase are currently used in telecommunication multiplexing, allowing for the transmission of data at several terabit rates. Moreover, during the past decades with the progress in the generation and detection of orthogonal spatial modes, e.g., Laguerre-Gauss and Hermite-Gauss, the communication channel capacity has been increased by a couple of orders of magnitude, allowing information to be transmitted much faster using spatial mode multiplexing [315].

The security of the current classical communication network is guaranteed by rigorous mathematical algorithms, e.g., Rivest–Shamir–Adleman (RSA), which is mainly based on the difficulties of finding prime factors of integer numbers using classical computers. However, algorithms—such as Shor's algorithm—implemented on a quantum computer would help to break these classical encryption techniques, and thus threatens our current classical encryption techniques. Quantum communication, e.g., quantum cryptography, employing laws of quantum physics provides approaches to monitoring a communication link and verifying the security threats. These methods are based on two main laws of quantum physics: the superposition principle, wherein a quantum entity can be in a superposition of two or more quantum states simultaneously; and the uncertainty principle, in which the measurement of conjugate quantities with arbitrarily high precision is not allowed [316]. The former resulted in a 'no-go' theorem, referred to as the no-cloning theorem, which directly indicates that a quantum state cannot be copied perfectly without introducing noise to both (multiple) copies [317]. Therefore, whenever a "quantum" message/key is shared, the attacker's (namely Eve's) presence introduces an impurity (noise) to the message/key, and by monitoring the noise, one can verify the security threat.

**Current and Future Challenges**

There have been significant advances in quantum communication, both at discrete and continuous variable regimes since the seminal quantum key distribution (QKD) proposal of Bennet and Brassard. Different QKD protocols based on a single photon, entangled photon pairs, attenuated coherent beams, and squeezed light are developed with proper security analysis [316]. Without considering the network architecture challenges, e.g., quantum repeaters and quantum memory, discrete and continuous variables QKD have their own advantages, difficulties, and challenges. Discrete variable QKD allows for long-range distance but limited key rates, while continuous variables have a limited range but provide higher key rates. Employing high-dimensional encryption in discrete variable QKD can potentially improve the key rate, and thus has received much attention over the last few decades, e.g., $d = 2^n$-dimensional encryption provides $n$-bits of information per sifted photon [62,318]. It has been shown that such high-dimensional encryption is more noise-tolerant and more sensitive to Eve's presence—see Figure 1.

For instance, the best quantum cloning machine introduces $\left(\frac{1}{2} - \frac{1}{d+1}\right)$ noise to the cloned wave functions; the introduced error is 0.17 for qubit ($d = 2$-dimensional) encryption, but increases to 0.25 for qutrit ($d = 3$-dimensional) encryption [319]. The classical methods using different DOF to perform multiplexing to increase classical communications would provide qudit states in the quantum regime. This includes complex photonics states with well-engineered spatial and temporal modes, polarisation, and frequency. Photonic polarisation (polarisation qubit) and time-bin (time-bin qudit) can be generated by means of electro-optic and/or optical devices at very high speeds (GHz to THz), which made polarisation and time-bin QKD a promising venue for technological advances. However, implementing both temporal and spatial modes in a practical QKD setup/network, due to the technical difficulties in the fast generation and detection, has hitherto remained a venue to explore and consider. Thus, developing novel linear and nonlinear approaches to generate, manipulate, and determine spatiotemporal modes at a high-speed rate will be highly rewarding but remains a challenging task for communication.

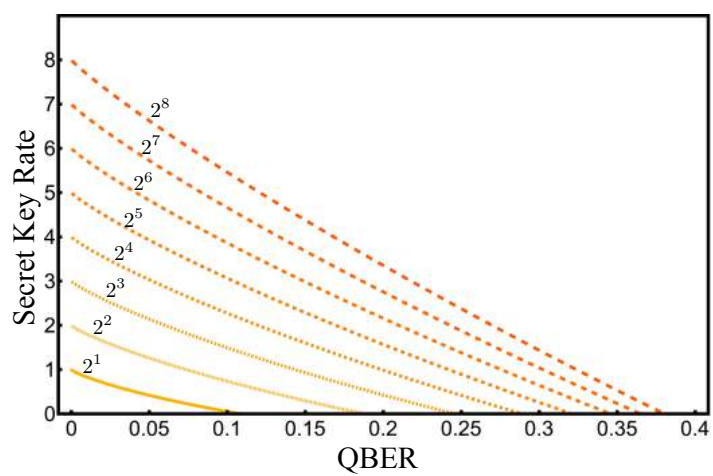

**Figure 1.** Secrete key rate as a function of Quantum Bit Error Rate (QBER) for high-dimensional BB84 protocols. When the QBER is zero, each sifted photon transmits $n = \log d = 2^n$ bits of information, where $n = \{1, ..., 8\}$. However, the key rate decreases exponentially with the noise, and no positive key is shared when the secrete key rate is zero. The QBER for qubit ($d = 2^1$), ququart ($d = 2^2$) and quoct ($d = 2^3$) is about 11%, 19%, and 25%, respectively, and asymptotically reaches 50% for $d \rightarrow +\infty$. This indicates that high-dimensional QKD is robust against noise while providing higher key rates in the QKD protocols.

Another promising venue, not only applicable to quantum communication but for photonics quantum computing, is to explore the generation of multi-photon high-dimensional entangled states [320]. In QKD, access to multi-photon high-dimensional entangled states grants certain protocols, e.g., device-independent QKD, enhanced security of the communication channel. Furthermore, high-dimensional quantum communication channels—fibre [321], free-space [59-60], underwater [61], and satellite [322] links—either need to be developed or fully characterized. For instance, one may need to extend 'vortex'- or 'twisted'-like fibres to support higher spatiotemporal quantum states without introducing significant distortion or crosstalk among the modes or develop dynamic real-time mode-distortion analysis in the current fibre network to recover the coherency among the transmitted qudit states. Of course, transmitting the photonics system in more complex media, such as turbulent and scattering channels or non-flat curved spacetime geometry [323], needs to be properly investigated theoretically and experimentally. This channel analysis will help us to use the desired quantum communication protocols, allowing one to handle certain noise thresholds, which without knowing the channel process matrix the communication key rate would decrease significantly.

**Concluding Remarks**

Engineering photonic quantum states provides access to high-dimensional vector spaces which can enhance the communication key rate. In addition, in the quantum regime, employing qudit QKD allows for the implementation of non-trivial and novel protocols that would enhance the security further or are more noise tolerant. There are three main obstacles or studies, in general, to be considered for the future of high-dimensional quantum communication: (1) linear or nonlinear methods to quickly manipulate/generate single photon, multi-photon entangled states, or in general, non-classical light; (2) explore different (QKD) protocols to optimise the key rate and noise tolerance after sifting as well as optimising the communication physical range, and (3) develop new fibres or methods to use current fibre infrastructure to support more spatiotemporal modes. Of course, the former includes different communication channels, e.g., ground-to-satellite, submersible-to-satellite, and intra-satellites, where the physics is less explored, and thus extensive experimental and theoretical studies must be conducted.

**Acknowledgements**

EK acknowledges the support of Canada Research Chairs, Ontario's Early Research Award, and NRC-uOttawa Joint Centre for Extreme Quantum Photonics (JCEP) via the High Throughput and Secure Networks Challenge Program at the National Research Council of Canada.

## 24. Simulating quantum systems with structured waves

*Filippo Cardano and Lorenzo Marrucci*

Università di Napoli Federico II

**Status**

Large quantum systems cannot be generally simulated through classical computers, as the number of required resources increases exponentially with the system size. As first suggested by Richard Feynman, a quantum system could be simulated more efficiently by another controllable quantum system. Nearly 50 years after this proposal, many quantum-simulating machines have been developed, ranging from proof-of-principle devices to rather sophisticated hardware that can handle tens (even hundreds) of qubits or bosonic modes and perform complex tasks. Besides the original motivation to overcome the computational complexity of large quantum systems, these simulators keep providing us with an extraordinary machinery to investigate quantum phenomena in controllable and accessible architectures [324].

Here we will briefly discuss how the accurate control of structured waves at the quantum level has contributed to this field, focusing on open issues and challenges, and trying to identify key technological steps that could boost the capabilities of current machines, or pave the way to novel approaches. To this end, let us consider the simulation of a single particle whose wavefunction $|\psi\rangle$ is defined in a discrete Hilbert space of dimension $N$, i.e. $|\psi\rangle = \sum_{i=1}^{N} c_i |\psi_i\rangle$, where $|\psi_i\rangle$ form a complete set of states and $c_i(t)$ are complex time-dependent coefficients evolving according to an Hamiltonian operator $H$. As an example, these states may correspond to the positions that an electron can take in a crystalline lattice (see Fig. 1(a)). In the simulation, the basis kets $|\psi_i\rangle$ can in turn be associated with distinct wave states, for example localized at different spatial positions (e.g. photons propagating in distinct waveguides within an integrated network). A viable alternative is provided by structured waves, corresponding to modes (or superposition of modes) having envelopes that are spatially overlapping, yet they feature a distinct internal mode structure that makes them orthogonal. Relevant examples of the latter approach are optical or atomic modes carrying quantized values of orbital angular momentum (OAM) or of transverse linear momentum, whose superposition leads to complex spatial patterns due to multi-mode interference.

Our ability to tailor the evolution of such structured waves, e.g. for the simulation of tight-binding Hamiltonians by coupling structured modes of atoms and photons via controlled momentum kicks (see Fig. 1(b,c)), have enabled the direct observation of many physical phenomena. Remarkable examples are optical analogues of the anomalous Quantum Hall effect [58], the formation of complex patterns in interacting atomic condensates [325], or uncharted effects at the boundary between quantum mechanics and general relativity [326]. Other relevant experiments will be mentioned in the following, yet an exhaustive review of relevant literature is out of the scope of this article. For the same reason, we will not discuss the case of temporally-structured waves, as for instance trains of optical pulses recently used to achieve demonstrate a quantum advantage in a photonic simulation [357].

**Current and Future Challenges**

Quantum simulators based on structured waves have not reached yet the level of complexity that would undermine their classical simulation. A first issue is that most platforms effectively simulate single-particle properties, often by using classical coherent fields. To increase the system complexity, and to observe genuine quantum features, it would be crucial to implement quantum states of many indistinguishable particles.

If one critical issue is associated with the quantum nature of the structured wave, another important aspect is related to the possibility of engineering complex Hamiltonians, with particular attention to the case of interacting systems. Strongly correlated many-body states can be nowadays realized in diverse simulators, thanks to the natural interactions that take place between particles in the simulator, such as for instance trapped atoms or ions. These powerful setups are currently much more expensive than photonic counterparts, yet the latter have only permitted simulation of mean-field approximations of many-body systems thus far, thanks to light propagation through structured nonlinear media [327].

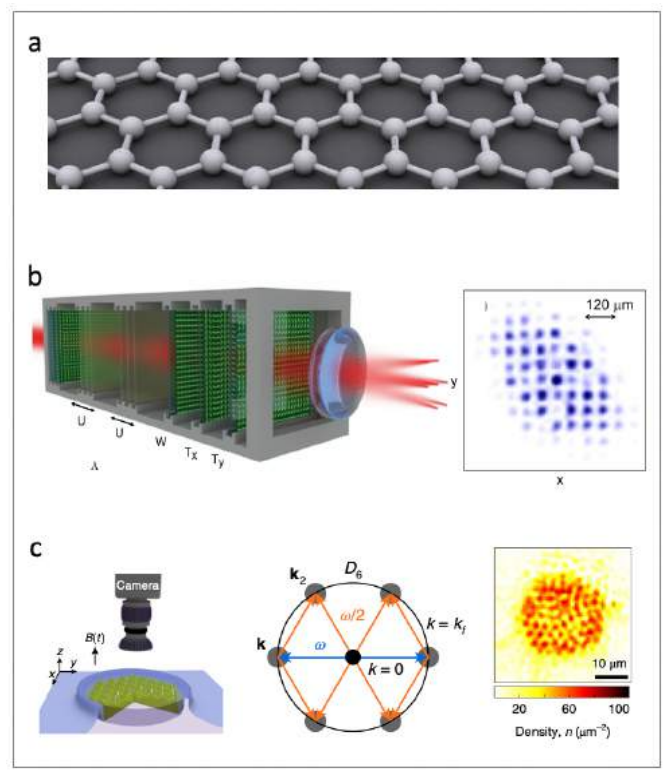

**Figure 1.** (a) Typical honeycomb lattice reproducing the position of carbon atoms in a graphene layer. (b) Liquid-crystal g-plates are arranged to give photons transverse momentum kicks to simulate 2D quantum walks [58]. In the focal plane of a lens light is distributed in spots forming a square lattice. (c) Atoms in a Bose-Einstein condensate are given linear momentum kicks by exploiting suitable optical pulses, according to the coupling scheme shown in the central part. The interference of these multiple waves results in a complex pattern, as shown to the right [325]. Figures sources: (a) from https://en.wikipedia.org/wiki/Graphene (CC BY-SA 3.0 license), (b) from Ref. [57], (c) from Ref. [325].

As discussed above, the interest in these simulators does not only depend on the complexity of their classical simulation, but it is also related to our ability to use them to study novel states of matter, possibly designed artificially to address a specific task. While structured waves have already proved key to the demonstration of a variety of topological effects, it would be crucial to remove barriers that nowadays force us to implement only a certain class of Hamiltonians. First, these are often translation-invariant, which prevents the simulation of finite systems with open boundaries, or random unitary evolutions that are at the basis of sampling experiments. Next, the simulators have some degree of tunability, but they are not completely reconfigurable, which is a crucial ingredient in view of the realization of universal simulators. Finally, when simulating out-of-equilibrium dynamics or when simply simulating a particle that explores large lattices, it is important to be able to control long temporal evolutions, preserving the coherence of the quantum state against decoherence mechanisms and avoiding losses in photonic experiments, or coherence decay between different atomic states.

**Advances in Science and Technology to Meet Challenges**

Generic simulators can be conceptually divided into three functional blocks: encoding, manipulating, retrieving.

   **Encoding –** Photonic simulations of complex quantum systems require multi-particle states made of indistinguishable photons. Important progress has been achieved in recent years, with state-of-the-art quantum-dot sources that can provide Fock states with ∼10 photons. While most implementations have focused on single-particle physics, which keeps providing new exotic and uncharted properties, we anticipate a growing use of multi-particle states. Recently, the use of squeezed quantum states with many photons has attracted broad attention, e.g. with the first demonstration of quantum supremacy in an optical processor, and promise exciting possibilities [328]. In general, setups based on structured waves also require being able to manipulate simultaneously many modes associated with diverse degrees of freedom (position/momentum, time/energy, polarization, atomic internal structure, etc.) to increase the number of synthetic dimensions [329].

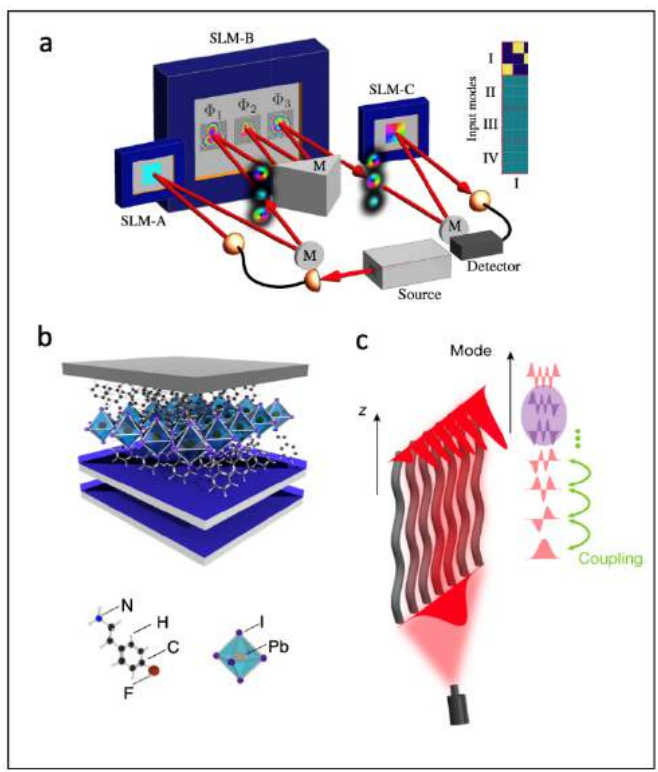

Figure 2.  (a) Multiple phase modulations allow one to mix suitably a set of modes carrying orbital angular momentum. (b) Perovskite crystals in a planar microcavity are arranged so that polaritons dynamics in the sample obey 2D topological Hamiltonians, with a carefully designed Berry curvature. (c) Simulation of a 2D topological insulator within a 1D array of optical waveguides. The second spatial dimension is provided by internal spatial modes, shown on the right, that are coupled thanks to the bending of the waveguides. Figure sources: (a) from Ref. [330], (b) from Ref. [332], (c) from Ref. [329].

   **Manipulating –** In the framework of non-interacting Hamiltonians, the possibility to explore novel dynamics will rely on our ability to controllably mix spatial modes of structured waves [330] with reconfigurable platforms (see Fig. 2(a)), e.g. as recently shown in Ref. [331] for simulating tunable spin-orbit Hamiltonians. Control of the spatial structure of materials embedded in optical microcavities (see Fig. 2(b)) is also enabling the simulation of topological Hamiltonians [332]. It will also be crucial to reduce optical losses for the simulation of multi-particle systems and long temporal evolutions, for instance by implementing the entire evolution with a reduced number of suitably engineered 3D devices, instead of a long sequence of 2D elements. The same concept could apply to atomic systems, by replacing pulses implementing single momentum kicks with more complex long-range excitations equivalent to a superposition of multiple kicks. When coming to the simulation of interacting systems, new approaches to engineering interaction in synthetic atomic dimensions will enable access to unexplored many-body physics [333]. On the optical side, it is well known that photons cannot easily mimic interacting systems. In this sense, developments in the engineering of photon-matter couplings will be crucial here, as recently shown in [334].

**Detection –** In a simulator hosting multi-particle quantum states, retrieving at any time the associated wavefunction would be an important feature. In photonic systems, it is often sufficient to count photons that are simultaneously in different modes, yet this is not equivalent to retrieving the whole quantum state. Challenges are both associated with the discrimination of multiple photons per mode and with managing systems made of many modes. Ideally, it would be desirable to have "quantum cameras" capable of discriminating the number of photons impinging on each pixel, determining both their spatial position and arrival-time, to allow a full retrieval of correlation maps.

## Concluding Remarks

The search of novel approaches to structuring wave propagation and confinement in atomic, photonic or hybrid systems will surely provide us with the possibility of simulating novel quantum states and engineer complex Hamiltonians. We conclude our analysis by focusing on recent hybrid approaches that mix localized modes (as, for instance, those confined in optical cavities) with the synthetic dimensions provided by structured waves. As an example, a proof-of-principle simulation of a 2D topological insulator has been recently performed in a 1D array of optical waveguides, by using different guided modes within the same waveguide as an additional synthetic dimension [329] (see Fig. 2(c), where the guided-mode internal structure is revealed). In this direction, all the techniques that have been developed to control structured waves are now ready to be embedded in simulators based on localized excitations, paving the way to experiments involving a much larger number of modes and simulating quantum systems in high spatial dimensions.

## Acknowledgments

We thank Pietro Massignan and Alexandre Dauphin for useful discussions and suggestions.

## 25. Artificial Intelligence for Structured Waves

*Mario Krenn and Florian Marquardt*

Max Planck Institute for the Science of Light

**Status**

The additional resources provided by the complexity of structured waves offer enormous potential for technological applications and new scientific research. However, this potential comes with a price. The increased complexity is challenging for human intuition and classical algorithmic approaches when dealing with the design or control of new systems or large-scale data analysis. The community recently started developing and employing methods from the domain of machine learning, or more broadly, artificial intelligence (AI), to overcome these new complications. These methods have seen revolutionary progress over the last decade, solving practical real-world challenges. Here, we will offer a big picture of the current applications to structured waves, followed by an outline of exciting opportunities that might lie ahead of us.

**AI for Data Analysis:** The experimental classification of structured waves can be challenging. An example is long-distance communication, where spatial structures are used to increase the information density of light. Those structures are strongly influenced by atmospheric turbulence, making it challenging to decode the information. Here, machine learning can help classify the modes even though they are distorted, as shown in a 143-kilometre turbulent link [335]. The combination of polarisation with spatial modes leads to highly complex vector vortex beams. Their classification with classical methods is expensive and can be enhanced by neural networks to classify the complex polarisation patterns with high quality [336]. Other works exploited neural networks to characterise spectral structures of beams and identify spectral instabilities upon propagation through optical fibres or improve the high-quality recognition of vortices in beams, for example by correcting aberration introduced by turbulence [358].

Rather than characterising the beams themselves, machine learning has also significantly advanced the spatial-waves-based analysis of objects. For example, the field of deep microscopy aims to advance the resolution of classical microscopy images solely by computation. A series of experiments showed how neural networks could improve single-pixel cameras by autonomously learning the ideal illumination patterns of the objects [337]. A very active field of research deals with wavefront shaping and imaging using complex media. Deep learning has become a powerful tool to replace classical, deterministic algorithms and directly learn from and act according to measured data [338].

**AI for Design:** Due to their complexity, the design of photonic or band-gap structures or quantum experiments can become infeasible for human researchers. Therefore, various AI approaches have been applied to this challenge in recent years. One example is experimental designs for high-dimensional multi-particle entangled systems [339]. This has led to numerous new experimental results involving the first observation of a multilevel GHZ state and the discovery of new entanglement and interference principles. An advanced and highly active field of research deals with designing new photonic nanostructures and metamaterials with AI. This field has recently been pioneered by gradient-based and deep learning methodologies [340]. In the search for new topological phenomena, AI systems have massively sped up the computation of the resulting properties directly from the geometric structure of photonic crystals [341], which can be used for rapid exploration and optimization. In a similar spirit, neural networks helped to predict ultrafast nonlinear dynamics in fibres [342].

**Current and Future Challenges**

For applying AI technologies in science in general, and for structured-wave research in particular, five critical points need to be considered:

**1. Data**: Large, high-quality labelled datasets are necessary for many ML approaches (supervised learning). Usually, even for simple problems, 100s or more examples are necessary. The dataset also needs to avoid any biases as neural networks are extremely good at exploiting those weaknesses. For example, the selection of the training examples should be random, from an appropriate distribution (and it is not always clear what the right distribution is). Furthermore, the data needs to be labelled. Hand-annotating 1000s of examples is infeasible in most situations. An alternative is to generate simulated or experimental data where labels, for example, correspond to experimental outcomes.

**2. Data from Simulations**: Experiments often are too expensive for generating a lot of training data; thus, simulators of the physical system are powerful. It is crucial that the simulators reflect the physical situations well, as neural networks will learn the biases of the simulator. Another issue might be that the simulator is very slow, making it difficult to generate a large amount of training data.

**3. Data from Experiments:** If experimental data can be collected for training data, biases need to be avoided as well. For example, the environment or parameters of the setup might change over time, which needs to be taken into account.

**4. False positives and negatives:** If neural networks are used in the experimental data analysis, they might make wrong predictions. For example, in NN-enhanced microscopy, the machine could predict an interesting effect in the image where there is none ("false positive"). This can be corrected by human intervention. More dramatic are situations where the machine overlooks an interesting effect in a microscopy image. This question is specifically important when the machine is used in a system for which there is no training data (out-of-distribution).

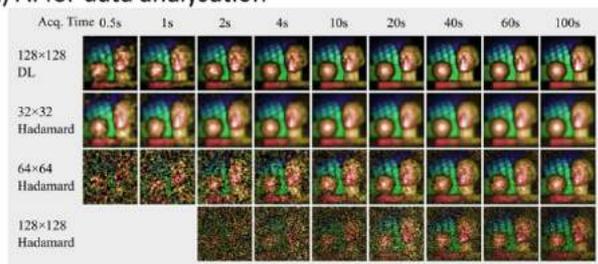

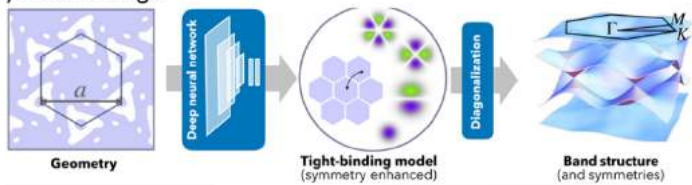

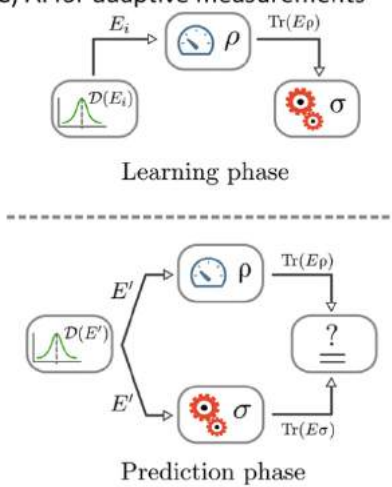

**Figure 1.** Three different applications of AI in the realm of structured waves. A) For data analysation. Here, the illumination patterns of a 3d single-pixel camera are constructed via deep neural networks and lead to significantly better results than classical methods. [337] B) For Design questions. Here, periodic nanostructures with interesting topological band structures are designed using deep neural networks [341]. C) For adaptive measurement. In the learning phase, the algorithm outputs a hypothesis of the state from randomly drawn measurements. Afterwards, the goal is to predict the outcome of a randomly chosen measurement. The "probably approximately correct model" method can speed up experimental quantum state estimation [343].

**Advances in Science and Technology to Meet Challenges**

Advances in technology can solve several of these challenges. The requirement of large datasets can be overcome by transfer learning. The system learns in simpler situations, and the pre-trained system requires only a small number of examples to adapt to the more challenging situation.

One promising direction is the application of AI for choosing the optimal next measurement to gain maximum information. Pioneering work on adaptive measurements for structured waves [343] considered the quantum state of a complicated 2-particle vector-vortex beam. Conventionally, the number of measurements scales exponentially with the qubit number for quantum state tomography. However, the authors applied an adaptive measurement procedure that allowed a "probably approximately correct" estimate of the quantum state in just a linear number of measurements.

In many situations, the decisions need to be fast (e.g. light through turbulence), which requires additional considerations. The requirement for rapid neural networks (for instance, in active control of experimental systems) can be achieved by hardwiring the neural network directly in FPGAs. One exciting alternative is the use of optics and photonics itself as a speed-up for AI, either as a co-processor or an end-to-end computing machine [344].

Many novel developments in photonic technologies, such as new modalities of sensors, could potentially benefit from the advanced capability of AI for data analysis. Similarly, we believe that a large degree of freedom in the automated control of experimental components (such as variable detection schemes or automated alignment via piezo-controlled mirrors) might enable fully AI-powered alignment and control systems of experimental setups. Even the experiments' design could be aided by AI [339-342].

Artificial intelligence is a culmination of very diverse, advanced computational algorithms that can be applied in a wide variety of fields—exemplified by the concrete examples above. Those techniques have been proven to be enormously powerful by computer scientists. The question is which new scientific and technological problems can we address with AI? In some way, we have a mighty hammer and need to identify suitable nails.

**Concluding Remarks**

With the emergence of AI, we have a powerful tool at our disposal with the potential to advance open questions in the study of complex structured waves. While numerous and diverse questions have already been addressed with AI, we believe that a vast amount of intriguing applications are yet to be explored. Finding new nails for this powerful hammer is a challenge that a community effort might best address. A great opportunity to do so is the biannual community conference ICOAM. A half-day online workshop with focused discussion groups on predefined topics could bring together domain experts on the theory and experiments of structured waves and researchers with expertise and a broad overview of AI technologies and their applications in science. AI approaches work well when large amounts of data are available, where huge search spaces hinder systematic scanning or where computationally very expensive functions are involved in the data analysis. Given the complexity of structured waves, we strongly expect exciting applications of AI to emerge from such collaborations.